\documentclass[aps, pra, superscriptaddress, reprint]{revtex4-2}
\usepackage{graphicx}
\usepackage{overpic}
\usepackage{ulem}
\usepackage{amsthm,amsfonts,amstext,amssymb,amsmath}
\usepackage{latexsym}
\usepackage[dvipsnames]{xcolor}
\usepackage{braket}
\usepackage{bbold}
\usepackage{dsfont}
\usepackage{bm}
\usepackage{placeins}
\usepackage{enumerate}
\usepackage{standalone}
\usepackage{float}
\usepackage{xcolor}
\usepackage{mathtools}
\usepackage{amssymb}
\usepackage{hyperref}
\hypersetup{colorlinks=true,
  linkcolor=blue,
  citecolor=blue,
  urlcolor =blue}
\usepackage{tcolorbox}

\renewcommand{\vec}[1]{\mathbf{#1}}

\newcommand*{\ketbra}[2]{\ensuremath{\ket{#1}\bra{#2}}}

\newcommand*{\tr}[2][]{\ensuremath{\textrm{Tr}_{#1}\left[ #2 \right]}}

\newtheorem{proposition}{Proposition}

\begin{abstract}

Quantum-enhanced sensors, which surpass the standard quantum limit (SQL) and approach the fundamental precision limits dictated by quantum mechanics, are finding applications across a wide range of scientific fields. This quantum advantage becomes particularly significant when a large number of particles are included in the sensing circuit. Achieving such enhancement requires introducing and preserving entanglement among many particles, posing significant experimental challenges. In this work, we integrate concepts from Floquet theory and quantum information to design an entangler capable of generating the desired entanglement between two paths of a quantum interferometer. We demonstrate that our path-entangled states enable sensing beyond the SQL, reaching the fundamental Heisenberg limit (HL) of quantum mechanics. Moreover, we show that a decoding parity measurement maintains the HL when specific conditions from Floquet theory are satisfied—particularly those related to the periodic driving parameters that preserve entanglement during evolution. We address the effects of a priori phase uncertainty, imperfect transmission, and other types of noises, showing that our method remains robust under realistic conditions. Finally, we propose a superconducting-circuit implementation of our sensor in the microwave regime, highlighting its potential for practical applications in high-precision measurements.
\end{abstract}

\begin{document}

\title{Bosonic Entanglement and Quantum Sensing from Energy Transfer in two-tone Floquet Systems}

\author{Yinan Chen}
\email{yinanc@caltech.edu}
\affiliation{Institute of Quantum Information and Matter, California Institute of Technology}
\affiliation{Department of Physics, California Institute of Technology, Pasadena CA 91125, USA}
\author{Andreas Elben}
\email{andreas.elben@psi.ch}
\affiliation{Institute of Quantum Information and Matter, California Institute of Technology}
\affiliation{Walter Burke Institute for Theoretical Physics, California Institute of Technology}
  \affiliation{Laboratory for Theoretical and Computational Physics and ETHZ-PSI Quantum Computing Hub, Paul Scherrer Institute, CH-5232 Villigen-PSI, Switzerland}
\author{Angel Rubio}
\affiliation{Max Planck Institute for the Structure and Dynamics of Matter, Luruper Chaussee 149, 22761 Hamburg, Germany}
\affiliation{Center for Computational Quantum Physics, Simons Foundation Flatiron Institute, New York, NY 10010 USA}
\author{Gil Refael}
\email{refael@caltech.edu}
\affiliation{Institute of Quantum Information and Matter, California Institute of Technology}
\affiliation{Department of Physics, California Institute of Technology, Pasadena CA 91125, USA}
\affiliation{AWS Center for Quantum Computing, Pasadena, CA, 91125, USA}
\date{\today}
\maketitle

\section{Introduction}

Quantum sensors are widely used today in both scientific research \cite{Aasi2013,PhysRevLett.123.231107,PhysRevLett.132.190001} and technical applications \cite{Aslam2023,Ahn2020,Chaste2012}. For an input with $N$ particles, traditional quantum-enabled sensors utilize uncorrelated (classical) states, which limits their sensitivity to the standard quantum limit (SQL), $\frac{1}{\sqrt{N}}$, dictated by the Heisenberg uncertainty principle \cite{Giovannetti2011}. In contrast, quantum-enhanced sensors employ entangled states of many particles, allowing them to surpass the SQL and reach the ultimate limit set by quantum mechanics, known as the Heisenberg limit (HL), $\frac{1}{N}$ \cite{doi:10.1126/science.1104149}.
Typically, in an interferometric setup, such quantum-enhanced sensors use entangled qubits \cite{PhysRevA.46.R6797,PhysRevA.54.R4649,Marciniak2022} or photons \cite{Demkowicz2015,PhysRevLett.99.070801} to probe an unknown parameter encoded in a quantum operation. Among these states, the Greenberger–Horne–Zeilinger (GHZ) state \cite{doi:10.1126/science.1104149,doi:10.1126/science.1097576} in qubit systems and the N$00$N state \cite{PhysRevLett.99.070801,doi:10.1080/00107510802091298} in optical systems—defined as $\ket{\text{N00N}}\equiv\frac{1}{\sqrt{2}}\left(\ket{N,0}+\ket{0,N}\right)$ over two photonic modes with a total of $N$ photons — was shown to precisely achieve the HL. Other types of entangled states, such as the entangled coherent state \cite{PhysRevLett.107.083601}, Bat state \cite{PhysRevLett.107.083601,10.1116/5.0026148,doi:10.1080/00107514.2010.509995}, and coherent light mixed with a squeezed vacuum state \cite{PhysRevLett.100.073601}, have also been demonstrated to asymptotically reach the HL at large photon numbers.
Recent experiments have demonstrated the generation of such entangled states of photons or phonons, typically with a particle number $N<10$. These states can be generated through phonon manipulation in trapped ion systems \cite{PhysRevLett.121.160502}, parametric down-conversion and post-selection \cite{Zhong2018,Mitchell2004,Xiang2011}, multiphoton entanglement
with deterministic single-photon sources \cite{Li2020}, mixing quantum and classical light \cite{doi:10.1126/science.1188172}, and employing variational optical non-linearities \cite{Munoz2024, Munoz2024arxiv}. However, developing efficient protocols to generate entangled states with a higher number of bosons and stabilizing the entanglement against experimental imperfections, such as photon loss and noise, remains a significant challenge.

The ability to engineer exotic states and phases of matter in controllable ways gave rise to extensive theoretical \cite{PhysRevB.79.081406,Lindner2011,PhysRevB.87.235131,PhysRevResearch.4.043060} as well as experimental \cite{doi:10.1126/science.1239834,Rechtsman2013,Peng2016} efforts. These often involve driving materials to induce novel behavior that lacks a static counterpart. For instance, spins driven adiabatically by fields with different frequencies give rise to quantized energy transfer between the drives, effectively realizing a temporal version of the quantum Hall effect \cite{PhysRevX.7.041008,PhysRevResearch.2.022023,PhysRevResearch.4.013169,PhysRevLett.119.123601,PhysRevX.3.031005,PhysRevX.6.021013,PhysRevB.96.155118,PhysRevB.93.245146}.
When treating the drives in such a scheme as quantized, their state evolves as well, and the interaction with a spin or other system elements could be used as a tool to engineer the quantum state of the drives \cite{PhysRevResearch.2.043411,PhysRevB.99.094311,PhysRevLett.128.183602}.

In this work, we leverage insights from Floquet engineering to design a \textit{Floquet quantum sensor} that uses periodic driving to prepare an entangled state suitable for a sub-SQL sensitivity. We relate the sensing properties of the engangled state to the quasi-energy spectrum of the Floquet system used to create it. We show that our protocol can generically produce entangled N-photon states at a time $T\sim N$, as opposed to the typical-quantum diffusion rate which results in $N$ entangled photons at times $T\sim N^2$. Also, we show that our protocol can produce entangled states capable of sub-SQL measurement resolution for phase estimation even with coherent states as initial inputs.

As a starting point, we consider a quantum system subject to two classical periodic drives. We demonstrate that the direction of energy transfer between the two drives is determined by the initial state of the system. Treating the drives quantum-mechanically, this allows to generate entanglement within two modes of bosons over time. We quantify this entanglement using quantum Fisher information (QFI) \cite{Helstrom1969,Carollo_2018,DiFresco2024}, which serves as a global lower bound on the sensitivity dictated by the quantum Cramér–Rao bound (qCRB) \cite{Pezz2018,Braunstein1994}. We show that the QFI of our path-entangled states, generated by energy transfer in Floquet systems, can be computed from the quasienergy band structure \cite{annurev:/content/journals/10.1146/annurev-conmatphys-031218-013423} of the Floquet Hamiltonian, and typically scales with the preparation time as $T^2$ up to $T=O(N)$, when the HL scaling $N^2$ is realized. Furthermore, we demonstrate that for drives with identical frequencies, a Mach-Zehnder interferometer (MZI) followed by a parity measurement \cite{10.1116/5.0026148,doi:10.1080/00107514.2010.509995} on the output bosons can read out the QFI with at least SQL precision. We establish conditions on the quasienergy bands necessary for maintaining such $T^2$ growth in the (classical) Fisher information using parity measurement, which at $T=O(N)$ enables readout with HL precision. Additionally, we address the effects of the a priori uncertainties in the parameter (phase) to be measured \cite{PRXQuantum.4.020333,Marciniak2022,PhysRevX.11.041045} as well as particle loss in imperfect transmissions \cite{PhysRevLett.102.040403,Escher2011,PhysRevA.80.013825}.
Our analysis suggests that our sensing protocol based on energy transfer can, in principle, be generalized to an arbitrarily large number of particles. Combined with the scalable readout by parity measurement, we anticipate that our sensing protocol can be realized in near-term experiments.

 The paper is structured as follows. In Sec.~\ref{Information theory}, we briefly review basic notions of quantum metrology. In Sec.~\ref{review}, we discuss the energy transfer in two-tone Floquet systems with classical driving fields. In Sec.~\ref{quantum polarization conversion}, we quantize the drives as bosonic quantum fields and demonstrate that path-entangled states with QFI approaching the HL at $T=O(N)$ can be generated, employing properties of the quasi-energy band structure. We apply our results to three examples of driven qubits. Next, in Sec.~\ref{parity as decoder}, we show that these path-entangled states can be used within a MZI with parity measurement  to reach a sensing precision given the  SQL and even the HL, if certain conditions satisfied. In Sec.~\ref{Loss}, we study the effect of a priori uncertainty in the sensing parameter and imperfect transmission (photon loss) in the interferometer. Finally, in Sec.~\ref{Experiment realization}, we elaborate on the realization of the Floquet sensor, including an entangler, a microwave beam splitter, a phase imprinter, and the parity measurement, within a superconducting circuit design.
 

\section{Quantum sensing and the quantum Fisher information}
\label{Information theory}

\subsection{Quantum Fisher information}

In this section, we review basic notions of quantum metrology and quantum estimation theory \cite{toth2014quantum}. We focus on the estimation of a single parameter (phase) $\theta$ encoded via a unitary transformation $\rho_0 \rightarrow \rho_\theta = e^{-i\theta A} \rho_0 e^{i\theta A}$ on a probe quantum state $\rho_0$ prepared in a quantum sensor.  Throughout the main text, we assume that the phase $\theta$ is \textit{unknown, but fixed} in repeated runs of the experiment. This setting can be analyzed within the quantum Fisher information framework \cite{toth2014quantum}. More generally, the parameter can be treated as a random variable with a probability distribution $P(\theta)$ within a Bayesian approach, see App.~\ref{global phase estimation}.

Our aim is to learn (estimate) the unknown parameter $\theta$ from measurements on $\rho_\theta$. We model such measurement with a POVM $\Pi$ specified by a set of positive Hermitian operators $\{\hat{\Pi}_{\mu}\}$ and corresponding outcomes $\pi_{\mu}$, satisfying $\sum\limits_{\mu=1}^{M}\hat{\Pi}_{\mu}=\hat{\mathds{1}}$, where $M$ is the number of possible measurement outcomes.
Here, the probability $p\left(\pi_{\mu}|\theta\right)$ to observe  outcome $\pi_{\mu}$, given an a priori value of $\theta$, is given by 
\begin{equation}
p\left(\pi_{\mu}|\theta\right)=\tr{\rho_\theta\hat{\Pi}_{\mu}}.
    \label{conditional probability}
\end{equation}
Finally, an estimator function $\theta_{\text{est}}$ is given such that for each outcome $\pi_{\mu}$, the estimated value of the a priori phase $\theta$ is given by $\theta_{\text{est}}\left(\pi_{\mu}\right)$, with mean value over repetitions $\overline{\theta}_{\text{est}}=\sum\limits_{\mu=1}^{M}\theta_{\text{est}}\left(\pi_{\mu}\right)p\left(\pi_{\mu}|\theta\right)$ and variance 
\begin{align}
\Delta{\theta}^2_{\text{est}}=\sum\limits_{\mu=1}^{M}\left(\theta_{\text{est}}\left(\pi_{\mu}\right)-\overline{\theta}_{\text{est}}\right)^{2}p\left(\pi_{\mu}|\theta\right).
\end{align}
The choice of the (classical) estimation function is crucial for the success of the estimation procedure. In the following, we concentrate on locally unbiased estimators which faithfully estimate $\overline{\theta}_{\text{est}} = \theta$ in an (at least infinitesimal) local environment around its fixed value, $\partial\theta_{\text{est}}\left(\pi_{\mu}\right)/\partial\theta = 1$.
In this case, the celebrated Cramer-Rao bound \cite{toth2014quantum} lower bounds the variance of the estimator
\begin{align}
    \Delta{\theta}^2_{\text{est}} \geq \frac{1}{F_\theta[\rho_0, A, \Pi]}.
\label{CRB}
\end{align}
Here, $F_\theta[\rho_0, A, \Pi]$ is the classical Fisher information 
\begin{equation}
    F_\theta[\rho_0, A, \Pi]=\sum\limits_{\mu=1}^{M}\left[\partial_{\theta}\log p\left(\pi_{\mu}|\theta\right)\right]^{2}p\left(\pi_{\mu}|\theta\right).
\label{classical Fisher information}
\end{equation}
We note that $F_\theta[\rho_0, A, \Pi]$ depends on the choice of measurement (the POVM) and probe state $\rho_\theta$. We can optimize  $F_\theta[\rho_0, A, \Pi]$ over all possible POVMs $\Pi$ to obtain the measurement that allows to estimate $\theta$ with minimal uncertainty (variance). The corresponding maximal Fisher information 
\begin{align}
        F_{q}[\rho_0, A]=\sup_{\Pi}F_\theta[\rho_0, A, \Pi]
\end{align}
is called the quantum Fisher information $F_q[\rho_0, A]$ \cite{toth2014quantum,Carollo_2020}. It can be computed explicitly and takes a \textit{$\theta$-independent} form
\begin{equation}
    F_{q}[\rho_0, A]=2\sum_{k,l, \lambda_{k}+\lambda_{l}>0}\frac{\left(\lambda_{k}-\lambda_{l}\right)^{2}}{\lambda_{k}+\lambda_{l}}\left|\bra{k}A\ket{l}\right|^{2},
\end{equation}
where $\lambda_{k}$ are the eigenvalues of the probe state $\rho_0$ with eigenvectors $\ket{k}$ \cite{toth2014quantum}.  For a pure probe state $\rho_0 = \ket{\psi_0}\bra{\psi_0}$ the QFI has a much simpler expression $    F_{q}[\rho_0, A]=4\left(\braket{A^2}-\braket{A}^{2}\right)$. Inserting into Eq.~\ref{CRB}, this yields the quantum Cramer-Rao bound
\begin{align}
     \Delta{\theta}^2_{\text{est}} \geq \frac{1}{F_q[\rho_0, A]},
\label{qCRB}
\end{align}
which fundamentally limits the sensing precision for parameter $\theta$ encoded via an operator $A$ into a quantum state $\rho_0$. For a quantum state $\rho_0$ of $N$ particles, the scaling of the quantum Fisher information with $N$ is of central importance: The standard quantum limit (SQL) $F_q[\rho_0,A]\propto N$ applies to states with only classically correlated (or uncorrelated) particles. It originates from the Heisenberg uncertainty relation for each particle. Entanglement (quantum correlations) between the particles allows us to surpass this and reach the Heisenberg limit (HL) $F_q[\rho_0,A]\propto N^2$.

A lot of research has thus been dedicated to devising and implementing probe states $\rho_0$ with large quantum Fisher information (see \cite{degen2017quantum,Pezz2018} and references therein). Additionally, the QFI involves a non-trivial optimization over all possible POVMs allowed by quantum mechanics, with the optimal POVM often not known or difficult to implement \cite{degen2017quantum,Pezz2018}. A second challenge concerns thus the realization of a measurement $\Pi$ such that the corresponding classical Fisher information is close to the optimal quantum Fisher information. Various examples of such sensors are known in the literature which we will discuss in the next subsection.

\subsection{Examples of quantum sensors}

\begin{figure*}
    \centering
    \includegraphics[width=1\linewidth]{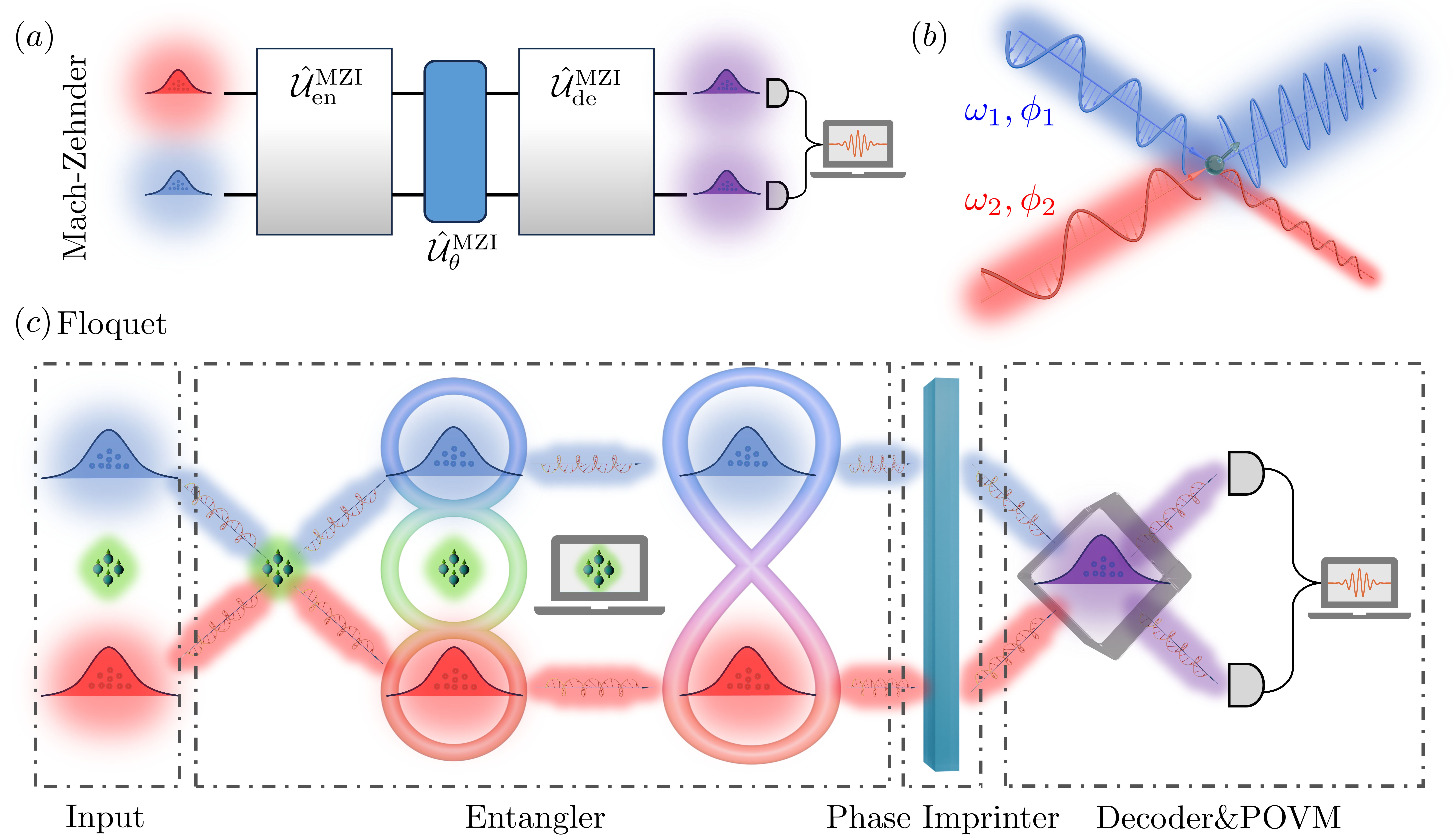}
    \caption{\textit{Quantum sensing from energy transfer}. (a) A circuit diagram of a Mach-Zehnder interferometer: Two modes of input bosons are first entangled by an entangler $\hat{\mathcal{U}}_{\text{en}}^{\text{MZI}}$. An unknown phase $\theta$ is then imprinted by a $\theta$-dependent unitary $\hat{\mathcal{U}}_{\theta}^{\text{MZI}}$. Finally, a decoding circuit $\hat{\mathcal{U}}_{\text{de}}^{\text{MZI}}$ together with a set of POVM measurements are applied to the probe state. An estimator (computer) deduces the value of $\theta$ based on the measurement outcome. (b) Energy transfer between two classical drives with commensurate frequencies $\frac{\omega_{1}}{\omega_{2}}\in\mathbb{Q}$ and initial phases $\phi_{1,2}$. Depending on the initial configuration of the qudit, energy is converted from one mode to the other. (c) Quantum sensor based on energy transfer in two-tone Floquet systems. The circuit differs from a Mach-Zehnder interferometer in the entangler: two modes of input bosons are now coupled to a finite-dimensional ancilla. The entangled output of bosons is then generated by projecting out the ancilla degrees of freedom.}
    \label{Floquet quantum sensor}
\end{figure*}

 Typical quantum sensors built from photons and atoms are the Mach-Zehnder interferometer (MZI) \cite{PhysRevA.59.1615,PhysRevB.82.155303,PhysRevLett.107.083601} and the Ramsey interferometer (RI) \cite{Marciniak2022,PRXQuantum.4.020333,PhysRevX.11.041045,10.1063/5.0121372}.  In these systems, the experimental quantum sensing protocol can be summarized as follows [Fig.~\ref{Floquet quantum sensor}.(a)]:
First, the probe state $\rho_0$ is prepared via an entangling operation $U_{\text{en}}$ starting from a simple initial (product) state of photons or atoms, $\rho_0 = U_{\text{en}}\rho_I U_{\text{en}}^\dagger$. Second, the phase is encoded $\rho_0 \rightarrow \rho_\theta = e^{-i\theta A} \rho_0 e^{i\theta A}$. Lastly, the POVM measurement is realized by a decoding operation $U_{\text{de}}\rho_\theta U_{\text{de}}^\dagger$ followed by projective measurement.\\

 In a Mach-Zehnder interferometer [Fig.~\ref{Floquet quantum sensor}.(b)], two modes of photons $\hat{a}_{1},\hat{a}_{2}$ are sent to a $50-50$ beam splitter represented by $\hat{\mathcal{U}}_{\text{en}}^{\text{MZI}}=e^{-i\hat{J}_{y}\frac{\pi}{2}}$ with $\hat{J}_{y}=\frac{i}{2}\left(\hat{a}_{1}^{\dagger}\hat{a}_{2}-\hat{a}_{2}^{\dagger}\hat{a}_{1}\right)$. The beam splitter separates the output photons to two interferometric paths, in which a phase $\theta$ is imprinted by a unitary operator $\hat{\mathcal{U}}_{\theta}^{\text{MZI}}=e^{-i\hat{J}_{z}\theta}$ with $\hat{J}_{z}=\frac{1}{2}\left(\hat{a}_{1}^{\dagger}\hat{a}_{1}-\hat{a}_{2}^{\dagger}\hat{a}_{2}\right)$. After imprinting the phase, the probe state is sent to another $50-50$ beam splitter given by $\hat{\mathcal{U}}_{\text{de}}^{\text{MZI}}=e^{-i\hat{J}_{x}\frac{\pi}{2}}$ with $\hat{J}_{x}=\frac{1}{2}\left(\hat{a}_{1}^{\dagger}\hat{a}_{2}+\hat{a}_{2}^{\dagger}\hat{a}_{1}\right)$. Finally, a photo-counting or parity measurement of the output photons are applied to decode the phase information. Note that in a Mach-Zehnder interferometer, $\hat{J}_{x,y,z}$ form the Schwinger representation of angular momentum, and hence is an example of the SU($2$) interferometer \cite{PhysRevA.33.4033}. To maximize the QFI $F_{q}= F_q\left[\rho_0,\hat{J}_{z}\right]$ corresponding to the generator $\hat{J}_{z}$, the (pure) probe state $\rho_0$ should maximize  the variance of $\hat{J}_{z}$, i.e.\ of the number difference of photons in the two modes. For an $N$-photon input, the probe state that maximizes such QFI is the N00N state $\ket{\text{N00N}}=\frac{1}{\sqrt{2}}\left(\ket{N}\otimes\ket{0}+\ket{0}\otimes\ket{N}\right)$, where $\ket{N}$ and $\ket{0}$ are the $N$-photon Fock state and the vacuum state, respectively. The N00N state yields a QFI $F_{q}=N^{2}$ that scales quadratically with the number of photons $N$. In comparison, in a classical (i.e.\ with classically non-entangled photons as probes) interferometer, $F_{q}$ can at most scale linearly as $N$ \cite{10.1116/5.0007577}. Entanglement between interferometric paths then improves the QFI by a factor of $N$. In the context of metrology, such entanglement between two interferometric paths is therefore expected to enhance the sensitivity of phase estimation.\\

In a Ramsey interferometer, the input state is instead a direct-product state $\ket{0}\otimes\ket{0}\otimes\cdot\cdot\cdot\otimes\ket{0}$ of $N$ $2$-level atoms. Accordingly, the entangler $\hat{\mathcal{U}}_{\text{en}}^{\text{RI}}=e^{-i\hat{S}_{y}\frac{\pi}{2}}$, phase imprinter $\hat{\mathcal{U}}_{\theta}^{\text{RI}}=e^{-i\hat{S}_{z}\theta}$, and the decoder $\hat{\mathcal{U}}_{\text{de}}^{\text{RI}}=e^{-i\hat{S}_{x}\frac{\pi}{2}}$ are realized via global rotations of the atoms $\hat{S}_{x,y,z}=\frac{1}{2}\sum_{j=1}^{N}\hat{\sigma}_{j}^{x,y,z}$ with $\hat{\sigma}_{j}^{x,y,z}$ the Pauli operators of the $j$-th atom. It is, again, clear from the construction that $\hat{S}_{x,y,z}$ form a representation of the SU($2$) group and hence is another example of the SU($2$) interferometer. The counterpart of the N00N state, which maximizes the QFI in the Mach-Zehnder interferometer, becomes the GHZ state defined as the macroscopically entangled state between atoms $\ket{\text{GHZ}}=\frac{1}{\sqrt{2}}\left(\ket{0}\otimes\cdot\cdot\cdot\otimes\ket{0}+\ket{1}\otimes\cdot\cdot\cdot\otimes\ket{1}\right)$, which gives an $N^{2}$-scaling QFI with respect to $\hat{S}_{z}$.\\

\subsection{This work: Floquet quantum sensor}
\label{This work: Floquet quantum sensor}
In the previous paragraphs, we discussed the MZI and RI, where metrologically useful quantum states are generated by applying entangling gates to the system’s photons and qubits, respectively. Alternatively, it was proposed to couple the system to ancillary degrees of freedom and use operations on the ancilla to entangle the system’s constituents \cite{PhysRevLett.116.220502,Maleki2019,Kang:23}. In line with this approach, we study a setup in this work involving two bosonic modes coupled to an ancilla qubit [Fig.~\ref{Floquet quantum sensor}.(b)]. 

Previous studies showed that in the limit of classical drives—i.e., an ancilla qubit driven by two periodic classical driving fields—the ancilla qubit can mediate an energy transfer between the drives, a process known as frequency conversion \cite{PhysRevX.7.041008,PhysRevB.108.064301}. In this work, we explore the full quantum system and demonstrate that the energy transfer between the classical drives, mediated by the ancilla qubit, leads to the generation of path-entangled states in the quantum system, also mediated by the ancilla qubit. Crucially, these path-entangled states are metrologically significant, with their QFI approaching the Heisenberg limit. We further show that this can be applied in an MZI-like interferometer to estimate phases with Heisenberg-limit sensitivity via a parity measurement under specific conditions.

In short, we propose a Floquet quantum sensor [Fig.~\ref{Floquet quantum sensor}.(c)] as follows: Two initially uncorrelated modes of bosons with, in total, $N$ bosons, for instance a product state of two coherent states or Fock states, first interact with a qubit (or qudit) for a time $T$. Following this interaction, a projective measurement on the qubit (or qudit) generates path-entangled bosonic modes as the output state. A phase  $\theta$  is then encoded using  $\hat{\mathcal{U}}_{\theta}^{\text{MZI}}$  as in the MZI. Finally, a readout circuit, comprising a decoder  $\hat{\mathcal{U}}_{\text{de}}^{\text{MZI}}$  followed by a parity measurement, is applied to the phase-encoded state. Under certain conditions on the initial states of the bosonic modes and their interaction with the qubit  (see below), this allows to estimate the phase $\theta$ with sensitivity $\Delta^2 \theta \sim 1/T^2$ improving quadratically in $T$ and reaching the HL at times $T=O(N)$.

\section{Energy transfer between  classical drives}
\label{review}

In this section, we discuss the energy transfer between two classical fields driving a single qubit acting as a mediator.  We assume the qubit is governed by static Hamiltonian $\hat{H}_{0}$  and subject to two periodic classical drives described by the coupling Hamiltonians $\hat{H}_{1,2}\left(\omega_{1,2}t+\phi_{1,2}\right)= \hat{H}_{1,2}\left[\omega_{1,2} (t+T_{1,2})+\phi_{1,2}\right]$ [Fig.~\ref{Floquet quantum sensor}.(b)]. Here, $\omega_{1,2}$ denote the frequencies, $T_{1,2}=2\pi/\omega_{1,2}$ the periods and  $\phi_{1,2}$  the initial phases of the drives.  The full Hamiltonian reads as
\begin{equation}\hat{H}\left(t\right)=\hat{H}_{0}+\hat{H}_{1}\left(\omega_{1}t+\phi_{1}\right)+\hat{H}_{2}\left(\omega_{2}t+\phi_{2}\right).
\label{two-drive model}
\end{equation}
 To obtain a Floquet system, we require that two frequencies $\omega_{1,2}$ are commensurate, i.e.\ that their ratio, $\omega_{1}/\omega_{2} \in \mathbb{Q}$ is a rational number. Consequently, the Hamiltonian $\hat{H}\left(t\right)$ exhibits a period $T_{\text{com}}$ (with $T_{\text{com}}$ the least common multiple of $T_{1,2}$) such that $\hat{H}\left(t+T_{\text{com}}\right)=\hat{H}\left(t\right)$.  

 Generally, the time evolution of the system is described by the $\hat{U}\left(t\right) = \mathcal{T}e^{-i\int_{0}^{t}\hat{H}\left(\tau\right)d\tau}$ where $\mathcal{T}$ is the time-ordering symbol. In the following, we are interested in the system at stroboscopic times $T = k T_{\text{com}}$ with $k$ an integer. We can express the qubit state conveniently in terms of the (quasi-)eigenenergies $\epsilon_n$ and eigenstates $\ket{n}$ with $n=1,2$ of the Floquet Hamiltonian $H_F \equiv i/T_{\text{com}} \log U(T_{\text{com}})$ such that ($\hbar=1$)
\begin{align}
    U(T) =  U(T_{\text{com}})^k = \sum_{n=1,2}\  e^{-i\epsilon_n k T_{\text{com}}} \ketbra{n}{n}.
\label{time-evolution unitary}
\end{align}

We are interested in the long-time energy transfer between the two classical drives, viewing the quantum system as a mediator. We study the power $P_j(t) = \dot{E_j}(t)\equiv \bra{\psi\left(t\right)}\frac{d\hat{H}_{j}\left(\omega_{j}t+\phi_{j}\right)}{dt}\ket{\psi\left(t\right)}$ absorbed from drive $j$ at time $t$. Here, $\ket{\psi(t)}= \hat{\mathcal{U}}(t) \ket{\psi(0)}$ is the quantum state evolved in time under $\hat{H}(t)$ with initial state $\ket{\psi(0)}$. In general, $P_{j}\left(t\right)$ varies over time. The average power absorbed by drive $j$ over a time window $[0,T]$ is given by the time average \cite{PhysRevB.108.064301} 
\begin{equation}
\begin{split}
\overline{P_{j}}\left(T\right)=&\frac{1}{T}\int_{0}^{T}dt\bra{\psi(t)}\frac{d\hat{H}_{j}}{dt}\ket{\psi(t)}\\=&\bra{\psi(0)}i\omega_{j}\frac{1}{T}\hat{\mathcal{U}}^\dagger\left(T\right)\partial_{\phi_{j}}\hat{\mathcal{U}}\left(T\right)\ket{\psi(0)}\\
    =&\omega_{j}\bra{\psi(0)}\hat{P}_{j}\left(T\right)\ket{\psi(0)}.
\end{split}
\label{power}
\end{equation}
\noindent Here $\hat{P}_{j}\left(T\right)\equiv\frac{i}{T}\hat{\mathcal{U}}^\dagger\left(T\right)\partial_{\phi_{j}}\hat{\mathcal{U}}\left(T\right)$ and we suppress the initial phases $\phi_{1,2}$ subscript.  For Floquet systems,  at $T=kT_{\text{com}}$ ($k \in \mathbb{N}$), we can derive  a compact form of $\overline{P_{j}}\left(T\right)$ for large $T$ (App.~\ref{classical power operator})
\begin{equation}
\hat{P}_{j}\left(T\right)=\sum\limits_{n=1,2}\ \left(\partial_{\phi_{j}}\epsilon_{n}\right)\ket{n}\bra{n}+O\left(\frac{1}{T}\right).
\label{power1}
\end{equation}
We denote the long-time limit as $\hat{P}_{j}\equiv\sum\limits_{n=1,2}\ \left(\partial_{\phi_{j}}\epsilon_{n}\right)\ket{n}\bra{n}$, the 'power operator'.  We note that this expression is specific for driven qubits (two-level systems). We derive corresponding expressions for qudits ($d$-level systems) in App.~\ref{classical power operator}. 
$\hat{P}_{j}$ is solely determined by properties of the Floquet Hamiltonian. Crucially, the transferred energy is however specified by the expectation value of  $\hat{P}_{j}$  with respect to the initial state of the system. This introduces controllability: by varying the initial state $\ket{\psi(0)}$, we can control the energy transfer between the two classical drives. 

We now demonstrate explicitly that it is indeed feasible to completely revert the energy transfer between the two drives by choosing the appropriate initial states of the driven quantum system. We focus first on the power operator  $\hat{P}_{1}$. We show in App.~\ref{classical power operator} that it is traceless, $\tr{\hat{P}_{1}} =0$ (equivalently, it also holds $\tr{\hat{P}_{2}} =0$). This ensures two opposite eigenvalues $\pm P$ of $\hat{P}_{1}$. Further denoting the corresponding eigenstates as $\ket{\pm P}$, this enables us to define two states $ \ket{0}, \ket{1}$ as

\begin{equation}
    \begin{split}
        \ket{0}=&\ket{P},\\
        \ket{1}=&\ket{-P}.
    \end{split}
\end{equation}
 
\noindent Equivalently, one have opposite eigenvalues of $\hat{P}_{2}$ with eigenstates $\ket{\pm P}'$ and define the corresponding states $\ket{0}'$ and $\ket{1}'$.
It follows from energy conservation $\omega_{1}\hat{P}_{1}+\omega_{2}\hat{P}_{2}=0$ (see App.~\ref{classical power operator}) that these states are closely related, $\ket{0}'=\ket{1}$ and $\ket{1}'=\ket{0}$. 
Thus, by preparing the qubit in state $\ket{0} $ ($\ket{1}$), we can reverse the power transferred from drive $1$ to drive $2$.

Finally, we discuss energy conservation in the driven system in more details. Due to its bounded Hilbert space, the qubit can only store a finite amount of energy. Therefore, over long times $T$, there is no net flow of energy in or out of the qubit.  The qubits acts as a mediator that enables energy exchange between the two drives. Mathematically, this poses a constraint on the quasienergy spectrum: $\epsilon_{n}=\epsilon_{n}(\omega_1,\omega_2,\phi_1,\phi_2)$ is only a function of $\frac{\phi_{1}}{\omega_{1}}-\frac{\phi_{2}}{\omega_{2}}$ \footnote{Note that $\omega_{1}\hat{P}_{1}+\omega_{2}\hat{P}_{2}=0$ directly leads to $(\omega_{1}\partial_{\phi_{1}}+\omega_{2}\partial_{\phi_{2}})\epsilon_{n}=0$.}, which is used in defining characteristic functions in Sec. \ref{parity as decoder}.

\section{Energy transfer and entanglement generation between quantum drives }

\label{quantum polarization conversion}
\subsection{Energy transfer between two quantum drives}

We now quantize the two classical driving fields and treat them as  two bosonic modes.

\subsubsection{Quantizing the drives}
We first rewrite the Hamiltonian Eq.~\eqref{two-drive model} to make its periodic time dependence more explicit
\begin{equation}
\begin{split}
\hat{H}\left(t\right)=&\hat{H}_{0}+\hat{H}_{1,o}\sin\left(\omega_{1}t+\phi_{1}\right)+\hat{H}_{1,e}\cos\left(\omega_{1}t+\phi_{1}\right)\\
&+\hat{H}_{2,o}\sin\left(\omega_{2}t+\phi_{2}\right)+\hat{H}_{2,e}\cos\left(\omega_{2}t+\phi_{2}\right)
\label{sin and cos}
\end{split}
\end{equation}
Here, we have expanded $\hat{H}_{j}\left(\omega_{j}t+\phi_{j}\right)$ in Eq.~\eqref{two-drive model} into  $\hat{H}_{j,o}\sin\left(\omega_{j}t+\phi_{j}\right)$ and  $\hat{H}_{j,e}\cos\left(\omega_{j}t+\phi_{j}\right)$. The subscript $e,o$ in the coefficients denote even and odd contributions. We assume that these drives are realized by bosonic fields. Typical examples are photons from optical cavities \cite{Macieszczak_2014}, bosons in superconducting circuits \cite{RevModPhys.93.025005}, phonons from lattice or mechanical vibrations \cite{RevModPhys.86.1391}, and magnons from spin waves \cite{Chumak2015}. The quantization of the drives is hence accomplished via the substitution $e^{i\left(\omega_{j}t+\phi_{j}\right)}\rightarrow \hat{a}_{j}$, with $\hat{a}_{j}$ the bosonic annihilation operator for the $j$-th drive satisfying $[\hat{a}_{j},\hat{a
}_j^\dagger]=1$ and $[\hat{a}^{(\dagger)}_{1},\hat{a
}_2^{(\dagger)}]=0$. With this identification, the two-tone Hamiltonian becomes a time-independent Hamiltonian for the combined system of qubit and quantum drives
\begin{equation}
\begin{split}
\hat{H}_{q}=&\hat{H}_{0}+\hat{H}_{B}+\hat{H}_{1,o}\frac{\hat{a}_{1}-\hat{a}_{1}^{\dagger}}{2i}+\hat{H}_{1,e}\frac{\hat{a}_{1}+\hat{a}_{1}^{\dagger}}{2}\\
&+\hat{H}_{2,o}\frac{\hat{a}_{2}-\hat{a}_{2}^{\dagger}}{2i}+\hat{H}_{2,e}\frac{\hat{a}_{2}+\hat{a}_{2}^{\dagger}}{2}.
\end{split}
\label{quantized model}
\end{equation}
We  have included a term $\hat{H}_{B}=\hat{\mathds{1}}_{0}\otimes\left(\omega_{1}\hat{a}_{1}^{\dagger}\hat{a}_{1}+\omega_{2}\hat{a}_{2}^{\dagger}\hat{a}_{2}\right)$ to account for the free evolution of two drives, where $\hat{\mathds{1}}_{0}$ is the identity operator on the qubit.
The Hilbert space becomes a direct product between $\mathcal{H}$ for the qubit and two Fock spaces $\mathcal{F}_{1,2}$ for quantum drives, $\mathcal{H}_{q}=\mathcal{H}\otimes\mathcal{F}_{1}\otimes\mathcal{F}_{2}$. An orthonormal basis of this space is denoted as   $\ket{\lambda}\otimes\ket{n_{1}}\otimes\ket{n_{2}}$ where $\left\{\ket{\lambda}\right\}_{\lambda=1,\dots,d}$ is an orthonormal basis of $\mathcal{H}$ and $\left\{\ket{n_{1,2}}\right\}_{n_{1,2}=0,1,\dots}$ the Fock states for $\mathcal{F}_{1,2}$, respectively.\\

\subsubsection{Approximation by a uniform lattice model}
\label{Approximation by a uniform lattice model}

In Sec.~\ref{review}, we have shown that a qubit driven by two \textit{classical} fields can serve a mediator for energy transfer. We now extend this to the quantized model. To this end, we note 
that for states satisfying the sub-Poissonian condition $\frac{\delta n_{1,2}}{\sqrt{n_{1/2,c}}}\ll 1$ with  mean $n_{1/2,c}\equiv\braket{\hat{a}_{1,2}^{\dagger}\hat{a}_{1,2}}$ and standard deviation $\delta n_{1,2}\equiv\sqrt{\braket{\left(\hat{a}_{1,2}^{\dagger}\hat{a}_{1,2}\right)^{2}}-\braket{\hat{a}_{1,2}^{\dagger}\hat{a}_{1,2}}^{2}}$,  the Hamiltonian \eqref{quantized model} can be approximated by a solvable Floquet lattice model \cite{PhysRevB.99.094311,PhysRevX.7.041008} 
\begin{equation}
\begin{split}
\hat{H}_{l}=&\hat{H}_{0}+\hat{H}_{B}+\hat{H}_{1,o}^{c}\frac{\hat{T}_{1}-\hat{T}_{1}^{\dagger}}{2i}+\hat{H}_{1,e}^{c}\frac{\hat{T}_{1}+\hat{T}_{1}^{\dagger}}{2}\\
&+\hat{H}_{2,o}^{c}\frac{\hat{T}_{2}-\hat{T}_{2}^{\dagger}}{2i}+\hat{H}_{2,e}^{c}\frac{\hat{T}_{2}+\hat{T}_{2}^{\dagger}}{2}.
\label{lattice model}
\end{split}
\end{equation}
Here, we neglected terms of error $O\left(\frac{\delta n_{1,2}}{\sqrt{n_{1/2,c}}}\right)$ and  defined the Fock space translation operators $\hat{T}_{1,2}\equiv\sum_{n_{1,2}}\ket{n_{1,2}-1}\bra{n_{1,2}}$ and $\hat{H}_{1/2,o/e}^{c}=\sqrt{n_{1/2,c}}\hat{H}_{1/2,o/e}$ are renormalized qubit operators. Here, we extend $n_{1,2}$ to $\pm\infty$ and impose a translation symmetry, which allows to change to a Fourier basis $\ket{\phi_{1,2}}=\int_{0}^{2\pi}dn_{1,2}e^{i\phi_{1,2}n_{1,2}}\ket{n_{1,2}}$ and switch to an interaction picture with respect to the bosonic Hamiltonian $\hat{U}\left(t\right)=e^{-i\hat{H}_{B}t}$. Now, the lattice model Eq.~\eqref{lattice model} decomposes into a direct sum of Bloch Hamiltonians $\hat{H}_{l}\left(t\right)=\bigoplus_{\bm{\phi}} \hat{H}_{\bm{\phi}}\left(t\right)\otimes\ket{\bm{\phi}}\bra{\bm{\phi}}$ with
\begin{equation}
\begin{split}
\hat{H}_{\bm{\phi}}\left(t\right)=& \hat{H}_{0}+ 
\hat{H}_{1,o}^{c}\sin\left(\omega_{1}t+\phi_{1}\right)+\hat{H}_{1,e}^{c}\cos\left(\omega_{1}t+\phi_{1}\right)\\
&+\hat{H}_{2,o}^{c}\sin\left(\omega_{2}t+\phi_{2}\right)+\hat{H}_{2,e}^{c}\cos\left(\omega_{2}t+\phi_{2}\right),
\end{split}
\label{Bloch Hamiltonian}
\end{equation}
with $\bm{\phi}\equiv\left(\phi_{1},\phi_{2}\right)$ and the corresponding $\ket{\bm{\phi}}\equiv\ket{\phi_{1}}\otimes\ket{\phi_{2}}$. We briefly remark on the form of $\hat{H}_{l}\left(t\right)$ in the interaction picture. First, the phase is uniquely defined only up to mod $2\pi$, and the corresponding states should be understood in a periodic way $\ket{\phi_{1,2}+2\pi}=\ket{\phi_{1,2}}$. Second, $\ket{\bm{\phi}}$ is motionless in the interaction picture, and hence the only dynamic evolution results from the (driven) system. For a given $\ket{\bm{\phi}}$, the system evolves under $\hat{H}_{0}+\hat{H}_{\bm{\phi}}\left(t\right)$, which is identical to the two-tone Floquet Hamiltonian Eq.~\eqref{sin and cos}. We hence conclude that the dynamics of each phase component is described by the two-tone Floquet Hamiltonian Eq.~\eqref{sin and cos}.

In the following study of energy transfer and entanglement generation, we assume the dynamics of the original model Eq.~\eqref{quantized model} being well-described by the lattice model Eq.~\eqref{lattice model} and leave the numerical justification to Sec.~\ref{examples and numerics}.

\subsubsection{Energy transfer}
 
We now analyze the dynamics under Eq.~\eqref{Bloch Hamiltonian}. We assume that there is no initial entanglement or correlation between the system and drives, i.e., we start with an initial product state \begin{equation}
\ket{\Psi\left(0\right)}=\ket{\psi\left(0\right)}\otimes\int\limits_{\text{BZ}}\bm{d^{2}\phi}f\left(\bm{\phi}\right)\ket{\bm{\phi}},
\label{input state}
\end{equation}
where $\ket{\Psi\left(0\right)}$ and $\ket{\psi\left(0\right)}$ denote the initial system state and initial qubit state and $\bm{d^{2}\phi}\equiv d\phi_{1}d\phi_{2}$ is the metric of the integral over the $1$-st Brillouin zone $\text{BZ}\equiv \left[0,2\pi\right)^{2}$. Finally, $f\left(\bm{\phi}\right)$ is the initial field distribution of the drives satisfying $\int\limits_{\text{BZ}}\bm{d^{2}\phi}\left|f\left(\bm{\phi}\right)\right|^{2}=1$. 

For each value of $\bm{\phi}$, the qubit evolves under a Floquet Hamiltonian $i\partial_{t}\ket{\psi\left(\bm{\phi},t\right)}=\hat{H}_{\bm{\phi}}\left(t\right)\ket{\psi\left(\bm{\phi},t\right)}$. As before, we want to quantify power transferred to the $j$-th drive with $j=1,2$:
\begin{equation}
\overline{P_{j}}\left(T\right)=\frac{\omega_{j}}{T}\left(\braket{\hat{a}_{j}^{\dagger}\hat{a}_{j}}_{0}-\braket{\hat{a}_{j}^{\dagger}\hat{a}_{j}}_{T}\right),
\label{average power}
\end{equation}
where $\braket{\square}_{0}=\braket{\Psi\left(0\right)|\square|\Psi\left(0\right)}$ and $\braket{\square}_{T}=\braket{\Psi\left(T\right)|\square|\Psi\left(T\right)}$ are expectation values of $\square$ at $t=0$ and $t=T$, respectively. We observe the dynamics at stroboscopic times $T=kT_{\text{com}}$ with $k=0,1,2,...$. In App.~\ref{quantum power appendix}, we show that the long-time limit $\lim_{T\rightarrow\infty}\overline{P_{j}}\left(T\right)=\omega_{j}\bra{\psi(0)}\hat{P}_{j}\left[f\right]\ket{\psi(0)}$ is determined by  the 'functional power operator'
\begin{equation}
\hat{P}_{j}\left[f\right]= \int\limits_{\text{BZ}}\bm{d^{2}\phi}  \; \left|f\left(\bm{\phi}\right)\right|^{2}\hat{P}_{j}(\bm{\phi}).
\label{general power operator}
\end{equation}
Here, $\hat{P}_{j}(\bm{\phi})=\sum\limits_{n=1,2}\ \left[\partial_{\phi_{j}}\epsilon_{n}(\bm{\phi})\right]\ket{n(\bm{\phi})}\bra{n(\bm{\phi})}$ is the power operator defined with respect to the $\bm{\phi}$-dependent Floquet Hamiltonian $\hat{H}_{\bm{\phi}}\left(t\right)$, analogous to Sec.~\ref{review}. Since $\hat{P}_{j}\left[f\right]$ is a weighted sum of power operators $\hat{P}_{j}(\bm{\phi})$, it directly inherits properties such as tracelessness $\tr{\hat{P}_{j}[f]}=0$ and energy conservation $ \omega_{1}\hat{P}_{1}\left[f\right]+\omega_{2}\hat{P}_{2}\left[f\right]=0$ (see Sec.~\ref{review} and App.~\ref{quantum power appendix}).

As its classical counterpart, the tracelessness of $\hat{P}_{j}\left[f\right]$ ensures two opposite eigenvalues $\pm P_{f}$  with eigenstates $\ket{\pm P}_{f}$, which enables us to define two states

\begin{equation}
    \begin{split}
        \ket{0}_{f}=&\ket{P}_{f},\\
        \ket{1}_{f}=&\ket{-P}_{f},
    \end{split}
\label{functional ancilla states}
\end{equation}

\noindent which physically results in opposite energy flow from the $j$-th drive as
$\bra{0}_{f}\hat{P}_{j}\left[f\right]\ket{0}_{f}+\bra{1}_{f}\hat{P}_{j}\left[f\right]\ket{1}_{f}=0$. We note that the functional power operator $\hat{P}_{j}\left[f\right]$ depends on the field distribution $f\left(\bm{\phi}\right)$ and so do its eigenvalues $\pm P_{f}$ and eigenstates $\ket{\pm P}_{f}$. As a result, different initial field distributions $f\left(\bm{\phi}\right)$ result in different states $\ket{0}_{f}$ and $\ket{1}_{f}$.\\

\subsection{Quadratically growing quantum Fisher information}

\label{entanglement generation}

\begin{figure*}
\centering
     \includegraphics[width=\textwidth]{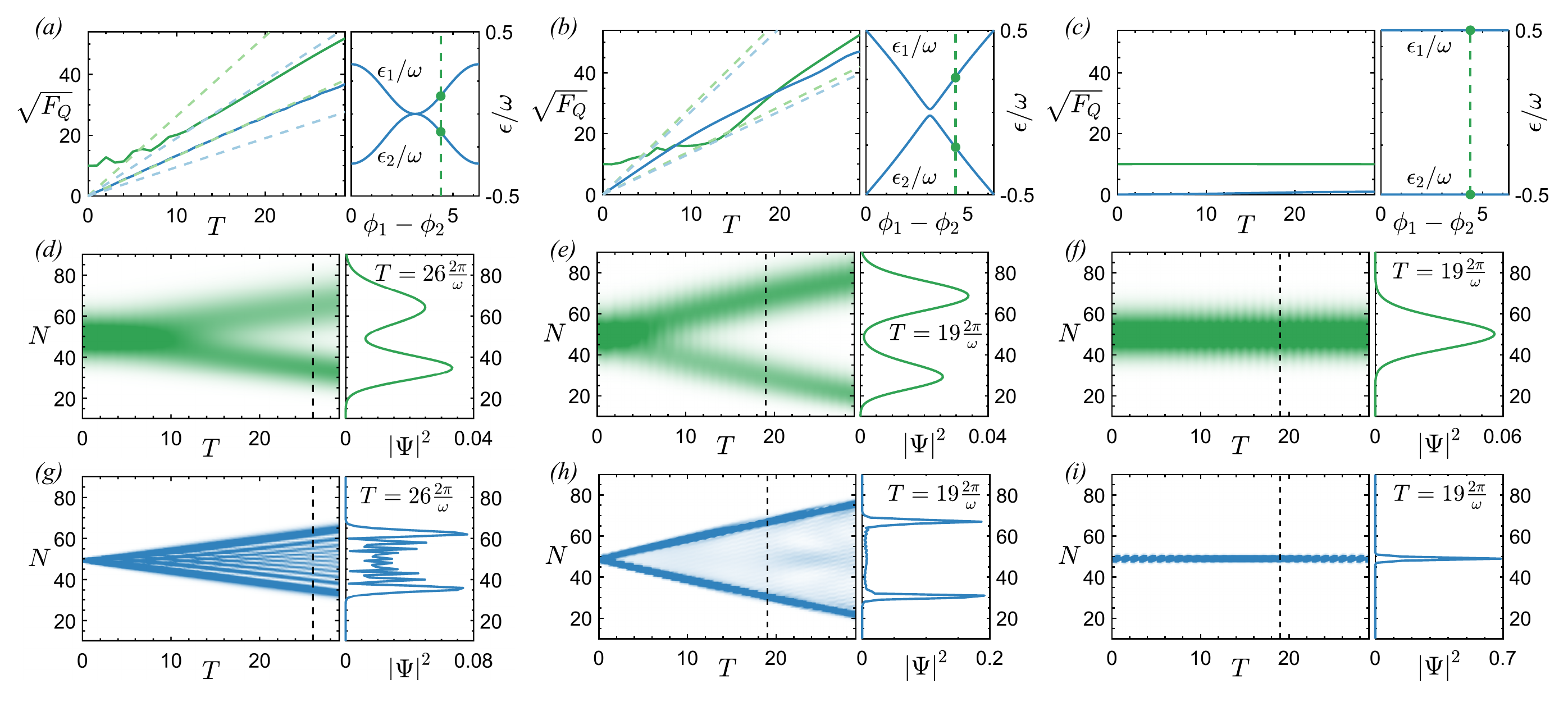}
        \caption{\textit{Quantum Fisher information and wavefunction profiles of qubit-drives systems.} (a)(d)(g) $2$-level system with circularly-polarized drives, (b)(e)(h) polarization conversion and (c)(f)(i) oscillating Zeeman field. In (a)-(c), we plot both $\sqrt{F_{Q}}\equiv\sqrt{F_{q}\left[ \ket{\text{PES}}_{T}, \hat{J}_{z}\right]}$ (left panels) and Floquet quasienergy bands (right panels) for three models. In the left panels, we plot $\sqrt{F_{Q}}$ for both coherent state (green solid) and Fock state (blue solid) input of the driving field. The corresponding bonds [upper and lower, see Eq.~\eqref{QFIbounds}] $P^{2}\left[f\right]$ are notated as green and blue dashed lines. In the right panels, we plot the quasienergy bands and indicate regions in the $1$-st Brillouin zone $\phi_{1}-\phi_{2}\in\left[0,2\pi\right)$, over which $f\left(\bm{\phi}\right)$ spreads for both coherent (green dashed lines and dots) and Fock (blue solid curves) inputs. In (d)-(i), we plot wavefunction profiles $\left|\Psi\right|^2$ of the second drive for both coherent ((d)-(f)) and  Fock ((g)-(i)) inputs as a function of time $T$ (left panels) and their cross sections (right panels) at a single moment indicated by the black dashed lines in the left panels. For all three models, we start with average $50$ bosons in each drive. For coherent-state inputs, we introduce an initial phase difference $\phi_{20}-\phi_{10}=0.6\pi$ between two drives. Other parameters in different models are given as follows: for the oscillating Zeeman field, $B_{0}=B_{1,2}$ $gB_{0}=\omega_{1,2}$; for the $2$-level system with circularly-polarized drives, $\omega=\frac{\omega_{0}}{4}$, $\mathcal{A}=\frac{\omega_{0}}{8}$; and for the polarization conversion, $\omega=\omega_{0}$, $\mathcal{A}=\frac{\omega_{0}}{2}$. The corresponding parameters in the quantized models can be obtained via the procedure given in the main text (see Sec.~\ref{quantum polarization conversion}). } 
        \label{QFI and wavefunction}  
\end{figure*}

 We have observed that flipping the qubit from $\ket{0}_{f}$ to $\ket{1}_{f}$ (and vice versa) changes the direction of energy flow accordingly. We now ask: how should we initialize the qubit so that useful entanglement for metrology can be generated at a later time?

Within the lattice approximation, we can address this question with the following statement:
\noindent Given the initial field distribution $f\left(\bm{\phi}\right)$ of the drives and a non-zero power operator $\hat{P}_{j}\left[f\right]$, a path-entangled state can be prepared by first initializing the ancilla qubit in $\ket{\beta,+}_{f}=\frac{1}{\sqrt{2}}\left(\ket{0}_{f}+e^{i\beta}\ket{1}_{f}\right)$, then evolving the qubit-drives system for some time $T=kT_{\text{com}}$ according to $\hat{H}_{l}$, and finally measuring the qubit in the initial state basis $\ket{\beta,+}_{f}, \ket{\beta,-}_{f}$. In what follows, we assume the outcome is $+$ and thus, the resulting state is given by
\begin{equation}\ket{\text{PES}}_{T}\equiv \frac{1}{\mathcal{N}_{T}}\hat{\Pi}_{\beta}e^{-i\hat{H}_{l}T}\left[\ket{\beta,+}_{f}\otimes\int\limits_{\text{BZ}}\bm{d^{2}\phi}f\left(\bm{\phi}\right)\ket{\bm{\phi}}\right],
\label{path-entangled state}
\end{equation}
\noindent where $\hat{\Pi}_{\beta}\equiv\ket{\beta,+}_{f}\bra{\beta,+}_{f}$ is the projection to the qubit state $\ket{\beta,+}_{f}$ and 'PES' is the abbreviation for 'path-entangled state'. $\mathcal{N}_{T}$ is a $T$-dependent normalization factor. We note that, as shown in Appendix~\ref{Postselection app}, the two outcomes ($+$ and $-$) of the qubit measurement are equally likely for long times $T$. If the outcome is $-$, it simply introduces a known phase shift in the sensing procedure, which can be corrected as we know the measurement result.  Thus, our protocol remains unaffected.

The QFI of $\ket{\text{PES}}_{T}$ with respect to $\hat{J}_{z}$ (recall the MZI) asymptotically exhibits a $T^2$ scaling with prefactor bounded by:

\begin{equation}
     \frac{1}{2}P^{2}\left[f\right]\leq\lim_{T\rightarrow\infty}\frac{1}{T^{2}}F_{q}\left[ \ket{\text{PES}}_{T}, \hat{J}_{z}\right]\leq 2P^{2}\left[f\right],
        \label{QFIbounds}
\end{equation}

\noindent with $\hat{J}_{z}=\frac{1}{2}\left(\hat{a}_{1}^{\dagger}\hat{a}_{1}-\hat{a}_{2}^{\dagger}\hat{a}_{2}\right)$ and the scalar bound

\begin{equation}
    P^{2}\left[f\right]\equiv\frac{1}{2}\int\limits_{\text{BZ}}\bm{d^{2}\phi}\left|f^{2}\left(\bm{\phi}\right)\right|\left\{\left[\left(\partial_{\Delta\phi}\right)\epsilon_{1}\right]^{2}+\left[\left(\partial_{\Delta\phi}\right)\epsilon_{2}\right]^{2}\right\},
\end{equation}
\noindent where $\Delta\phi\equiv\phi_{1}-\phi_{2}$ and $\partial_{\Delta\phi}\equiv\partial_{\phi_{1}}-\partial_{\phi_{2}}$. We note that the $T \to \infty$ limit is formally valid only in the lattice approximation [cf discussion below Eq.~\eqref{lattice model}]. For any finite initial number of photons $N=n_{1,c}+n_{2,c}$, the QFI stops growing at $T = \mathcal{O}(N)$, which is consistent with the fundamental Heisenberg limit (see also Sec.~\ref{lattice discussion})\\

While the proof is left to App.~\ref{QFI appendix}, we clarify here the terminology used and the physical implications of the statement. The term 'path-entangled state' is borrowed from optical interferometry \cite{Jin2013}, where two spatially distinguishable photonic modes enter different interferometric arms or paths. Introducing (mode) entanglement between two modes  yields 'path entanglement'. Proper path entanglement can endow the interferometer with HL sensitivity. Conversely, however, HL sensitivity (expressed in the QFI) does not necessarily imply path entanglement, unlike in qubit systems \cite{PhysRevLett.127.260501,Pezz2018,PhysRevLett.126.080502}. Examples of this are single-mode cat states $\frac{1}{2}\left[\ket{\alpha}\pm\ket{-\alpha}\right]$ \cite{DODONOV1974597}, squeezed states \cite{PhysRevLett.123.231107}, etc. Nevertheless, we demonstrate in App~\ref{mode entanglement appendix} that path entanglement in $\ket{\text{PES}}_{T}$ significantly enhances the quantum Fisher information. The generalization and optimization of the quantum Fisher information for a d-level ancilla are discussed in App.~\ref{optimization and qudits}.

Within the lattice approximation [Eq.\eqref{lattice model}], this quadratic scaling $\bigl(F_{q} \propto T^2\bigr)$ can be understood as the \textit{ballistic} ($\propto T^2$) expansion of a wavefunction in a quantum random walk---contrasting with the \textit{diffusive} ($\propto T$) spreading seen in classical random walks~\cite{Kempe_2003,Venegas_Andraca_2012}. Under standard assumptions of a local graph with local jumps, $T^2$ is indeed the fastest quantum-mechanically allowed spreading possible in the corresponding lattice model.

Finally, since the power operator $\hat{P}_{j}\left[f\right]$ and the bound $P^{2}[f]$ on the QFI are completely determined by the Floquet spectrum $\left\{\epsilon_{n}, \ket{n}\right\}$ with initial field distributions $f(\bm{\phi})$, the sensing ability of $\ket{\text{PES}}_{T}$ can be assessed before the experiment. In Sec.\ref{parity as decoder}, we will extend this observation to a specific choice of POVM measurement, and determine the maximal measurement sensitivity from the classical Floquet spectrum for Eq.~\eqref{Bloch Hamiltonian}.

\subsection{Numerical examples}
\label{examples and numerics}
\subsubsection{Example I: Two identical circularly-polarized
 drives}

The first non-trivial model is a $2$-level system under circularly-polarized drives:

\begin{equation}
\begin{split}
    \hat{H}_{\bm{\phi}}\left(t\right)=&\frac{\omega_{0}}{2}\hat{\sigma}_{z}+\mathcal{A}\left[\cos\left(\omega t+\phi_{1}\right)+\cos\left(\omega t+\phi_{2}\right)\right]\hat{\sigma}_{x}\\
    &+\mathcal{A}\left[\sin\left(\omega t+\phi_{1}\right)+\sin\left(\omega t+\phi_{2}\right)\right]\hat{\sigma}_{y}.
\end{split}
\label{2-level with circularly-polarized drives}
\end{equation}

\noindent The drives' frequencies $\omega$, amplitudes $\mathcal{A}$, and polarization are identical. The effective Hamiltonian in the rotating frame is given as (up to a unitary transform due to different choice of starting time $\left[t_{0}, t_{0}+T_{\text{com}}\right)$):

\begin{equation}
    \hat{H}_{\text{eff}}=\frac{\omega_{0}-\omega}{2}\hat{\sigma}_{z}+2\mathcal{A}\cos\left(\frac{\phi_{1}-\phi_{2}}{2}\right)\hat{\sigma}_{x},
\end{equation}

\noindent with quasienergies (back in the lab frame) 
\begin{equation}
\epsilon_{1/2}=\frac{1}{2}\left[\omega\pm\sqrt{\left(\omega_{0}-\omega\right)^{2}+16\mathcal{A}^{2}\cos^{2}\left(\frac{\phi_{1}-\phi_{2}}{2}\right)}\right].
\label{quasienergies1}
\end{equation}

\noindent Eq.~\eqref{quasienergies1} has a non-trivial dispersion relation with respect to $\phi_{1}-\phi_{2}$, which, for a general field distribution $f\left(\bm{\phi}\right)$, exhibits non-vanishing power $\hat{P}_{j}\left[f\right]$ and quadratically growing quantum Fisher information $F_{q}\sim P^{2}\left[f\right]T^{2}$. An exception comes with a symmetric field distribution $f\left(\phi_{1},\phi_{2}\right)=f\left(\phi_{2},\phi_{1}\right)$, where the power operator vanishes $\hat{P}_{j}\left[f\right]=0$, while $P^{2}\left[f\right]$ remains non-zero in general. In Fig.~\ref{QFI and wavefunction}.(b), (e), (h), we plot the time dependence of the QFI for $\ket{\text{PES}}_{T}$ and the wavefunctions $\left|\Psi\right|^2(N_{2})\equiv\braket{\ket{N_{2}}\bra{N_{2}}}_{T}$ of the second mode for different inputs. The quasienergy bands given in the right panel of Fig.~\ref{QFI and wavefunction}
.(b) are dispersive except at $\phi_{1}-\phi_{2}=0, \pi$ (it is also gapless at $\pi$ for $\omega=\omega_{0}$). $F_{q}$ is given in the left panel and is well-bounded by $\frac{1}{2}P^{2}\left[f\right]T^{2}$ and $2P^{2}\left[f\right]T^{2}$ represeneted by four dashed lines (green lines for coherent-state input with $\left|f(\bm{\phi})\right|^{2}=\delta(\phi_{1}-\phi_{10},\phi_{2}-\phi_{20})$, where $\phi_{10},\phi_{20}$ are the initial phases of the drives, and blue lines for Fock-state input with $\left|f\left(\bm{\phi}\right)\right|^{2}=\frac{1}{4\pi^{2}}$). 

Typically, energy transfer and the QFI are analogous to the motion of waves: the mean group velocity $\overline{\partial_{k}\epsilon\left(k\right)}$ quantifies the energy transfer, while the square mean of the group velocity $\overline{\left[\partial_{k}\epsilon\left(k\right)\right]^2}$quantifies the QFI. In Fig.~\ref{QFI and wavefunction}.(e), the wavefunction splits into up- and down-branches for a coherent-state input with an initial phase difference $\phi_{20}-\phi_{10}=0.6\pi$ due to a non-vanishing $\hat{P}_{j}\left[f\right]$. In comparison, in Fig.~\ref{QFI and wavefunction}.(h), the wavefunction of a twin-Fock-state input diffuses into a 'Bat state' at later times due to a vanishing $\hat{P}_{j}\left[f\right]$ but a non-zero $P^{2}\left[f\right]$ for $\left|f\left(\bm{\phi}\right)\right|^{2}=\frac{1}{4\pi^{2}}$.

\subsubsection{Example II: Two circularly-polarized
 drives with opposite chirality}

 \label{polarization conversion model}

Flipping the sign of $\sin\left(\omega t+\phi_{2}\right)$ in Eq.~\eqref{2-level with circularly-polarized drives} and shifting the time-independent part $\frac{\omega_{0}}{2}\hat{\sigma}_{x}$, the model now describes energy transfer between two circularly-polarized drives with opposite chiralities (polarization conversion):

\begin{equation}
\begin{split}
    \hat{H}_{\bm{\phi}}\left(t\right)=&\frac{\omega_{0}}{2}\hat{\sigma}_{x}+\mathcal{A}\left[\cos\left(\omega t+\phi_{1}\right)+\cos\left(\omega t+\phi_{2}\right)\right]\hat{\sigma}_{x}\\
    &+\mathcal{A}\left[\sin\left(\omega t+\phi_{1}\right)-\sin\left(\omega t+\phi_{2}\right)\right]\hat{\sigma}_{y}.
\end{split}
\label{polarization conversion}
\end{equation}

\noindent The effective Hamiltonian of Eq.~\eqref{polarization conversion}, as well as its quasienergies and quasieigenstates, can only be obtained numerically. Fortunately, some core properties of the spectrum can be obtained by looking at the symmetry of the Floquet Hamiltonian. Note that $\hat{\sigma}_{x}\hat{H}_{\phi_{1},\phi_{2}}\left(t\right)\hat{\sigma}_{x}=\hat{H}_{\phi_{2},\phi_{1}}\left(t\right)$, which swaps the subscript $\phi_{1},\phi_{2}\rightarrow\phi_{2},\phi_{1}$. This directly leads to $\epsilon_{1,2}\left(\phi_{1},\phi_{2}\right)=\epsilon_{1,2}\left(\phi_{2},\phi_{1}\right)$, $\ket{1\left(\phi_{2},\phi_{1}\right)}=\hat{\sigma}_{x}\ket{1\left(\phi_{1},\phi_{2}\right)}$ and $\ket{2\left(\phi_{2},\phi_{1}\right)}=\hat{\sigma}_{x}\ket{2\left(\phi_{1},\phi_{2}\right)}$. The $\hat{\sigma}_{x}$ operation hence matches quasienergies and quasieigenstates at $\left(\phi_{1},\phi_{2}\right)$ and $\left(\phi_{2},\phi_{1}\right)$. Consequently, for a symmetric field distribution $f\left(\phi_{1},\phi_{2}\right)=f\left(\phi_{2},\phi_{1}\right)$, the functional power operator also experiences a sign flip under $\hat{\sigma}_{x}$ operation: $\hat{\sigma}_{x}\hat{P}_{j}\left[f\right]\hat{\sigma}_{x}=-\hat{P}_{j}\left[f\right]$. The initial state of the ancilla can be hence prepared as $\ket{\beta,+}_{f}=\ket{\uparrow/\downarrow}$ with $\hat{\sigma}_{x}\ket{\uparrow/\downarrow}=\pm\ket{\uparrow/\downarrow}$ eigenstates of $\hat{\sigma}_{x}$. 

More properties are obtained by numerically solving the quasienergies, which are given in the right panel of Fig.~\ref{QFI and wavefunction}.(c). First, the bands are gapless at $\phi_{1}=\phi_{2}$ with a finite dispersion, as the quasienergies are defined modulo $\omega$, and hence $\pm0.5\omega$ are identified. Second, the bands are linear about $\phi_{1}-\phi_{2}=0, \pi$, which results in $P^{2}\left[f\right]$ approximately independent of the initial distribution $f\left(\bm{\phi}\right)$. This is shown in Fig.~\ref{QFI and wavefunction}.(c), where the QFI for the coherent-state and Fock-state inputs grows at a similar rate and the dashed lines representing the bounds overlap. 

Another consequence of the linearity is that wavefunctions do not disperse as the second derivative vanishes $\partial_{\Delta\phi}^{2}\epsilon_{1,2}=0$ except at $\phi_{1}-\phi_{2}=\pi$ with a finite gap opening. In Fig.~\ref{QFI and wavefunction}.(f), (i), wavefunctions for both coherent-state and Fock-state inputs are translated into up- and down-branches and maintain their initial shapes for each branch, analogous to soliton motion in waves. This model, therefore, functions like a quantum beam splitter that evenly splits the inputs without introducing unwanted superpositions. Specifically, a Fock-state input results in a N00N-like output [Fig.\ref{QFI and wavefunction}.(i)], whereas the previous model produces more superpositions [Fig.\ref{QFI and wavefunction}.(h)] for the same input.

\subsubsection{Example III: Two-tone oscillating Zeeman field}

Consider a spin-1/2 particle interacting with a two-tone Zeeman field $\bm{B}\left(t\right)=\left[B_{0}+B_{1}\cos\left(\omega_{1}t+\phi_{1}\right)+B_{2}\cos\left(\omega_{2}t+\phi_{2}\right)\right]\hat{z}$ with commensurate frequencies $\omega_{1,2}$ along $z$-axis:

\begin{equation}
    \hat{H}\left(t\right)=-g\bm{B}\left(t\right)\cdot\hat{\bm{S}},
\end{equation}

\noindent with $g$ the coupling strength and $\hat{\bm{S}}=\left(\hat{S}_{x},\hat{S}_{y},\hat{S}_{z}\right)$ the spin vector. The effective Hamiltonian is given as:

\begin{equation}
    \hat{H}_{\text{eff}}=-gB_{0}\hat{S}_{z},
\end{equation}

\noindent which is completely independent of initial phases $\bm{\phi}=\left(\phi_{1},\phi_{2}\right)$. Since the Hamiltonian commutes with itself at different times $\left[\hat{H}\left(t\right),\hat{H}\left(t'\right)\right]=0$. The $\bm{\phi}$-independence makes both $\hat{P}_{j}\left[f\right]$ and $P^{2}\left[f\right]$ vanish for any field distribution $f\left(\bm{\phi}\right)$. As a result, there is neither energy transfer between the drives nor entanglement generation when quantizing the drives. In Fig.~\ref{QFI and wavefunction}.(a), (d), (g), we plot the time dependence of the QFI and wavefunctions for different initial states of the drives. The flat bands in the right panel of Fig.~\ref{QFI and wavefunction}.(a) now results in $T$-independent QFI [Fig.~\ref{QFI and wavefunction}.(a), left panel] and wavefunctions [Fig.~\ref{QFI and wavefunction}.(d), (g)]. In other words, both energy transfer and growing sensitivity require a dispersive quasi-energy spectrum.\\

\begin{figure}
\centering
     \includegraphics[width=0.34\textwidth]{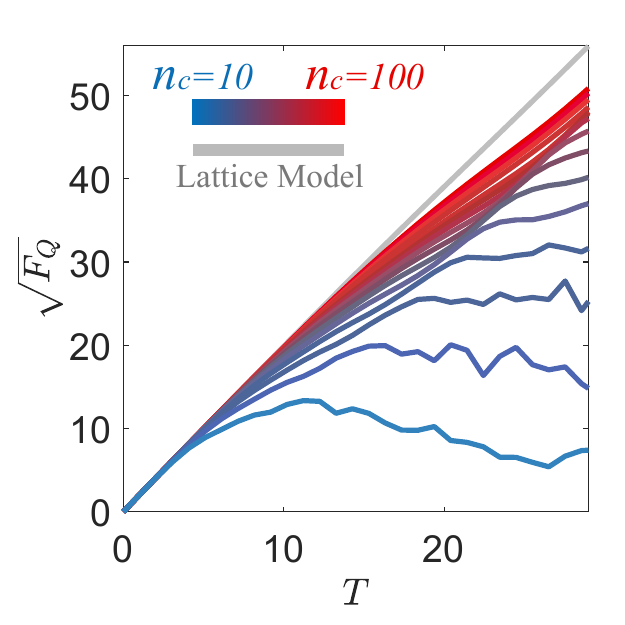}
        \caption{\textit{Quality of the lattice approximation.} 
In this figure, we plot the quantum Fisher information $\sqrt{F_{Q}}$ as function time $T$ for Fock-state inputs with different initial photon numbers per mode, ranging from $n_{c}=10$ to $n_{c}=100$. 
As $n_c$ increases, $\sqrt{F_{Q}}$ converges to the linear growth (gray line) predicted by the lattice model.}
        \label{Boundary effect}  
\end{figure}

\subsubsection{Discussion of the lattice approximation}\label{lattice discussion}

Throughout our study of the dynamics and entanglement generation, we assume that the Floquet lattice model [Eq.~\eqref{Bloch Hamiltonian}] provides a good approximation to the quantized-drives model [Eq.~\eqref{quantized model}] under the condition
\[
   \frac{\delta n_{1,2}}{\sqrt{n}_{1/2,c}} \ll 1
\]
throughout the time evolution. At later times, however, this condition breaks down because the growing $\Delta J_{z}$ causes $\delta n_{1,2}$ to become comparable to $n_{1/2,c}$ [Figs.~\ref{QFI and wavefunction}(e), (f), (h), (i)].

To study the convergence to the lattice model more quantitatively as $n_{c}$ increases, Fig.~\ref{Boundary effect} shows the growth of the QFI,
$F_{q}\bigl[\ket{\text{PES}}_{T}, \hat{J}_{z}\bigr]$,
for Fock-state inputs with different initial photon numbers (colored lines). Here, we use the Hamiltonian from Eq.~\eqref{polarization conversion}. For comparison, the growth of the QFI under the corresponding lattice model is plotted in gray. For the original model [Eq.~\eqref{quantized model}], the $T^2$ scaling of the QFI arises from the energy transfer between the two modes (cf App~\ref{QFI appendix}), which necessarily ends at $T = \mathcal{O}(N)$ for an initial population of $N$ bosons (by this time, one of the modes is nearly empty). Hence, the $\propto T^2$ growth in the QFI only persists up to $T = \mathcal{O}(N)$. Indeed, for each colored line, the QFI ceases to grow at $T=O(N)$. Consistent with this picture, our numerical simulations show that, for generic POVMs, the Fisher information also ceases to grow once $T$ reaches $\mathcal{O}(N)$.

Therefore, to attain the full Heisenberg-level sensitivity of $\sim N^2$, we require a POVM that retains the $T^2$ scaling of the Fisher information up to $T = \mathcal{O}(N)$. In Sec.~\ref{parity as decoder}, we show that, under suitable conditions on the Floquet band structure, parity measurements fulfill this role.

However, as the initial photon number increases, the QFI dynamics under Eq.~\eqref{polarization conversion} aligns more closely, and for longer times, with its lattice approximation. We observe a similar behaviour for the coherent-state input. Therefore, we conclude that the early-time dynamics of the quantized-drives model [Eq.~\eqref{quantized model}] is well described by Eq.~\eqref{Bloch Hamiltonian} as a direct sum of two-tone Floquet Hamiltonians [Eq.~\eqref{sin and cos}].

\section{Decoding a phase with parity measurements}

\label{parity as decoder}

\subsection{General requirements on drive frequencies and band structure}

We have shown that PESs with QFI scaling as $T^2$ up to $T = \mathcal{O}(N)$ can be generated in preparation time $T$ using various Floquet models of the form Eq.~\eqref{quantized model}. This raises a natural question: can such $T^2$ scaling of the QFI be preserved under measurements accessible to current experiments? In this section, we aim to identify general conditions under which models of the form Eq.~\eqref{quantized model} yield classical Fisher information (FI) $\propto T^2$ using experimentally feasible measurements. For quantized-drive models, this implies Heisenberg-limited (HL) sensitivity scaling as $N^2$ at $T = \mathcal{O}(N)$. Importantly, our definition of the HL refers to the $N^2$ scaling with respect to the number of photons $N$ in the probe - and we study the preparation time to realize states with HL scaling -, in contrast to definitions based on the \textit{interrogation time} used in Refs.~\cite{Zhou2018,Dong2022,Wang2023}.

In optical interferometry, one widely used measurement is the parity of an output mode of an interferometer \cite{10.1116/5.0026148}, and is shown to approach sub-SQL sensitivity for various path-entangled states \cite{PhysRevLett.107.083601,doi:10.1080/00107514.2010.509995,PhysRevA.82.013831,PhysRevLett.104.103602}. In principle, the parity measurement is scalable to large number of photons with recent experimental success \cite{Sun2014,Cohen:14}. We thus concentrate in this section on the use of the parity measurement. 

As described in Sec.~\ref{This work: Floquet quantum sensor}, we adopt a parity measurement on the second output mode $\hat{\Pi}\equiv\left(-1\right)^{\hat{a}_{2}^{\dagger}\hat{a}_{2}}$ after a Mach-Zehnder interferometer as our POVM measurement [Fig.~\ref{Floquet quantum sensor}.(c)]. By the end of this section, we will filter out Floquet models Eq.~\eqref{quantized model} capable of generating PES's that exhibit sub-SQL sensitivity from a single-shot parity measurement. Recall the phase imprinter and decoder in a Mach-Zehnder interferometer is given as $\hat{\mathcal{U}}_{\theta}^{\text{MZI}}=e^{-i\hat{J}_{z}\theta}$ and $\hat{\mathcal{U}}_{\text{de}}^{\text{MZI}}=e^{-i\hat{J}_{x}\frac{\pi}{2}}$, respectively. The parity operator, under these transforms, becomes

\begin{equation}
\hat{\Pi}_{\theta}=\hat{\mathcal{U}}_{\theta}^{\text{MZI},\dagger}\ \hat{\mathcal{U}}_{\text{de}}^{\text{MZI},\dagger}\ \hat{\Pi}\ \hat{\mathcal{U}}_{\text{de}}^{\text{MZI}}\ \hat{\mathcal{U}}_{\theta}^{\text{MZI}}=e^{2i\hat{J}_{z}\left(\theta-\frac{\pi}{2}\right)}\hat{S},
\label{parity operator}
\end{equation}

\noindent where $\hat{S}=\sum\limits_{n,m=1}^{\infty}\ket{n}\bra{m}\otimes\ket{m}\bra{n}$ is the SWAP operator over two input bosonic modes. Redefining the phase $\Tilde{\theta}\equiv\theta-\frac{\pi}{2}$, the parity operator has a simple dependence on $\Tilde{\theta}$ as $\hat{\Pi}_{\Tilde{\theta}}=e^{2i\hat{J}_{z}\Tilde{\theta}}\hat{S}$. The derivation of Eq.~\eqref{parity operator} is left to App.~\ref{parity operator appendix}.\\

From classical information theory (Sec.~\ref{Information theory}), the  variance of an unbiased estimator $V_{\text{ub}}(\Tilde{\theta})$ is lower-bounded by the inverse of the classical Fisher information $F_{\Tilde{\theta}}$ [Eq.~\eqref{classical Fisher information}], which, for the parity operator $\hat{\Pi}_{\Tilde{\theta}}$, is given by \cite{10.1116/5.0026148}
\begin{equation}
F_{\Tilde{\theta}}=\frac{\left|\partial_{\Tilde{\theta}}\braket{\Pi_{\Tilde{\theta}}}_{\text{PES}}\right|^2}{1-\braket{\Pi_{\Tilde{\theta}}}_{\text{PES}}^{2}},
\label{parity sensitivity}
\end{equation}
 where $\braket{\square}_{\text{PES}}\equiv\bra{\text{PES}}_{T}\square\ket{\text{PES}}_{T}$ the expectation value of $\square$ with respect to the PES.\\

For $\omega_{1}\neq\omega_{2}$, we expect Eq.~\eqref{parity sensitivity} to be unable to reach either the SQL or the HL. We leave the analytic arguments to App.~\ref{characteristic functions appendix}.
Instead, here we provide the physical intuition behind this observation: in an SU($2$) interferometer, all unitary operators commute with the total number operator, which is the total spin $\hat{S}^{2}=\hat{S}_{x}^{2}+\hat{S}_{y}^{2}+\hat{S}_{z}^{2}$ in a Ramsey interferometer and the total number of photons $\hat{N}=\hat{a}_{1}^{\dagger}\hat{a}_{1}+\hat{a}_{2}^{\dagger}\hat{a}_{2}$ in a Mach-Zehnder interferometer. In other words, each operation in the sensing circuit preserves the angular momentum or total number of particles. Superposition and interference can hence only exist between states with identical angular momentum or number of particles. In the language of functional power operators, the conservation of particle number is translated into $\hat{P}_{1}\left[f\right]+\hat{P}_{2}\left[f\right]=0$. This is consistent with the energy conservation $\omega_{1}\hat{P}_{1}\left[f\right]+\omega_{2}\hat{P}_{2}\left[f\right]=0$ if and only if two frequencies are identical $\omega_{1}=\omega_{2}$. We thus find that a necessary condition for approaching the SQL or HL is having drives with identical frequencies.\\

For $\omega_{1}=\omega_{2}=\omega$, in general, we show in App.~\ref{characteristic functions appendix} that the asymptotic behaviors of $\braket{\Pi_{\Tilde{\theta}}}_{\text{PES}}$ and $\partial_{\Tilde{\theta}}\braket{\Pi_{\Tilde{\theta}}}_{\text{PES}}$ are, instead, governed by
$\braket{\Pi_{\Tilde{\theta}}}_{\text{PES}}=O(\frac{1}{\sqrt{T}})$ and $\partial_{\Tilde{\theta}}\braket{\Pi_{\Tilde{\theta}}}_{\text{PES}}= O\left(\sqrt{T}\right)$, which endow $F_{\Tilde{\theta}}=O(T)$ that approaches the SQL ($\sim N$) at $T=O(N)$. Although the QFI of path-entangled states $\ket{\text{PES}}_{T}$ exhibit the $T^2$ scaling in general, the parity measurement $\braket{\Pi_{\Tilde{\theta}}}_{\text{PES}}$ can only extract this phase information with precision given by the SQL. This is because the time evolution appears in exponentials $e^{-i\epsilon_{n}\left(\bm{\phi}\right)T}$ [Eq.~\eqref{time-evolution unitary}] oscillating fast at $T\gg1$ and wiping out a significant amount of information as most of them average to 0 in $\braket{\hat{\Pi}_{\Tilde{\theta}}}$.\\

\begin{figure}
    \centering
    \includegraphics[width=0.6\linewidth]{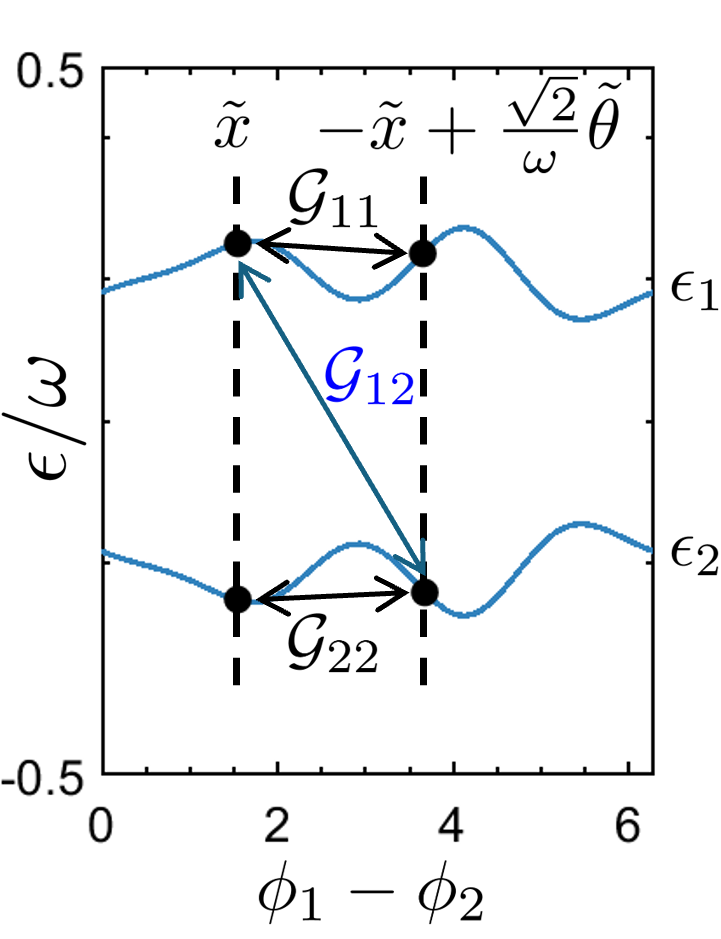}
    \caption{\textit{A Schematic illustration of the characteristic functions.} For the qubit model described in Eq.~\eqref{specific model}, one has two quasienergy bands symmetric about $\epsilon=0$. Consequently, two classes of characteristic functions can be defined: intraband, $\mathcal{G}_{11}$ and $\mathcal{G}_{22}$, and interband, $\mathcal{G}_{12}$ and $\mathcal{G}_{21}$. Physically, they measure the difference in quasienergies within and between bands. The band structure of Eq.~\eqref{specific model} is obtained for $\omega=\omega_{0}$ and $\mathcal{A}=\frac{\omega_{0}}{2}$.}
    \label{CharacteristicFunction}
\end{figure}

To overcome such fast oscillation in the long time limit, stationary phases (also known as critical points) of the exponentials have to be degenerate, giving a 'steady region' of $\epsilon_{n}$ in which the exponentials vary slowly. This, as derived in App.~\ref{characteristic functions appendix}, can be formulated by the following condition:

\begin{equation}
    \partial^{2}_{\Tilde{x}} \mathcal{G}_{nm}(\Tilde{x},\Tilde{\theta})|_{\Tilde{x}=\Tilde{x}_{0}}=0,\quad\partial_{\Tilde{\theta}} \mathcal{G}_{nm}(\Tilde{x}_{0},\Tilde{\theta})\neq0,
\label{2nd condition}
\end{equation}

\noindent where 

\begin{equation}
\mathcal{G}_{nm}(\Tilde{x},\Tilde{\theta})\equiv\epsilon_{n}\left(\Tilde{x}\right)-\epsilon_{m}(-\Tilde{x}+\frac{\sqrt{2}}{\omega}\Tilde{\theta})
\label{characteristic g function}
\end{equation}

\noindent is a measure of energy difference between band $n,m=1,2$ at $\Tilde{x}$ and $-\Tilde{x}+\frac{\sqrt{2}}{\omega}\Tilde{\theta}$, respectively. $\Tilde{x}\equiv\frac{1}{\sqrt{2}\omega}\left(\phi_{1}-\phi_{2}\right)$ is the time delay (recall, due to energy conservation, the quasi energies only depend on a single parameter $\tilde x$) between two drives and $\Tilde{x}_{0}$ the critical point defined by $\partial_{\Tilde{x}}\mathcal{G}_{nm}(\Tilde{x},\Tilde{\theta})|_{\Tilde{x}=\Tilde{x}_{0}}=0$. An asymptotic quadratic enhancement $F_{\Tilde{\theta}}\sim T^2$ at $\Tilde{\theta}$ is obtainable from the parity measurement on $\ket{\text{\normalfont PES}}_{T}$ only if Eq.~\eqref{2nd condition} has solution $\Tilde{x}_{0}$ for at least one choice of $n,m$. Here, we assume that $\Tilde{x}_{0},-\Tilde{x}_{0}+\frac{\sqrt{2}}{\omega}\Tilde{\theta}$ lie within the support of $f(\bm{\phi})$. We hence dub $\mathcal{G}_{nm}$ characteristic functions (of quasienergy bands) determining the sensitivity from parity measurement. A schematic illustration of the characteristic functions $\mathcal{G}_{nm}$ for a 2-level ancilla can be found in Fig.~\ref{CharacteristicFunction}.

\begin{figure*}
\centering
\includegraphics[width=\textwidth]{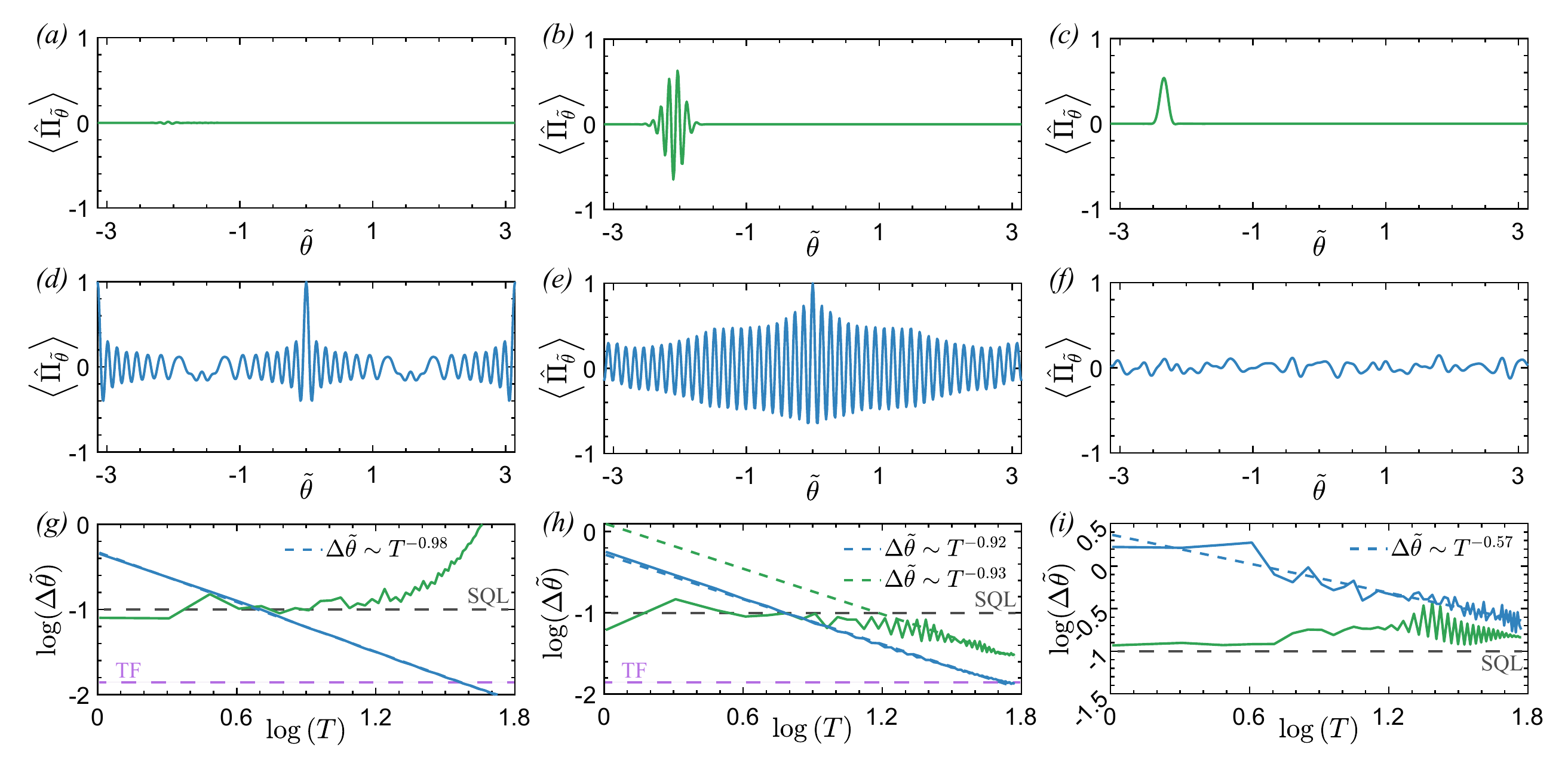}
        \caption{\textit{Parity measurements and corresponding sensitivity for three models:} (a)(d)(g) $2$-level system with circularly-polarized drives, (b)(e)(h) polarization conversion and (c)(f)(i) the specific model given in Eq.~\eqref{specific model}. In (a)(b)(c), we plot the expectation value of the parity operator $\hat{\Pi}_{\Tilde{\theta}}$ as a function of $\Tilde{\theta}$ for coherent-state inputs. In (d)(e)(f), the same quantity is plotted for Fock-state inputs. The optimal sensitivity from the parity measurement for both inputs and all three models is given in (g)(h)(i), where the green (blue) curves and dashed green (blue) lines represent the sensitivity of coherent (Fock) inputs and their asymptotic scaling, respectively. For all three models, we start with average $50$ bosons in each drive, with the SQL given as $\Delta\Tilde{\theta}=10^{-1}$ (dashed-gray line), the HL given as $\Delta\Tilde{\theta}=10^{-2}$ and the parity sensitivity for the twin-Fock state indicated as dashed-violet line for comparison. For coherent-state inputs, we introduce an initial phase difference $\phi_{20}-\phi_{10}=0.6\pi$ between two drives. Significantly, the coherent-state input for the polarization model yields a sub-SQL precision in (h), which is further illustrated in Fig. \ref{Uncertainty_and_loss}.(a). Other parameters in different models are given as follows: for the specific model given in Eq.~\eqref{specific model}, $\omega=\omega_{0}$, $\mathcal{A}=\frac{\omega_{0}}{2}$; for the $2$-level system with circularly-polarized drives, $\omega=\frac{\omega_{0}}{4}$, $\mathcal{A}=\frac{\omega_{0}}{8}$; and for the polarization conversion, $\omega=\omega_{0}$, $\mathcal{A}=\frac{\omega_{0}}{2}$.}
        \label{Paritysensitivity}    
\end{figure*}

\subsection{Numerical examples}

\subsubsection{Example I: Two circularly-polarized drives with same chiralities}
\label{Example I: Two identical circularly-polarized drives}
A simple way to produce degenerate critical points, as defined in Eq.~\eqref{2nd condition}, is requiring a symmetric band structure about some value of $\Tilde{\theta}$. An example of symmetric band structures can be seen in our previous $2$-level system with identical circularly polarized drives [Eq.~\eqref{2-level with circularly-polarized drives}]. In this model, two bands meet at $\phi_{1}-\phi_{2}=\pi$ while being symmetric about $\phi_{1}-\phi_{2}=0, \pi$, which gives $\mathcal{G}_{11}(\Tilde{x},0)=\mathcal{G}_{22}(\Tilde{x},0)=0$ and $\mathcal{G}_{11}(\Tilde{x},\pi)=\mathcal{G}_{22}(\Tilde{x},\pi)=0$ for all $\Tilde{x}\in\left[-\pi/\sqrt{2}\omega,\pi/\sqrt{2}\omega\right)$.\\ 

Consequently, a generic field distribution $f\left(\bm{\phi}\right)$ is expected to exhibit an asymptotic scaling $\Delta\Tilde{\theta}\sim T^{-1}$ at $T\gg T_{\text{com}}$ for $\Tilde{\theta}=0, \pi$ [Fig.~\ref{Paritysensitivity}.(b): from numerics, we observe that with a Fock state input the sensitivity scales approximately as $\Delta\Tilde{\theta}\sim T^{-0.98}$ at $\Tilde{\theta}=0, \pi$]. However, for other values of $\Tilde{\theta}$, critical points of $\mathcal{G}_{nm}(\Tilde{x},\Tilde{\theta})$ become non-degenerate and hence fail to approach the sub-SQL sensitivity.

There is one subtlety for coherent-state inputs: a symmetric coherent-state input $f\left(\phi_{1},\phi_{2}\right)=f\left(\phi_{2},\phi_{1}\right)$ requires a degenerate critical point at $\Tilde{\theta}=0$ (App.~\ref{characteristic functions appendix}), but as the power $\partial_{\Delta\phi}\epsilon_{1,2}$ is vanishing at the same point, it fails to satisfy the second condition in Eq.~\eqref{2nd condition} and fails to exhibit $\Delta\Tilde{\theta}\sim T^{-1}$. On the other hand, an asymmetric distribution $\left|f\left(\bm{\phi}\right)\right|^{2}=\delta\left(\phi_{1}-\phi_{10},\phi_{2}-\phi_{20}\right)$ for coherent-state inputs requires a degenerate critical point of $\mathcal{G}_{nm}(\Tilde{x},\Tilde{\theta})$ at $\Tilde{\theta}=\phi_{10}-\phi_{20}$ (App.~\ref{characteristic functions appendix}), where $\phi_{10,20}$ are (different) initial phases of the coherent drives. This is only obtainable for $\phi_{10}-\phi_{20}=\pi$, which again leads to vanishing power as $\partial_{\Delta\phi}\epsilon_{1,2}|_{\Delta\phi=\pi}=0$. As a result, this model [Eq.~\eqref{2-level with circularly-polarized drives}] fails to reach the HL for any coherent-state input with the parity measurement. Thus, for a coherent-state input, we find somewhat counterintuitively that the QFI grows quadratically while the parity measurement is not sensitive to the encoded phase [Fig.~\ref{Paritysensitivity}.(b), (h)].

\subsubsection{Example II: Two circularly-polarized drives with opposite chiralities}
\label{Example II: Two circularly-polarized drives with opposite chiralities}
Two ways to recover the $\Delta\Tilde{\theta}\sim T^{-1}$ scaling for coherent-state inputs are as follows: $(1)$ a degenerate critical point at $\Tilde{\theta}=0, \pi$, while maintaining a finite dispersion $\partial_{\Delta\phi}\epsilon_{1,2}\neq0$, or $(2)$ requiring linear and symmetric dispersion about some value of $\phi_{1}-\phi_{2}$. For $(1)$, one can simply choose a symmetric distribution or a $\pi$-phase difference for the coherent-state input, and the non-vanishing power ensures an asymptotic $\Delta\Tilde{\theta}\sim T^{-1}$ scaling. For $(2)$, the restriction on $\Tilde{\theta}$ is further lifted, since linearity and symmetry allow $\mathcal{G}_{nm}(\Tilde{x},\Tilde{\theta})$ with degenerate critical points for any $\Tilde{x}$ and $-\Tilde{x}+\frac{\sqrt{2}}{\omega}\Tilde{\theta}$ as long as they both lie in such linear and symmetric regions (recall Dirac points). Both cases can be found in our previous polarization conversion model [Eq.~\eqref{polarization conversion}], where the band structure is symmetric and linear about both $\Tilde{\theta}=0$ and $\Tilde{\theta}=\pi$, and has a Dirac point at $\Tilde{\theta}=0$, resulting in a non-vanishing power $\partial_{\Delta\phi}\epsilon_{1,2}|_{\Delta\phi=0}$. For $\Tilde{\theta}=\pi$, the power vanishes while retaining its symmetry and linearity about this point. Consequently, the model follows $\Delta\Tilde{\theta}\sim T^{-1}$ for coherent-state inputs even for a general asymmetric distribution $f\left(\bm{\phi}\right)$ that requires degenerate critical points at $\Tilde{\theta}\neq 0,\pi$ [Fig.~\ref{Paritysensitivity}.(c), (i): an initial phase difference $\phi_{20}-\phi_{10}=0.6\pi$ is chosen for the coherent-state input]. To benchmark this $T^{-1}$ behavior and to show the validity of the lattice model, in Fig. \ref{Uncertainty_and_loss}.(a), we plot the sensitivity $\Delta\Tilde{\theta}$ of the parity measurement for coherent-input states with different photon numbers $n_{c}$. As $n_{c}$ increases from $50$ to $80$,  the sensitivity follows more closely the $T^{-1}$ scaling predicted by the lattice model (gray line). For Fock-state inputs, this lifted restriction on $\Tilde{\theta}$ also poses less constraints on the a priori knowledge of $\Tilde{\theta}$ and results in a wider range of sub-SQL sensitivity in the spectrum [Fig.~\ref{Paritysensitivity}.(f)].\\

\subsubsection{Example III: a specific driven qubit model}
Finally, we consider the sensitivity of the parity measurement for a specific model given as 

\begin{equation}
\begin{split}
    \hat{H}_{\bm{\phi}}\left(t\right)=&\frac{\omega_{0}}{2}\hat{\sigma}_{z}+\mathcal{A}\left[3\sin\left(\omega t+\phi_{1}\right)\hat{\sigma}_{y}+\sin\left(\omega t+\phi_{2}\right)\hat{\sigma}_{x}\right]\\
    &+\mathcal{A}\left[\cos\left(\omega t+\phi_{1}\right)+2\cos\left(\omega t+\phi_{2}\right)\right]\hat{\sigma}_{z},
\end{split}
\label{specific model}
\end{equation}

\noindent whose quasienergy bands are given in Fig.~\ref{CharacteristicFunction} and do not exhibit any degeneracy or symmetry. Specifically, its characteristic functions $\mathcal{G}_{nm}(\Tilde{x},\Tilde{\theta})$ only have non-degenerate critical points for any value of $\Tilde{\theta}$. Consequently, the HL is inaccessible for any type of inputs and the non-degenerate critical points can only approach the SQL [Fig.~\ref{Paritysensitivity}.(c), (f), (i)].\\

\section{Particle loss and a finite phase range}
\label{Loss}

\begin{figure*}
\centering
\includegraphics[width=0.93\linewidth]{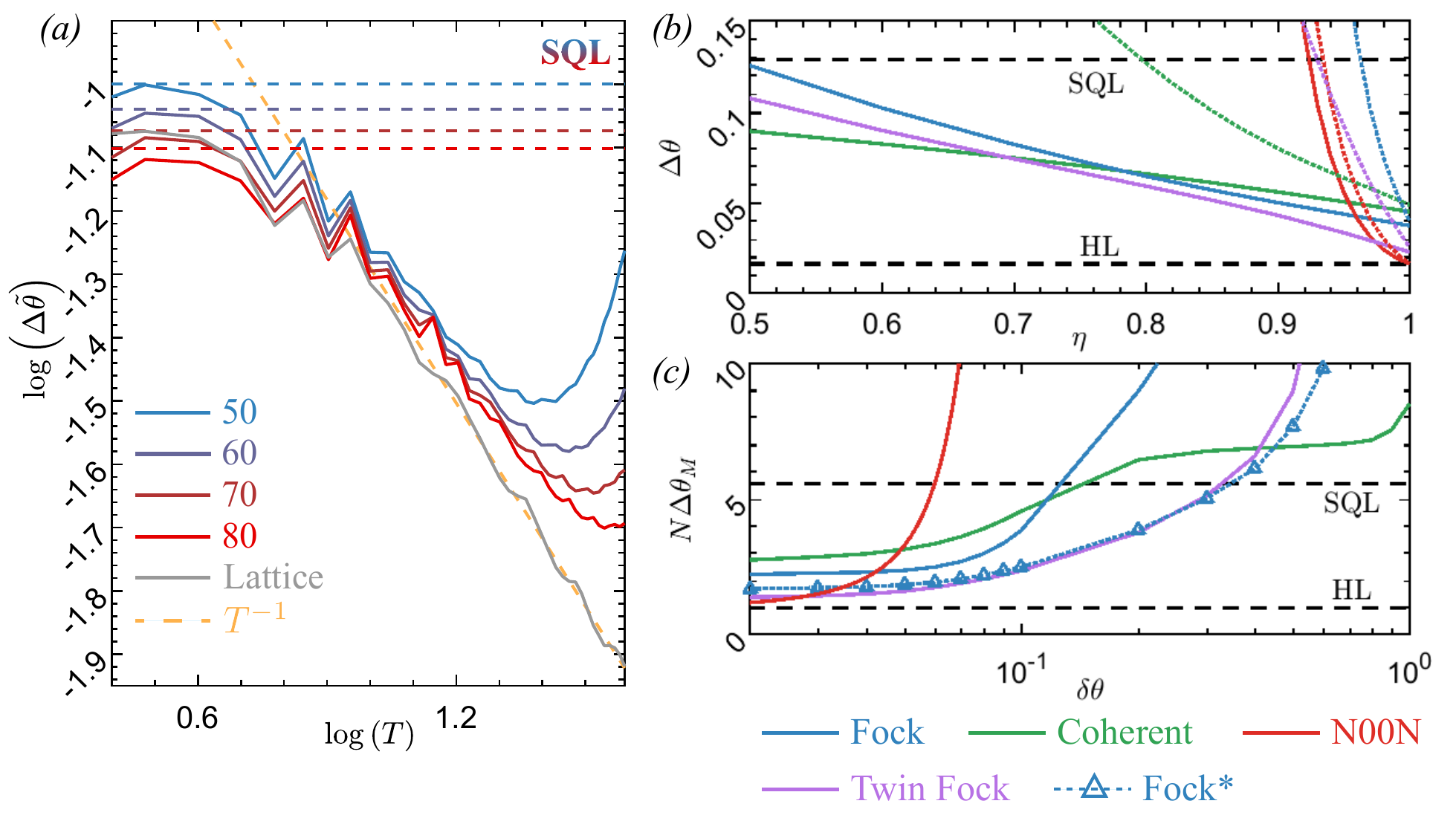}
        \caption{\textit{Sensitivity with and without particle loss and a priori phase uncertainties.} (a) Sensitivity of the parity measurement for coherent-state inputs. We plot the time evolution of $\Delta\Tilde{\theta}$ for different initial photon numbers per mode, ranging from $n_{c}=50$ to $n_{c}=80$. The sensitivity $\Delta\Tilde{\theta}(T)$ for the lattice model (gray line) is given for comparison. We also show the SQL (blue-dashed and red-dashed lines) for each value of $n_{c}$ as well as the $\Delta\Tilde{\theta}\sim T^{-1}$ (yellow-dashed line) for the lattice model. (b) The sensitivity $\Delta\theta$ from both the qCRB (solid) and parity measurement (dashed) for coherent-state (green) and Fock-state (blue) inputs. We also plot $\Delta\theta$ for the N$00$N state (red) and the twin-Fock state (violet) for comparison. The entangler is chosen as the polarization conversion model with $\omega=\omega_{0}$, $\mathcal{A}=\frac{\omega_{0}}{2}$ and $30$ bosons in each mode initially. For the N00N state, we choose $N=60$, while for the twin-Fock state, we have $\ket{TF}=e^{i\frac{\pi}{4}\left(\hat{a}_{1}^{\dagger}\hat{a}_{2}+\hat{a}_{2}^{\dagger}\hat{a}_{1}\right)}\ket{\frac{N}{2}}\otimes\ket{\frac{N}{2}}$ with $N=60$. (c) The improvement $\Delta\theta_{m}$ in Eq.~\eqref{improvement} as a function of the a priori phase uncertainty $\delta\theta$ for coherent (green) and Fock (blue) inputs. The behaviors of $\Delta\theta_{m}$ for the N$00$N state (red) and for the twin-Fock state (violet) are shown for comparison. The model is the same as in (a), but with $15$ bosons in each mode initially. For the N00N state and the twin-Fock state, we choose $N=30$. We also plot the improvement $\Delta\theta_{M}$ for the Fock input (dashed-$\Delta$) from the Hamiltonian Eq.~\eqref{2-level with circularly-polarized drives}. It shows a significant improvement to that for the twin-Fock state.}
        \label{Uncertainty_and_loss}    
\end{figure*}

In this section, we discuss the robustness of the PES's against loss in the interferometer and a finite (a priori) phase range $\delta\theta$. The mathematical tools employed are the Kraus representation for lossy dynamics and the Bayesian approach (App.~\ref{global phase estimation}) for dealing with a finite (a priori) phase range.\\

\subsection{Effects of particle loss}

The first source of error arises from the potential loss of particles in the interferometer. We use Kraus operators \cite{PhysRevA.80.013825,PhysRevLett.102.040403} to simulate such loss. To be specific, we consider an interferometer with one lossy arm ($\hat{a}_{1}$) and one lossless arm ($\hat{a}_{2}$). We expect qualitatively similar behaviors when both arms are lossy. In the Kraus representation, the state after a lossy transmission line with transmissivity $\eta\in\left[0,1\right]$ is given by:

\begin{equation}
    \hat{\rho}=\sum_{j=0}^{+\infty}\hat{K}_{j}\hat{\rho}_{0}\hat{K}_{j}^{\dagger},
\end{equation}
\noindent with $\hat{\rho}_{0}=\ket{\text{PES}}_{T}\bra{\text{PES}}_{T}$ and $\hat{K}_{j}\equiv\left(1-\eta\right)^{\frac{j}{2}}\eta^{\frac{1}{2}\hat{a}_{1}^{\dagger}\hat{a}_{1}}\hat{a}_{1}^{j}/\sqrt{j!}$. Noting that $\eta=1$ represents a perfect transmission with no boson loss while $\eta=0$ stands for a complete loss of bosons. The simulation is made for the polarization conversion model with $30$ input bosons in each drive with identical frequencies $\omega_{1}=\omega_{2}=\omega_{0}$ and renormalized amplitudes $\mathcal{A}=\frac{\omega_{0}}{2}$. For the coherent-state input, we specify an initial phase difference $\phi_{20}-\phi_{10}=0.6\pi$. The conversion time is chosen as $T=32T_{\text{com}}$. In Fig.~\ref{Uncertainty_and_loss}.(b), we plot the sensitivity $\Delta\theta$ both from the QFI (solid lines) and parity measurement (dashed lines) as a function of the transmissivity $\eta$ for coherent-state inputs (green), Fock-state inputs (blue), the twin-Fock state (violet) and the N00N state (red). Typically, the QFIs for coherent-state and Fock-state inputs exhibit robustness against loss and present sub-SQL sensitivity even at $\eta=0.5$, showing a similar behavior as the twin-Fock state. When $\eta<0.7$, the coherent-state input starts to outperform the twin-Fock state. In terms of the parity measurement, this advantage starts to show up for $\eta<0.97$. As comparison, the N00N state is much more fragile to loss and already goes beyond the SQL at $\eta\approx0.91$. This is consistent with the observation that entangled coherent states and $\ket{N-m:m}\equiv\frac{1}{\sqrt{2}}\left(\ket{N-m,m}+\ket{m,N-m}\right)$ with $1\leq m\leq N-1$ are, in general, more robust than the N00N state against loss \cite{PhysRevLett.102.040403}. The parity measurement, however, appears more sensitive to the lossy interferometer for all three kinds of states. Specifically, the parity measurement for the Fock-state input and N00N state performs worse than the SQL for $\eta<0.93$. The coherent-state input is more robust and its super-resolution survives up to $\eta\approx0.8$. However, this is still much worse than the QFI itself. As a result, we conclude that the PES's we generate are robust against loss themselves, while the parity measurement is less able to extract the QFI when there is significant loss.\\

When the loss is small, however, all PES's converge to almost the same sub-SQL sensitivity as a result of the linearity of the band [Fig.~\ref{QFI and wavefunction}.(c)]. However, such sub-SQL sensitivity only applies for specific values of $\Tilde{\theta}$ for a general field distribution $f\left(\bm{\phi}\right)$. In Fig.~\ref{Paritysensitivity}.(c), (f), we plot the expecation value of the parity operator $\braket{\hat{\Pi}_{\Tilde{\theta}}}$ for coherent-state inputs (c) and Fock-state inputs (f) at $\eta=1$ as a function of $\Tilde{\theta}$. For coherent-state inputs, the sub-SQL sensitivity only appears within a finite range of $\Tilde{\theta}$, while Fock-state inputs are sub-SQL in a wider range. As seen earlier, this is an interplay between degenerate critical points of $\mathcal{G}$ and initial field distributions $f\left(\bm{\phi}\right)$. Consequently, when the transmission is perfect, one is, potentially, able to tailor the input $f\left(\bm{\phi}\right)$ such that the range of $\Tilde{\theta}$ exhibiting sub-SQL sensitivity is maximized.

\subsection{Effects of a finite phase range}
\label{Effects of a priori uncertainty}

Another source of error comes from a finite (a priori) phase range 
$\delta\theta$. This phase uncertainty arises when a priori knowledge of its value is lacking. The quantity of interest now becomes the posterior variance, $\Delta_{\text{p}}\theta^{2}$ given in Eq.~\eqref{MSE cost function}, which one aims to minimize (App.~\ref{global phase estimation}). Given the probe state as $\ket{\psi}_{\text{p}}$ ($\hat{\rho}_{\text{p}}$ for a mixed state) and the phase imprinting unitary as $\hat{\mathcal{U}}_{\theta}$, one then needs to minimize $\Delta_{\text{p}}\theta^{2}$ over all possible POVM measurements. This challenging task is facilitated by the work done in \cite{Macieszczak_2014}, which narrows the infinitely many possililities down to projective measurements $\left\{\ket{\mu}\bra{\mu}\right\}$ with outcome $\mu$ satisfying $\sum\limits\ket{\mu}\bra{\mu}=\hat{\mathds{1}}_{0}$. The projective measurement can hence be described by a hermitian operator

\begin{equation}
    \hat{S}=\sum\limits_{\mu}\mu\ket{\mu}\bra{\mu}.
\end{equation}

\noindent The task now becomes finding the optimal $\hat{S}$ that minimizes $\Delta_{\text{p}}\theta^{2}$. Denoting the probe state after the phase imprinter as $\hat{\rho}_{\theta}=\hat{\mathcal{U}}_{\theta}\hat{\rho}_{\text{p}}\hat{\mathcal{U}}^{\dagger}_{\theta}$, such a minimization is realized by choosing \cite{Rubio_2019}

\begin{equation}
\hat{S}=2\sum_{k,l,\lambda_{k}+\lambda_{l}>0}\frac{\bra{k}\overline{\rho}\ket{l}}{\lambda_{k}+\lambda_{l}}\ket{k}\bra{l},
\end{equation}

\noindent where $\lambda_{k}$ and $\ket{k}$ are eigenvalues and eigenstates of the $\theta$-averaged mixed state $\rho\equiv\int d\theta P\left(\theta\right)\hat{\rho}_{\theta}$ and $\overline{\rho}\equiv\int d\theta P\left(\theta\right)\hat{\rho}_{\theta}\ \theta$. The upper bound $\Delta\theta_{m}^{2}$ of optimization from the measurement is now given as the difference between the inverse of the minimal posterior variance $\Delta_{\text{p}}\theta^{2}$ and the a priori classical CRB of the a priori phase distribution $\frac{1}{F_{0}}$ as \cite{PhysRevX.11.041045}:

\begin{equation}
    \frac{1}{\Delta\theta_{m}^{2}}=\frac{1}{\Delta_{\text{p}}\theta^{2}}-\frac{1}{F_{0}}.
\label{improvement}
\end{equation}

\noindent The HL and SQL are now defined with respect to $\Delta\theta_{m}$. Numerically, we again choose the polarization conversion model Eq.~\eqref{polarization conversion} with $15$ bosons in each drive with identical frequencies $\omega_{1}=\omega_{2}=\omega_{0}$ and renormalized amplitudes $\mathcal{A}=\frac{\omega_{0}}{2}$. As before, we specify an initial phase difference $\phi_{20}-\phi_{10}=0.6\pi$ for the coherent-state input. The probe state $\ket{\text{PES}}_{T}$ is obtained at $T=8T_{\text{com}}$. In Fig.~\ref{Uncertainty_and_loss}.(c), we plot $\Delta\theta_{m}$ as a function of the a priori uncertainty $\delta\theta$ for coherent-state inputs (green), Fock-state inputs (blue), the twin-Fock state (violet) and the N00N state (red) upon assuming a Gaussian distribution

\begin{equation}
P\left(\theta\right)=\frac{1}{\sqrt{2}\delta\theta}\exp{\left(-\frac{\theta^{2}}{2\delta\theta^{2}}\right)}
\end{equation}

\noindent of the uncertain phase \cite{PhysRevX.11.041045}. For small a priori uncertainty $\delta\theta<0.1$, the PES's exhibit sub-SQL sensitivity and go beyond the SQL when $\delta\theta>0.2$ for both coherent-state and Fock-state inputs. Meanwhile, the N00N state is less effective to the finite a priori distribution $\delta\theta$ and performs worse than the SQL for $\delta\theta>0.06$. In comparison, the twin-Fock state exhibits a better resilience to $\delta\theta$. When switching the entangler to Eq.~\eqref{2-level with circularly-polarized drives}, we observe a significant improvement in robustness against phase uncertainties for the same Fock-state input (blue $-\Delta-$), comparable to the twin-Fock state. However, they both require high-Fock-state inputs (which are challenging for current experiments) and leave the sub-SQL region for $\delta\theta>0.35$. The PES's generated by our Floquet entangler are hence ineffective when trying to measure a phase that has a non-local a priori uncertainty. The advantage that entangled states have when measuring a minuscule phase shift require some prior knowledge of the vicinity scrutinized, but this results in poor  performance when such knowledge is absent (cf Ref. \cite{Marciniak2022}).\\

\subsection{Other sources of noise and imperfections}

For other types of noise and imperfections that may compromise the performance of the Floquet sensor, we refer the reader to Appendix~\ref{Other noise}. In particular, we consider dephasing noise on the qubit, phase noise in the driving mode, and the effects of imperfect driving-frequency tuning on the parity measurement. 

We remark that this work considers mainly uncorrelated noise. It is well known that such noise typically degrades metrological performance and leads to the well-established no-go theorem \cite{Demkowicz-Dobrzański2012}. Nevertheless, recent research indicates that correlated noise can, in some cases, be harnessed to control and even prolong entanglement between qubits \cite{Zou2024,PhysRevA.97.042109}. Although this type of noise lies beyond the scope of the present study, we believe it holds significant potential for future advances in quantum-enhanced metrology.

\section{Experimental considerations}

\label{Experiment realization}

In this section, we propose an experiment that realizes our Floquet quantum sensor. After discussing general experimental considerations, we will propose a superconducting circuit realization. As discussed in Sec.~\ref{entanglement generation}, the prerequisite of the path entanglement is a dispersive quasienergy band structure that depends on initial phases $\phi_{1},\phi_{2}$ of two drives. Given the quasienergies evaluated and periodic in the driving frequency $\omega_{1,2}$, the amplitudes of drives should hence be comparable to these frequencies. In the quantized model [Eq.~\eqref{lattice model}], this requires the normalized coupling $\hat{H}_{1/2,0/e}^{c}$ to be comparable to $\omega_{1,2}$, posing significant constraints when the drives are conventional optical cavities (Fabry-Perot cavity, microsphere cavity, etc), where the frequency ($>100\text{THz}$) is much larger than their typical coupling ($<1\text{GHz}$) to matter \cite{RevModPhys.85.623,doi:10.1126/science.287.5457.1447,PhysRevA.57.R2293}. Energy transfer in such cavities typically requires $N>10^{10}$ particles for enhancement.\\

To lower this bound on the number of particles, we suggest using microwave cavities (coplanar waveguide resonator, LC resonator, etc.) with significantly smaller frequencies $\omega\sim 1-10\text{GHz}$ \cite{Niemczyk2010,PhysRevLett.105.237001}, which couple to superconducting qubits in a controllable way. Based on our protocol for generating PES's, the interaction time between the qubit and cavities needs to be shorter than the relaxation time $t_{1}$ and dephasing time $t_{2}$ of the qubit (given the $Q$-factor of the two resonators is high enough). Consequently, the frequency of the resonators is lower-bounded by the inverse of the coherence time as $\omega\gg\frac{1}{t_{1,2}}$ (this also implies that qubits with long coherence times are preferred). Assuming these conditions are satisfied, we now discuss a potential realization of Floquet entanglers in superconducting circuits.

We next show how the examples used in this work can be realized in a superconducting framework. Again, our discussion falls into two regimes: classical drive and quantum drives. We discuss the realization of these regimes separately.

The realization of Eq.~\eqref{2-level with circularly-polarized drives} is straightforward: for classical drives, the superconducting qubit (for example, a transmon), described by $\frac{\omega_{0}}{2}\hat{\sigma}_{z}$, is coupled to both a microwave current source $I(t)=\mathcal{A}\left[\cos(\omega t+\phi_{1})+\cos(\omega t+\phi_{2})\right]$ and a microwave voltage source $V(t)=\mathcal{A}\left[\cos(\omega t+\phi_{1})+\cos(\omega t+\phi_{2})\right]$; for quantum drives, the same superconducting qubit is coupled to each resonator both inductively $\hat{I}=g\left(\hat{a}_{1,2}+\hat{a}_{1,2}^{\dagger}\right)\hat{\sigma}_{x}$ and capacitively $\hat{V}=ig\left(\hat{a}_{1,2}-\hat{a}_{1,2}^{\dagger}\right)\hat{\sigma}_{y}$. These interactions can be achieved by an inductor and a capacitor, respectively.

For the polarization conversion model [Eq.~\eqref{polarization conversion}], the realization, instead, requires some engineering. For the classical drives, we apply both current and voltage drives on the qubit as [Fig.~\ref{SC element}.(b)]

\begin{equation}
    \hat{H}\left(t\right)=\frac{\omega_{q}}{2}\hat{\sigma}_{z}+I\left(t\right)\hat{\sigma}_{x}+V\left(t\right)\hat{\sigma}_{y},
\label{SC classical}
\end{equation}

\noindent with the current drive $I\left(t\right)=2\text{Re}\left\{e^{i\omega_{q}t}\left[\frac{\omega_{0}}{2}+\mathcal{A}\cos\left(\omega t+\phi_{1}\right)+\mathcal{A}\cos\left(\omega t+\phi_{2}\right)\right]\right\}$ and the voltage drive $V\left(t\right)=2\mathcal{A}\cdot\text{Re}\left\{e^{i\omega_{q}t}\left[\sin\left(\omega t+\phi_{1}\right)-\sin\left(\omega t+\phi_{2}\right)\right]\right\}$. By moving to a rotating frame with $\hat{U}\left(t\right)=e^{-i\frac{\omega_{q}t}{2}\hat{\sigma}_{z}}$ and a following rotating-wave approximation \cite{PhysRevLett.126.163602}, the interaction Hamiltonian becomes the Floquet one in Eq.~\eqref{polarization conversion}.

The same technique applies when the drives are quantum, where the qubit and cavities are coupled in a time-dependent way

\begin{equation}
\begin{split}
    \hat{H}\left(t\right)=&\frac{\omega_{q}}{2}\hat{\sigma}_{z}+\omega\left(\hat{a}_{1}^{\dagger}\hat{a}_{1}+\hat{a}_{2}^{\dagger}\hat{a}_{2}\right)\\
    &+2\mathcal{A}\cos\left(\omega_{q}t\right)\left[\frac{\hat{a}_{1}-\hat{a}^{\dagger}_{1}}{2i}-\frac{\hat{a}_{2}-\hat{a}^{\dagger}_{2}}{2i}\right]\hat{\sigma}_{y}\\
    &+2\cos\left(\omega_{q}t\right)\left[\frac{\omega_{0}}{2}+\mathcal{A}\frac{\hat{a}_{1}+\hat{a}^{\dagger}_{1}}{2}+\mathcal{A}\frac{\hat{a}_{2}+\hat{a}^{\dagger}_{2}}{2}\right]\hat{\sigma}_{x}.
\end{split}
\label{SC realization}
\end{equation}

\noindent Similarly, by moving to the rotating frame with $\hat{U}\left(t\right)$ and keeping non-oscillating terms only, one obtains Eq.~\eqref{polarization conversion} with quantum drives in the interaction picture. We now show how such a Hamiltonian can potentially be realized in a superconducting circuit. We split our discussion on elements [Fig.~\ref{SC element}.(a)] in our Floquet sensor and leave a detailed analysis of the superconducting circuit to App.~\ref{superconducting circuit appendix}. See Fig.~\ref{SC element}.(c) for a simplified version.

\subsection{Input: Superconducting qubit and cavities}

\begin{figure}
    \centering
    \includegraphics[width=\linewidth]{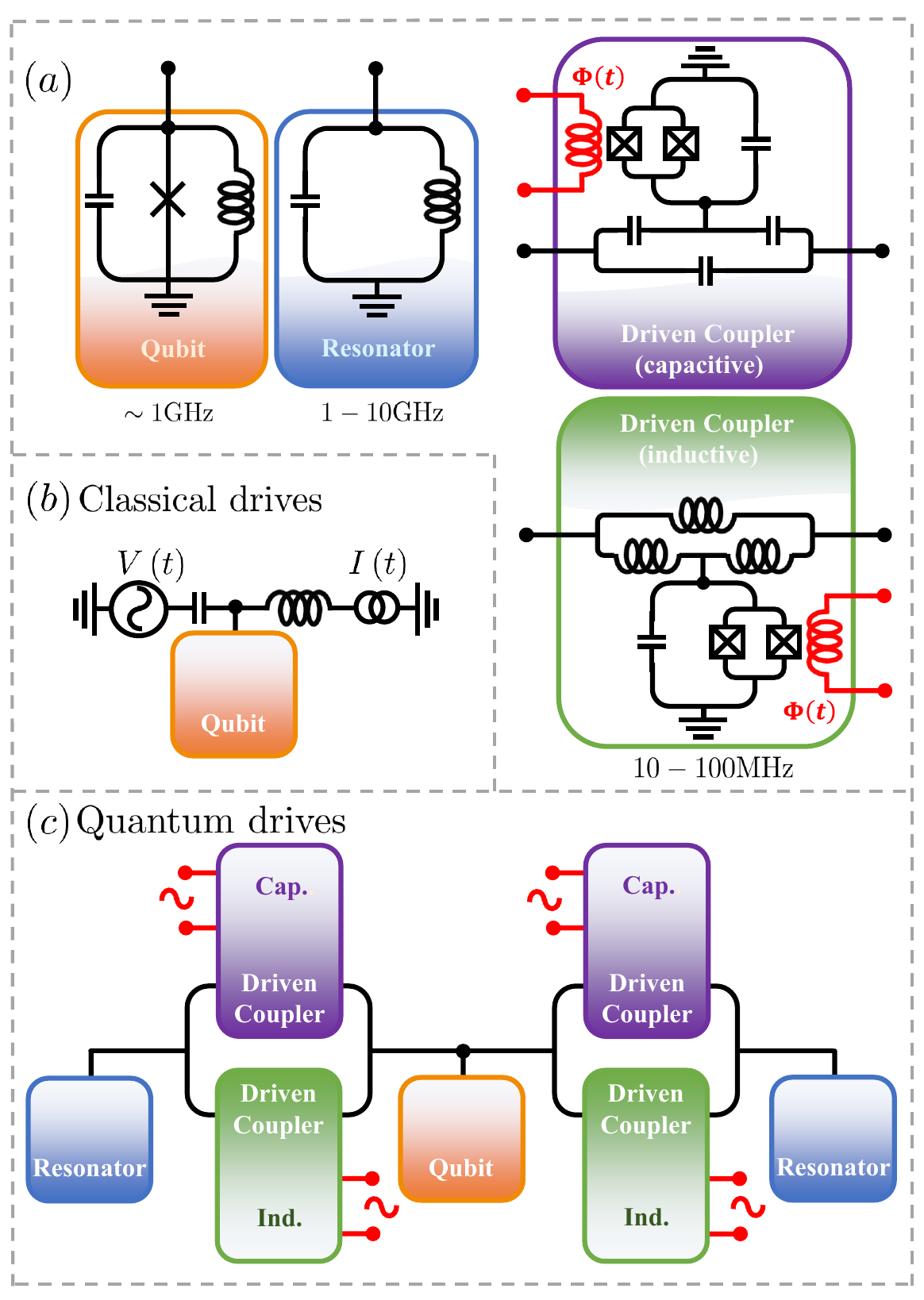}
    \caption{\textit{Superconducting circuit realization for Eq.~\eqref{SC realization}.} (a) A table of superconducting elements in the circuit: the circuit diagram includes a flux qubit, an LC resonator, and two types of driven (tunable) couplers. The tunability arises from an indirect coupling to an idler mode, whose frequency is adjusted by the external flux $\Phi\left(t\right)$. The energy scales of these elements in current experiments are listed below. (b) A schematic circuit implementing Eq.~\eqref{SC classical} where the qubit is driven by a voltage source $V\left(t\right)$ and a current source $I\left(t\right)$. Proper grounding of the qubit is assumed. (c) A schematic circuit for Eq.~\eqref{SC realization}, where the qubit is coupled to two resonators via tunable couplers (variable capacitors and inductors), with typical realizations shown in (a). Proper grounding of all circuit elements is again assumed.
   }
    \label{SC element}
\end{figure}

Superconducting-circuit qubits [Fig.~\ref{SC element}.(a)] are formed by truncating Josephson-junction based oscillators to the lowest two energy levels with gaps ranging from $1\text{GHz}<\omega_{q}<20\text{GHz}$ \cite{RevModPhys.85.623,PRXQuantum.2.040204}, which are assumed to be separated from higher excited levels \cite{PRXQuantum.2.040204}. For two-tone Floquet Hamiltonians, excitation to higher energy levels changes the energy transfer, and hence the path entanglement, completely. With this in mind, qubits with large anharmonicity are preferred. We hence suggest flux or fluxonium-like designs, which provide a large energy gap ($>20\text{GHz}$) between two lowest energy levels and higher ones \cite{PhysRevB.60.15398} and coherence times as long as $100\sim400\mu s$ \cite{Pop2014,PhysRevX.9.041041}.

We couple the qubit to two cavities. In order to obtain dispersive quasienergy bands with relatively small number of bosons, we suggest microwave cavities with close frequencies and coupling strengths. Two such examples are the coplanar waveguide resonator ($\omega\approx1-10\text{GHz}$, $\frac{g}{2\pi}>300\text{MHz}$ and $Q>10^{8}$) and the LC resonator ($\omega\approx1-10\text{GHz}$, $\frac{g}{2\pi}=0.1-1\text{GHz}$ and $Q\sim10^{3}$) in Fig.~\ref{SC element}.(a).

\subsection{Entangler: Tunable coupler and qubit drive}
The most nontrivial step in shaping Eq.~\eqref{SC realization} is the time-dependent coupling $\mathcal{A}\cos\left(\omega_{q}t\right)$ between the qubit and resonators. This requires variable capacitors and inductors whose capacitance $C(t)=C_{0}\cos\left(\omega_{q}t\right)$ and inductance $L(t)=L_{0}\cos\left(\omega_{q}t\right)$ are tuned periodically in a superconducting environment. There are existing ways to realize such variable couplings \cite{Moskalenko2022,PhysRevLett.98.177001,PhysRevLett.112.123601,PhysRevApplied.14.024070,PhysRevApplied.10.054062,vandenBrink2005}, one of which is indirectly coupling the qubit and resonators by an idler qubit/cavity \cite{PhysRevApplied.14.024070,PhysRevApplied.19.064043}, whose frequency is significantly larger than those of the qubit and resonators. The idler qubit/cavity remains in its ground state. However, by changing an external flux, one tunes its frequency in a controllable way, which effectively gives a time-dependent coupling after integrating out such idler degrees of freedom [Fig.~\ref{SC element}.(a)]. The same principle underlies other tunable couplers like Delft couplers and Gmon couplers.

Another time dependence in Eq.~\eqref{SC realization} is in the term $\omega_{0}\cos\left(\omega_{q}t\right)\hat{\sigma}_{x}$, which gives rise to the qubit energy in the rotating frame. This is typically realized by coupling the qubit to a current source $I_{d}(t)$ inductively or a voltage $V_{d}(t)$ source capacitively (App.~\ref{superconducting circuit appendix}).

Finally, we estimate the number of bosons one needs to use for a given design of the circuit in order to have a dispersive spectrum. As mentioned earlier, this requires that the normalized amplitudes of the drives [Eq.~\eqref{lattice model}] comparable to the frequency of the bosons $\sqrt{N}\mathcal{A}\approx\omega$, which gives a rough number of bosons one need to have in the experiments as $N\approx\left(\frac{\omega}{\mathcal{A}}\right)^{2}$. Consequently, for an effective coupling $\mathcal{A}\approx100\text{MHz}$ and resonator frequency $\omega\approx1\text{GHz}$ [Fig.~\ref{SC element}.(a)], one needs roughly $>100$ bosons in the cavity. To further decrease the number of bosons, stronger couplings and cavities with longer wave-lengths are suggested. On the other hand, for two-tone Floquet models with simpler couplings and are realizable without moving to a rotating frame [for example, Eq.~\eqref{2-level with circularly-polarized drives}], direct couplings between the qubit and resonators can reach to as large as $\frac{g}{2\pi}>800\text{MHz}$ \cite{PhysRevLett.105.237001} that lowers the bound on the number of bosons significantly.

\subsection{Qubit measurement}

The projective measurement of the ancilla qubit can be implemented in three steps. First, the qubit is disconnected from the two resonators, either passively by detuning their frequencies or actively by introducing a tunable coupler that is quickly and precisely adjusted to zero coupling \cite{Moskalenko2022,PhysRevLett.98.177001,PhysRevLett.112.123601,PhysRevApplied.14.024070,PhysRevApplied.10.054062,vandenBrink2005,PhysRevApplied.19.064043}. Such a process typically uses a fast control pulse lasting less than $10\ \text{ns}$, and the evolution of the qubit during the decoupling can be further compensated by a phase gate that accounts for the accumulated phase. Next, a single-qubit unitary gate is applied to rotate the state from $\ket{\pm s}$ to the computational basis. Finally, the qubit is read out via dispersive coupling to another resonator, a process that can be completed in as short as $100$ ns with high fidelity \cite{Chen2023}.

\subsection{Phase imprinter}

The phase imprinter and parity measurement for a cavity mode can both be implemented by coupling to another qubit dispersively via $\hat{H}_{dis}=-\chi\hat{a}^{\dagger}\hat{a}\hat{\sigma}_{z}^{'}$ \cite{Wang2019,Sun2014}. Here, by 'dispersive', we do not refer to dispersive quasienergy bands, but rather a change in the frequency of the cavity when coupling to a qubit. We use a different notation for the Pauli matrices $\hat{\sigma}_{x,y,z}^{'}$ to distinguish this qubit from the ancilla qubit. The phase operation can be generated by evolving the cavity-qubit system for some time $\tau$, yielding $\theta=-\chi\tau$. Here, a measurement on $\theta$ can be viewed as one on $\tau$, given that the coupling strength $\chi$ is known, or as one on $\chi$, if the evolution time $\tau$ is known. 

\subsection{Parity measurement}

The parity measurement is implemented on one output mode of the beam splitter [Eq.~\eqref{parity operator}]. A standard beam splitting Hamiltonian $\hat{H}_{BS}=g_{BS}\left(\hat{a}_{1}^{\dagger}\hat{a}_{2}+\hat{a}_{2}^{\dagger}\hat{a}_{1}\right)$ between two resonators is directly realized by coupling two resonators via a capacitor or inductor up to a time $t=\frac{\pi}{4g_{BS}}$. Moreover, recent methods allow for improved control of the coupling $g_{BS}(t)$ (which may be time-dependent), either through the use of Josephson junctions as tunable couplers~\cite{Lu2023}, or via parametric processes based on three-wave or four-wave mixing~\cite{Chapman2023}. These approaches enable operations on a timescale of $100$~ns with high on-off ratios exceeding $10^5$, and are typically $10^3$ times faster than the qubit decoherence time.

After the beam splitter, a parity measurement can be realized by first coupling a probe qubit for $t=\frac{\pi}{\chi}$, where a contolled-phase gate $\hat{C}_{\pi}=\ket{g}\bra{g}+e^{i\pi\hat
{a}^{\dagger}\hat{a}}\ket{e}\bra{e}$ is realized, with $\ket{g}$ and $\ket{e}$ representing the ground and excited states of the qubit. Next, by inserting $\hat{C}_{\pi}$ between two $\frac{\pi}{2}$ pulses in a Ramsey-type interferometer, one encodes odd and even parities to the $\ket{g}$ and $\ket{e}$ states of the qubit \cite{Sun2014}. A different realization of the parity measurement using no-photon measurements can be found in \cite{Cohen:14}.

\section{Conclusion \& Outlook}

In this work, we demonstrate that energy transfer in two-tone Floquet systems can be utilized to generate path-entangled states (PES's) that exhibit sub-SQL in phase estimation. Classically, such energy transfer between drives with commensurate frequencies is described by power operators derived from the quasienergies and quasieigenstates of the Floquet Hamiltonians. In the quantum regime, a quantized-drives model is approximated by a lattice model, which is a direct sum of classical Floquet models with different initial phases. By initializing the qubit (qudit in App.~\ref{optimization and qudits}) properly, we obtain path-entangled states with phase sensitivity approaching the HL. For drives on a single qubit, we also give explicit lower- and upper-bounds on the corresponding QFI, which physically describes the square mean of energy transfer. We show by numerics that the above conclusions drawn from the lattice model are applicable to the original quantized-drives model.

For a standard parity readout in optical quantum sensors and superconducting platforms, we show that the path-entangled states of two driving modes with identical frequencies can reach the Heisenberg limit if its classical Floquet model has quasi-energy bands with degenerate critical points. We show both analytically and numerically that failing to satisfy either of these conditions (identical frequencies and degenerate critical points) leads to phase estimation with at most the SQL sensitivity. There is hence an open question of extracting the phase information (to the HL) encoded in driving modes with different frequencies or from non-degenerate quasienergy bands.

The effects of a finite a priori distribution of the phase and the robustness of our quantum sensor against loss are manifested in the study of uncertainty reduction by measurements and imperfect transmissions in the interferometer. Our numerical results show that the path-entangled states generated in our Floquet entangler are generally ineffective to an a priori phase distribution but more robust against imperfect transmissions. However, the parity measurement is less robust against loss in the interferometer, consistent with earlier observations that such measurement is no longer optimal for lossy interferometers \cite{PhysRevA.78.063828}. Therefore, another natural future task is to find a better and more \textit{experimental-friendly} decoding scheme in the presence of loss and a priori uncertainties. Ongoing efforts in these directions can be found in recent works \cite{Zhou2018,PRXQuantum.4.020333,PhysRevA.109.042406}.

It is now realized that, besides interferometric designs, some single-mode quantum states also exhibit phase sensitivity beyond the standard quantum limit (SQL) \cite{PhysRevA.94.023840, Wang2019}. We expect that our sensor design based on energy transfer can naturally apply to this new type of sensor by treating only one drive quantum-mechanically while the other remains classical \cite{PhysRevB.99.094311}. It is important to note that for single-mode quantum sensing, different phase encoding and decoding schemes are required. Beyond quantum metrology, we anticipate that our two-tone Floquet models can potentially be utilized in state preparation and engineering. An example of this is provided by the polarization conversion model [Eq.~\eqref{polarization conversion}], where the non-diffusive wavefunctions from the linearity of the bands suggest potential high-Fock state generation [Fig.~\ref{QFI and wavefunction}.(i)]. We thus expect that state engineering can be further facilitated by band engineering and energy transfer in Floquet systems.

As a final comment, while we use a single qubit as a mediator for energy transfer and entanglement generation, driving a quantum many-body system can give rise to rich correlations and entanglement \cite{Berdanier_2018,Berdanier2017}. Along these lines, quantum critical states and non-equilibrium steady states \cite{Carollo2018,Carollo_2019,Giovanni2022} have been shown to possess metrological advantages. We leave the exploration of these directions to future work.

\begin{acknowledgments}
We thank Tuvia Geven and Frederick Nathan for insightful discussions. We acknowledge the support
of the Institute for Quantum Information and Matter, an NSF Physics Frontiers Center (PHY-2317110). AE acknowledges funding by the German National Academy of Sciences Leopoldina under the grant number LPDS 2021-02 and by the Walter Burke Institute for Theoretical Physics at Caltech. AR  was supported by the Cluster of Excellence ‘CUI: Advanced Imaging of Matter’ of the Deutsche Forschungsgemeinschaft (DFG) (EXC 2056 and SFB925), and the Max Planck-New York City Center for Non-Equilibrium Quantum Phenomena. G.R. is grateful to AFOSR MURI program, under agreement number FA9550-22-1-0339, as well as to the Simons Foundation. Part of this work was done at the Aspen Center for Physics, which is supported by the
NSF grant PHY-1607611.
\end{acknowledgments}

\appendix
\onecolumngrid
\section{Energy transfer and power operators  -  classical driving fields} 
\label{classical power operator}
In this appendix, we prove useful properties of the power operator in a quantum system driven by two classical fields. As in the main text, we assume that our system is a two-tone Floquet model defined in a finite-dimensional Hilbert space of dimension $d$. It is governed by a Hamiltonian of the form
\begin{equation}\hat{H}\left(t\right)=\hat{H}_{0}+\hat{H}_{1}\left(\omega_{1}t+\phi_{1}\right)+\hat{H}_{2}\left(\omega_{2}t+\phi_{2}\right),
\end{equation}
with $H_0$ a static contribution and $\hat{H}_{i}\left(\omega_{i}t+\phi_{i}\right)$ periodically driven parts with drive frequencies $\omega_i$ and initial phases $\phi_i$ ($i=1,2$). In order to obtain a Floquet system, we assume that the frequencies are commensurate, i.e.\ $\omega_1/\omega_2 \in \mathbb{Q}$ is rational, and there exists a time $T_{\text{com}}$ (the least common multiple of $T_1 = 2\pi/\omega_1$ and $T_2= 2\pi/\omega_2$) such that $\hat{H}\left(t\right) = \hat{H}\left(t+T_{\text{com}}\right)$. We further assume that $H_i$ are differentiable  (operator-valued)  functions of $\phi_i$ ($i=1,2$).

First, we show that, at stroboscopic times $T=kT_{\text{com}}$ ($k\in \mathbb{Z}$), the power operator $\hat{P}_j(T)$ takes the compact form as $\hat{P}_{j}\equiv\sum\limits_{n=1}^{d}\ \left(\partial_{\phi_{j}}\epsilon_{n}\right)\ket{n}\bra{n}$. To this end, we first  rewrite the unitary evolution operator $\hat{\mathcal{U}}\left(T\right) = \mathcal{T}e^{-i\int_{0}^{T}\hat{H}\left(t\right)dt}$ in its eigenbasis
\begin{equation}
    \hat{\mathcal{U}}\left(T\right)=\sum_{n=1}^{d}e^{-i\epsilon_{n}T}\ket{n}\bra{n}
\end{equation}
where $\epsilon_{n}$ and $\ket{n}$ are quasi-eigenenergies and quasi-energies eigenstates of the effective Floquet Hamiltonian $\hat{H}_{eff} = i/ T_{\text{com}}\log\hat{\mathcal{U}}\left(T_{\text{com}}\right) $. We note that both $\epsilon_{n}=\epsilon_{n}(\phi_{1,2},\omega_{1,2})$ and $\ket{n}=\ket{n(\phi_{1,2},\omega_{1,2})}$ are dependent on the phases $\phi_{1,2}$ and frequencies $\omega_{1,2}$, but are independent of $T$. The power operator can hence be written as

\begin{align}
    \hat{P}_j(T)=&\frac{i}{T}\hat{\mathcal{U}}^\dagger\left(T\right)\partial_{\phi_{j}}\hat{\mathcal{U}}\left(T\right)\\ =&\sum_{n=1}^{d}\left(\partial_{\phi_{j}}\epsilon_{n}\right)\ket{n}\bra{n}+\frac{i}{T}\sum_{n=1}^d\left(1-e^{2i\epsilon_{n}T}\right)\ket{n}\bra{n}\partial_{\phi_{j}}\left(\ket{n}\bra{n}\right).
    \label{qudit power operator}
\end{align}

\noindent Noting that $\| \left(1-e^{2i\epsilon_{n}T}\right)\ket{n}\bra{n}\partial_{\phi_{j}}\left(\ket{n}\bra{n}\right)\|$ is bounded by a time-independent constant, we recover  $\hat{P}_{j}$, as defined in the main text, in the limit $T\rightarrow\infty$.

Second, we prove that  $\tr{\hat{P}_{j}}=0$. This follows from its definition via

\begin{align}
\tr{\hat{P}_{j}}=&i\lim_{T\rightarrow\infty}\frac{1}{T}\tr{\hat{\mathcal{U}}^\dagger\left(T\right)\partial_{\phi_{j}}\hat{\mathcal{U}}\left(T\right)}\\
=& \lim_{T\rightarrow\infty}\frac{1}{\omega_j T}\int_{0}^{T}dt\ \tr{\hat{\mathcal{U}}^\dagger\left(t\right)\frac{d\hat{H}_{j}}{dt}\hat{\mathcal{U}}\left(t\right)}\\
=&\lim_{T\rightarrow\infty}\frac{1}{\omega_{j}T}\int_{0}^{T}dt\ \tr{\frac{d\hat{H}_{j}}{dt}}\\
=&\lim_{T\rightarrow\infty}\frac{1}{\omega_{j}T}\tr{\hat{H}_{j}(T)-\hat{H}_{j}(0)}\\
=&0.\label{A4}
\end{align}

Finally, we show  $\sum_{j}\omega_{j}\hat{P}_{j} = 0$, which physically is a restatement of the energy conservation: the system can only restore a finite amount of energy. Therefore, over long periods of time $T$, there is no net flow of energy in or out. To see this, we again start from the definition of $\hat{P}_{j}$ and expand as: 
\begin{align}
\sum_{j}\omega_{j}\hat{P}_{j}=&\lim_{T\rightarrow\infty}\frac{i}{T}\sum_{j}\hat{\mathcal{U}}^\dagger\left(T\right)\omega_{j}\partial_{\phi_{j}}\hat{\mathcal{U}}\left(T\right)\\
=&\lim_{T\rightarrow\infty}\frac{1}{T}\int_{0}^{T}dt\ \hat{\mathcal{U}}^\dagger\left(t\right)\left[\sum_{j}\frac{d\hat{H}_{j}}{dt}\right]\hat{\mathcal{U}}\left(t\right)\\
=&\lim_{T\rightarrow\infty}\frac{1}{T}\int_{0}^{T}dt\ \hat{\mathcal{U}}^\dagger\left(t\right)\frac{d\hat{H}\left(t\right)}{dt}\hat{\mathcal{U}}\left(t\right)\\
=&\lim_{T\rightarrow\infty}\frac{1}{T}\hat{\mathcal{U}}^\dagger\left(t\right)\hat{H}\left(t\right)\hat{\mathcal{U}}\left(t\right)|_{0}^{T}-\lim_{T\rightarrow\infty}\frac{1}{T}\int_{0}^{T}dt\ \frac{d\hat{\mathcal{U}}^\dagger\left(t\right)}{dt}\hat{H}\left(t\right)\hat{\mathcal{U}}\left(t\right)-\lim_{T\rightarrow\infty}\frac{1}{T}\int_{0}^{T}dt\ \hat{\mathcal{U}}^\dagger\left(t\right)\hat{H}\left(t\right)\frac{d\hat{\mathcal{U}}\left(t\right)}{dt}\\
=&\lim_{T\rightarrow\infty}\frac{1}{T}\hat{\mathcal{U}}^\dagger\left(t\right)\hat{H}\left(t\right)\hat{\mathcal{U}}\left(t\right)|_{0}^{T}-i\lim_{T\rightarrow\infty}\frac{1}{T}\int_{0}^{T}dt\ \hat{\mathcal{U}}^\dagger\left(t\right)\left[\hat{H}\left(t\right),\hat{H}\left(t\right)\right]\hat{\mathcal{U}}\left(t\right)\nonumber \\
=&0,
\end{align}
where from the third to fourth line we have integrated by parts, and from the last second to last line we have used the fact that $\hat{H}\left(t\right)$ commutes with itself at any time $t$.

\section{Energy transfer and power operators  -  quantized driving fields}
\label{quantum power appendix}

In this part, we compute the long-time limit of the functional power operator $\hat{P}_{j}\left[f\right]$. As in the main text, we express operators and states in the (Fourier-transformed) phase space and move to the interaction picture defined by a unitary transform $\hat{U}\left(t\right)=e^{-i\hat{H}_{B}t}$. In the interaction picture, the dynamics of the combined system of qudit and quantum drives $i\partial_{t}\ket{\Psi\left(t\right)}=\hat{H}_{l}(t)\ket{\Psi\left(t\right)}$
is decomposed into that of each phase component $i\partial_{t}\ket{\psi\left(\bm{\phi},t\right)}=\hat{H}_{\bm{\phi}}\left(t\right)\ket{\psi\left(\bm{\phi},t\right)}$ [Eq.~\eqref{Bloch Hamiltonian} and discussion below]. At stroboscopic moments $T=kT_{\text{com}}$, with $k$ an integer, such time evolution can be expressed in terms of Floquet modes as:

\begin{equation}
\ket{\psi\left(\bm{\phi},T\right)}=\sum\limits_{n=1}^{d}c_{n}e^{-i\epsilon_{n}T}\ket{n},
\label{system dynamics}
\end{equation}
\noindent where $\epsilon_{n}$ and $\ket{n}$ are quasi-eigenenergies and quasi-energies eigenstates of the effective Floquet Hamiltonian $\hat{H}_{eff} = i/ T_{\text{com}}\log\hat{\mathcal{U}\left(T_{\text{com}}\right)} $. $c_{n}=\braket{n|\psi\left(\bm{\phi},0\right)}$ are initial overlaps between $\ket{\psi\left(\bm{\phi},0\right)}$ and $\ket{n}$. In this work, we assume the initial state being a direct product between the qudit and driving modes, giving $\ket{\psi\left(\bm{\phi},0\right)}=\ket{\psi\left(0\right)}$ at $t=0$. However, it should be clear that all other quantities $c_{n}$, $\epsilon_{n}$, and $\ket{n}$ depend on $\bm{\phi}$ implicitly and we drop such dependence to simplify the notation. The stroboscopic dynamics of the full state $\ket{\Psi\left(T\right)}$ becomes a superposition of $\ket{\psi(\bm{\phi},T)}\otimes\ket{\bm{\phi}}$ as:

\begin{equation}
\ket{\Psi\left(T\right)}=\sum\limits_{n=1}^{d}\ \int\limits_{\text{BZ}}\bm{d^{2}\phi}f\left(\bm{\phi}\right)c_{n}e^{-i\epsilon_{n}T}\ket{n}\otimes\ket{\bm{\phi}}.
\end{equation}

\noindent We next evaluate $\overline{P}_{j}\left(T\right)$ for the evolved state:

\begin{equation}
    \overline{P}_{j}\left(T\right)=\frac{1}{T}\left(\braket{\hat{a}_{j}^{\dagger}\hat{a}_{j}}_{0}-\braket{\hat{a}_{j}^{\dagger}\hat{a}_{j}}_{T}\right).
\label{quantum power1}
\end{equation}

\noindent Here, we use $\braket{\square}_{0}=\braket{\Psi\left(0\right)|\square|\Psi\left(0\right)}$ and $\braket{\square}_{T}=\braket{\Psi\left(T\right)|\square|\Psi\left(T\right)}$ to abbrievate expectation values of $\square$ at $t=0$ and $t=T$, respectively. Noting that the number operator $\hat{a}_{j}^{\dagger}\hat{a}_{j}$ of the $j$-th drive, as the canonical quadrature of the $j$-th phase $\phi_{j}$, becomes $\hat{a}_{j}^{\dagger}\hat{a}_{j}=-i\partial_{\phi_{j}}$ in the phase space, we can calculate Eq.~\eqref{quantum power1} term by term:

\begin{align}
\braket{\hat{a}_{j}^{\dagger}\hat{a}_{j}}_{0}=&\int\limits_{\text{BZ}}\bm{d^{2}\Bar{\phi}d^{2}\phi}f^{*}\left(\bm{\Bar{\phi}}\right)f\left(\bm{\phi}\right)\braket{\psi\left(\bm{\Bar{\phi}},0\right)|\psi\left(\bm{\phi},0\right)}\braket{\bm{\Bar{\phi}}|\frac{\partial_{\phi_{j}}}{i}|\bm{\phi}}\\
=&i\int\limits_{\text{BZ}}\bm{d^{2}\phi}f^{*}\left(\bm{\phi}\right)\partial_{\phi_{j}}f\left(\bm{\phi}\right)+i\int\limits_{\text{BZ}}\bm{d^{2}\phi}\left|f\left(\bm{\phi}\right)\right|^{2}\braket{\psi\left(\bm{\Bar{\phi}},0\right)|\partial_{\phi_{j}}|\psi\left(\bm{\phi},0\right)}\\
=&i\int\limits_{\text{BZ}}\bm{d^{2}\phi}f^{*}\left(\bm{\phi}\right)\partial_{\phi_{j}}f\left(\bm{\phi}\right),
\end{align}

\noindent where from the $1$-st to $2$-nd line we have integrated by parts, and from the $2$-nd to $3$-rd line we have used that $\partial_{\phi_{j}}\ket{\psi\left(\bm{\phi},0\right)}=0$ given the assumption (see Sec.~\ref{quantum polarization conversion}) that the initial state is a  product state between the system and drives.

We next calculate the second contribution $\braket{\hat{a}_{j}^{\dagger}\hat{a}_{j}}_{T}$ in Eq.~\eqref{quantum power1}. After integrating by parts, it takes exactly the same form as the expansion for $\braket{\hat{a}_{j}^{\dagger}\hat{a}_{j}}_{0}$:

\begin{align}
\braket{\hat{a}_{j}^{\dagger}\hat{a}_{j}}_{T}
=&i\int\limits_{\text{BZ}}\bm{d^{2}\phi}f^{*}\left(\bm{\phi}\right)\partial_{\phi_{j}}f\left(\bm{\phi}\right)+i\int\limits_{\text{BZ}}\bm{d^{2}\phi}\left|f\left(\bm{\phi}\right)\right|^{2}\braket{\psi\left(\bm{\phi},T\right)|\partial_{\phi_{j}}|\psi\left(\bm{\phi},T\right)}.
\label{number at T}
\end{align}

\noindent Notably, the first term comes from the initial field distribution of the drives and cancels exactly the same contribution from $\braket{\hat{a}_{j}^{\dagger}\hat{a}_{j}}_{0}$. The non-trivial contribution comes from the second integral in Eq.~\eqref{number at T}:

\begin{equation}
\Delta N_{j}\left(T\right)=i\int\limits_{\text{BZ}}\bm{d^{2}\phi}\left|f\left(\bm{\phi}\right)\right|^{2}\braket{\psi\left(\bm{\phi},T\right)|\partial_{\phi_{j}}|\psi\left(\bm{\phi},T\right)},
\end{equation}

\noindent which measures the difference between the average boson number $\braket{\hat{a}_{j}^{\dagger}\hat{a}_{j}}_{T}$ at moment $t=T$ and initial boson number $\braket{\hat{a}_{j}^{\dagger}\hat{a}_{j}}_{0}$ at $t=0$. At stroboscopic times, expanding $\ket{\psi\left(\bm{\phi},T\right)}$ in terms of quasi-eigenenergies and quasi-energies eigenstates [Eq.~\eqref{system dynamics}], we can rewrite the average power input from the $j$-th drive as:

\begin{align}
    \overline{P_{j}}\left(T\right)=&-\frac{\Delta N_{j}\left(T\right)}{T}\\
    =&\sum\limits_{n=1}^{d}\ \int\limits_{\text{BZ}}\bm{d^{2}\phi}\left|f\left(\bm{\phi}\right)\right|^{2}\left|c_{n}\right|^{2}\partial_{\phi_{j}}\epsilon_{n}-\frac{i}{T}\sum\limits_{n=1}^{d}\ \int\limits_{\text{BZ}}\bm{d^{2}\phi}\left|f\left(\bm{\phi}\right)\right|^{2}c_{n}^{*}\partial_{\phi_{j}}c_{n}\\
    &-\frac{i}{T}\sum_{m,n=1}^{d}\ \int\limits_{\text{BZ}}\bm{d^{2}\phi}\left|f\left(\bm{\phi}\right)\right|^{2}c_{m}^{*}c_{n}e^{-i\left(\epsilon_{n}-\epsilon_{m}\right)T}\braket{m|\partial_{\phi_{j}}|n}.
\end{align}

\noindent Noting that both $\|\sum\limits_{n=1}^{d}\ \int\limits_{\text{BZ}}\bm{d^{2}\phi}\left|f\left(\bm{\phi}\right)\right|^{2}c_{n}^{*}\partial_{\phi_{j}}c_{n}\|$ and $\|\sum_{m=1}^{D}\sum\limits_{n=1}^{d}\ \int\limits_{\text{BZ}}\left|f\left(\bm{\phi}\right)\right|^{2}c_{m}^{*}c_{n}e^{-i\left(\epsilon_{n}-\epsilon_{m}\right)T}\braket{m|\partial_{\phi_{j}}|n}\|$ are bounded by $T$-independent constants, the contributions from them in $\overline{P_{j}}\left(T\right)$ henceforth scales as $O\left(\frac{1}{T}\right)$ and can be ignored when taking the limit $T\rightarrow\infty$. We hence conclude that $\overline{P_{j}}\left(T\right)$ has a finite, well-defined limit:

\begin{align}
    \lim_{T\rightarrow\infty}\overline{P_{j}}\left(T\right)=&\sum\limits_{n=1}^{d}\ \int\limits_{\text{BZ}}\bm{d^{2}\phi}\left|f\left(\bm{\phi}\right)\right|^{2}\left|c_{n}\right|^{2}\partial_{\phi_{j}}\epsilon_{n}\\
    =&\bra{\psi\left(0\right)}\hat{P}_{j}\left[f\right]\ket{\psi\left(0\right)},
\end{align}

\noindent with the functional operator defined as $\hat{P}_{j}\left[f\right]\equiv\sum\limits_{n=1}^{d}\ \int\limits_{\text{BZ}}\bm{d^{2}\phi}\left|f\left(\bm{\phi}\right)\right|^{2}\left(\partial_{\phi_{j}}\epsilon_{n}\right)\ket{n}\bra{n}$. Physically, it can be viewed as a superposition of the power operators $\hat{P}_{j}=\sum\limits_{n=1}^{d}\left(\partial_{\phi_{j}}\epsilon_{n}\right)\ket{n}\bra{n}$ for the two-tone Floquet Hamiltonian, whose coefficient $\left|f\left(\bm{\phi}\right)\right|^{2}$ is given by the initial field distributions in two drives. Without any ambiguity, we henceforth call it the 'functional power operator' for quantized driving fields.\\

The tracelessness and energy conservation of $\hat{P}_{j}\left[f\right]$ then follow easily from those of $\hat{P}_{j}$, which are proven in App.~\ref{classical power operator}.

\section{Projective measurement and post-selection}\label{Postselection app}

\begin{figure*}[h]
\centering
     \includegraphics[width=0.91\textwidth]{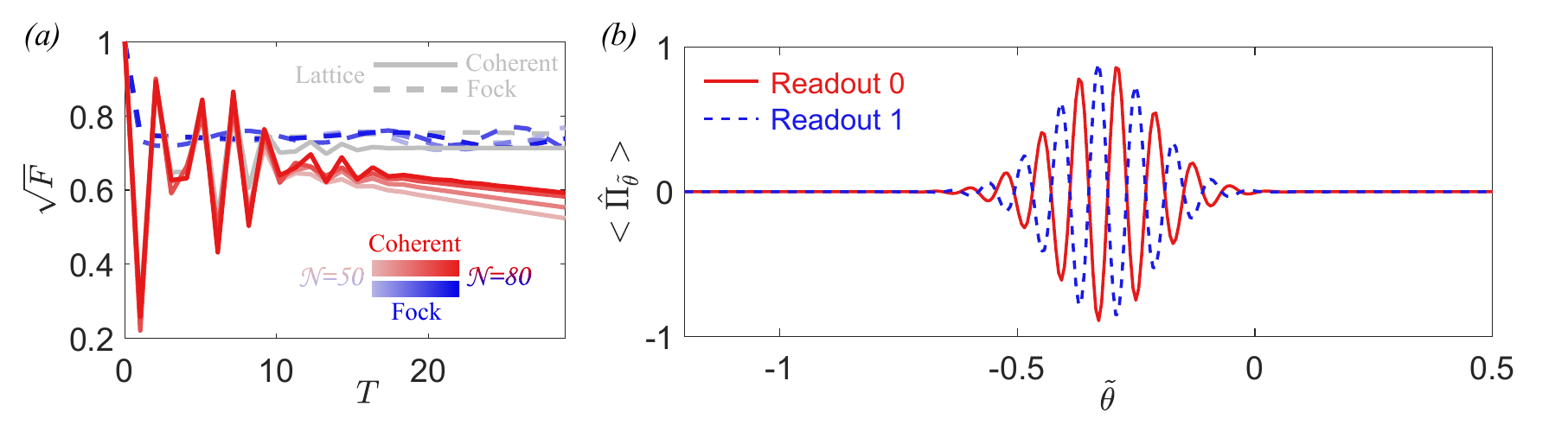}
        \caption{\textit{Fidelity of projective measurement and parity measurement for different readout.} (a). Fidelity between $\rho_{s}(T)$ and the initial state $\rho_{s}(0)$ for Eq.~\eqref{polarization conversion} with different inputs. We show the Uhlmann fidelity $F(T)$ for both the lattice (gray) and original (colored) models with different initial boson numbers. For the lattice model, as expected, both the Fock-state (dashed-gray) and the coherent-state (gray) inputs converges to $F\approx0.5$, indicating a maximally mixed state. For the original models, the Fock-state inputs (dashed-blue) for all values of $N$ converge fastly to $F=0.5$, while for the coherent-state inputs (solid-red) the fidelity $F(T)$ decays slowly at larger $T$. This decay is less significant for larger initial boson numbers, and the agreement between the original model and its lattice approximation becomes better. (b). Parity measurement of the coherent-state input under different qubit readout outcomes. The parity undergoes a phase flip for readout 0 (projection onto the initial qubit state $\ket{s}$) and readout 1 (projection onto $\ket{-s}$). We choose the lattice version of Eq.~\eqref{polarization conversion} and start with 50 bosons per mode. The result is obtained for a conversion time $T=80\frac{\pi}{\omega}$.} 
        \label{Fidelity}  
\end{figure*}

 In this appendix, we explain why our protocol does not require post-selection and, therefore, is independent of the outcome of the projective measurement. 
 
To proceed, we assume a lattice model with linear Floquet bands and $f(\bm{\phi})-$independent group velocity $\partial_{\phi}\epsilon=\pm P$ for $\ket{0}_{f},\ket{1}_{f}$, respectively. In the main text, we show that the polarization conversion model [Eq.~\eqref{polarization conversion}] satisfies these conditions.

To be concrete, we start with a Fock-state input $\ket{s}\otimes\ket{\frac{N}{2}}\otimes\ket{\frac{N}{2}}$. Here $\ket{s}=\frac{1}{\sqrt{2}}\left(\ket{0}_{f}+\ket{1}_{f}\right)$. The state evolves later into a cat-like state $\frac{1}{\sqrt{2}}\left(\ket{0}_{f}\otimes\ket{\frac{N}{2}-PT}\otimes\ket{\frac{N}{2}+PT}+e^{i\phi}\ket{1}_{f}\otimes\ket{\frac{N}{2}+PT}\otimes\ket{\frac{N}{2}-PT}\right)$. Tracing out two bosonic modes, the reduced density matrix for the ancilla qubit becomes a maximally mixed one
\begin{equation}
    \rho_{s}=\frac{1}{2}\left(\ket{0}_{f}\bra{0}_{f}+\ket{1}_{f}\bra{1}_{f}\right),
    \label{maximally mixed state}
\end{equation}
independent of the choice of the basis. Consequently, the probability of projecting onto the initial state $\ket{s}$ (and actually onto any direction) is given by
\begin{equation}
    P=\tr{\rho_{s}\ket{s}\bra{s}}=0.5.
\end{equation}
This result is not prestigious for a Fock input. In fact, as long as the interaction time $T$ is long enough that there is no overlap between two oppositely translated wavefunctions (originating from opposite energy conversion), the reduced density matrix $\rho_{s}$ is always a maximally mixed one. Hence, within this lattice model, we expect that this probability $P=0.5$ holds for even more general inputs, including the coherent-state input. 

In Fig. \ref{Fidelity}.(a), we numerically show the Uhlmann fidelity between the reduced density matrix $\rho_{s}(T)$ and the input state $\rho_{s}(0)=\ket{s}\bra{s}$
\begin{equation}
    F(T)=\tr{\sqrt{\sqrt{\rho_{s}(T)}\rho_{s}(0)\sqrt{\rho_{s}(T)}}}^{2}.
\end{equation}
Physically, this Uhlmann fidelity shows the probability of obtaining $\ket{s}$ in the projective measurement. For the lattice model (gray and dashed-gray lines), this fidelity converges to $F=0.5$ at later times for both the Fock-state input (dashed-gray) and the coherent-state input (solid-gray), consistent with the prediction that $\rho_{s}(T)$ becomes the maximally mixed state given in Eq. \eqref{maximally mixed state}. For the original model (colored and dashed-colored lines), the Uhlmann Fidelity for the Fock-state input converges fastly to $F=0.5$ for inputs with initial boson numbers ranging from $N=50$ to $N=80$. For the coherent-state input, at earlier time, $F$ is in good agreement with the lattice approximation, which deviates and decays slowly at later times. However, this deviation is less significant and shows a clear convergence to the lattice model as one starts with larger boson numbers. We view this as one more example that validates the lattice approximation at larger $N's$ (cf Fig.\ref{Boundary effect}).

Importantly, we note that a projection in $\ket{-s}=\frac{1}{\sqrt{2}}\left(\ket{0}_{f}-\ket{1}_{f}\right)$ also leads to a growth in the quantum Fisher information as well as an improvement in $T^{-1}$ for the sensitivity $\delta\Tilde{\theta}$ of the parity measurement. This can be explicitly seen as follows: For the evolved state $\ket{\Psi(T)}=\frac{1}{\sqrt{2}}\left(\ket{0}_{f}\otimes\ket{\frac{N}{2}-PT}\otimes\ket{\frac{N}{2}+PT}+e^{i\phi}\ket{1}_{f}\otimes\ket{\frac{N}{2}+PT}\otimes\ket{\frac{N}{2}-PT}\right)$, a projection onto the initial state $\ket{s}=\frac{1}{\sqrt{2}}\left(\ket{0}_{f}+\ket{1}_{f}\right)$ or its compliment $\ket{-s}=\frac{1}{\sqrt{2}}\left(\ket{0}_{f}-\ket{1}_{f}\right)$ only differs in a $\pi$ phase in the second superposition as
\begin{align*}
    \ket{s}\bra{s}\cdot\ket{\Psi(T)}&=\frac{1}{\sqrt{2}}\ket{s}\otimes\left(\ket{\frac{N}{2}-PT}\otimes\ket{\frac{N}{2}+PT}+e^{i\phi}\ket{\frac{N}{2}+PT}\otimes\ket{\frac{N}{2}-PT}\right),\\
    \ket{-s}\bra{-s}\cdot\ket{\Psi(T)}&=\frac{1}{\sqrt{2}}\ket{-s}\otimes\left(\ket{\frac{N}{2}-PT}\otimes\ket{\frac{N}{2}+PT}-e^{i\phi}\ket{\frac{N}{2}+PT}\otimes\ket{\frac{N}{2}-PT}\right),
\end{align*}
which is a sign flip for the sensing task. In Fig.\ref{Fidelity}.(b), we plot the expectation value of the parity measurement for the coherent-state input with different projective readout: 0 for the projection onto the initial qubit state $\ket{s}$ and 1 for the projection onto $\ket{-s}$. It can be seen that they share exactly the same sensitivity, differing only by a sign flip $\braket{\hat{\Pi}_{\Tilde{\theta}}}\rightarrow -\braket{\hat{\Pi}_{\Tilde{\theta}}}$. This sign flip is known from the readout of the projective measurement and can be offset in a local phase estimation. We therefore believe that our protocol is less probabilistic compared to other non-linear methods \cite{Mitchell2004,Xiang2011,Zhong2018,Li2020}.

\section{Bounds on the quantum Fisher information}
\label{QFI appendix}

In this appendix, we prove that the quantum Fisher information of the path-entangled state Eq.~\eqref{path-entangled state} grows quadratically in time.  We start with a general discussion on an ancilla qudit with dimension $d$, and specify to an ancilla qubit $d=2$ later. 

For a d-dimensional Floquet system, the power operator for the $j$-th drive is given as $\hat{P}_{j}=\sum\limits_{n=1}^{d}\left(\partial_{\phi_{j}}\epsilon_{n}\right)\ket{n}\bra{n}$ [Eq.~\eqref{qudit power operator}], from which the functional power operator follows as

\begin{equation}
    \hat{P}_{j}\left[f\right]=\sum\limits_{n=1}^{d}\ \int\limits_{BZ}\bm{d^{2}\phi}\left|f\left(\bm{\phi}\right)\right|^{2}\left(\partial_{\phi_{j}}\epsilon_{n}\right)\ket{n}\bra{n}.
\end{equation}

\noindent The tracelessness of $\hat{P}_{j}\left[f\right]$ now ensures two disjoint sets $\mathcal{W_{\pm}}\left[f\right]$ of eigenvalues, containing positive and negative eigenvalues, respectively. Denoting eigenvalues of $\mathcal{W_{\pm}}\left[f\right]$ as $\left\{P_{\pm,k}\left[f\right]\right\}$, the tracelessness can be written as $\sum_{k=1}^{W_{+}\left[f\right]}P_{+,k}\left[k\right]+\sum_{l=1}^{W_{-}\left[f\right]}P_{-,l}\left[l\right]=0$ where $W_{\pm}\left[f\right]$ are cardinalities of $\mathcal{W_{\pm}}\left[f\right]$. Further denoting the eigenstates of $P_{\pm,k}\left[f\right]$ as $\ket{\pm,k}_{f}$, we define the initial qudit state as

\begin{equation}
    \begin{split}
        \ket{\beta,+}_{f}=&\frac{1}{\sqrt{W_{+}\left[f\right]+W_{-}\left[f\right]}}\left[\sum_{k=1}^{W_{+}\left[f\right]}e^{i\beta_{k}^{\left(+\right)}}\ket{+,k}_{f}+\sum_{l=1}^{W_{-}\left[f\right]}e^{i\beta_{l}^{\left(-\right)}}\ket{-,l}_{f}\right].\\
    \end{split}
\label{ancilla states}
\end{equation}

\noindent where $e^{i\beta_{k}^{\left(\pm\right)}}$ are arbitrary phase factors that satisfy $\bra{\beta,+}_{f}\hat{P}_{j}[f]\ket{\beta,+}_{f}=0$. We therefore have $D=W_{+}\left[f\right]+W_{-}\left[f\right]-1$ independent phases $\left\{\beta_{1}^{(+)},...,\beta_{W_{+}\left[f\right]-1}^{(+)},\beta_{1}^{(-)},...,\beta_{W_{-}\left[f\right]}^{(-)}\right\}$. Following the discussion in Sec.~\ref{entanglement generation}, we define the path-entangled state as the output of three successive procedures: 1. prepare the input ancilla state as $\ket{\beta,+}_{f}$, according to an initial field distribution $f\left(\bm{\phi}\right)$; 2. evolve the ancilla-drives system according to the quantized-drives model $\hat{\mathcal{U}}\left(T\right)= e^{-i\hat{H}_{l}T}$ for some $T=kT_{\text{com}}$ [Eq.~\eqref{Bloch Hamiltonian}]; 3. project the ancilla degrees of freedom back to the input $\ket{\beta,+}_{f}$ defined by the projection operator $\hat{\Pi}_{\beta}=\ket{\beta,+}_{f}\bra{\beta,+}_{f}$. At stroboscopic moments, this procedure yields:

\begin{equation}
\ket{\text{PES}}_{T}= \frac{1}{\mathcal{N}_{T}}\hat{\Pi}_{\beta}\sum\limits_{n=1}^{d}\ \int\limits_{\text{BZ}}\bm{d^{2}\phi}f\left(\bm{\phi}\right)u_{n}e^{-i\epsilon_{n}T}\ket{n}\otimes\ket{\bm{\phi}},
\label{path-entangled state qudit}
\end{equation}

\noindent where $u_{n}=\braket{n|\beta,+}_{f}$ is the overlap between the input $\ket{\beta,+}_{f}$ and the $n$-th quasienergy eigenstate of the effective Floquet Hamiltonian [Eq.~\eqref{Bloch Hamiltonian}]. The normalization factor can be calculated by requiring $\bra{\text{PES}}_{T}\hat{\mathds{1}}\ket{\text{PES}}_{T}=1$:

\begin{align}
\mathcal{N}_{T}^{2}=&\sum_{m,n=1}^{d}\ \int\limits_{\text{BZ}}\bm{d^{2}\phi}\left|f\left(\bm{\phi}\right)\right|^{2}u^{*}_{m}u_{n}e^{-i\left(\epsilon_{n}-\epsilon_{m}\right)T}\bra{m}\hat{\Pi}_{\beta}\ket{n}\\
=&\sum_{m,n=1}^{d}\ \int\limits_{\text{BZ}}\bm{d^{2}\phi}\left|f\left(\bm{\phi}\right)\right|^{2}\left|u^{*}_{m}u_{n}\right|^{2}e^{-i\left(\epsilon_{n}-\epsilon_{m}\right)T}\label{Nt3}\\
=&\sum\limits_{n=1}^{d}\ \int\limits_{\text{BZ}}\bm{d^{2}\phi}\left|f\left(\bm{\phi}\right)\right|^{2}\left|u_{n}\right|^{4}+O\left(\frac{1}{\sqrt{T}}\right).\label{C5}
\end{align}

\noindent where in the last line we have used the stationary phase approximation upon assuming non-degenerate $\epsilon_{n}-\epsilon_{m}$ on the support of $f\left(\bm{\phi}\right)$, which scales as $O\left(T^{-\frac{w}{2}}\right)$ for a $w$-dimensional integral. Mathematically, such non-degenerate condition requires a non-vanishing Hessian $\text{Hess}\left[\epsilon_{n}-\epsilon_{m}\right]\neq0$. We note that we obtain $1/\sqrt{T}$ in Eq.~\eqref{C5} although expecting $1/T$ from naively using the stationary phase approximation of a (seemingly) two-dimensional integral. To see why the two-dimensional integral is effectively just one-dimensional, note the energy conservation property of the power operators $\omega_{1}\hat{P}_{1}+\omega_{2}\hat{P}_{2}=0$ gives a constraint on the quasienergies as:

\begin{equation}
    \left(\omega_{1}\partial_{\phi_{1}}+\omega_{2}\partial_{\phi_{2}}\right)\epsilon_{n}=0
\end{equation}

\noindent for $n=1,...,d$. As a result, the quasienergies $\epsilon_{n}\left(\Tilde{x}\right)$ are only functions of $\Tilde{x}\equiv\frac{1}{\sqrt{2}}\left(\frac{\phi_{1}}{\omega_{1}}-\frac{\phi_{2}}{\omega_{2}}\right)$ and do not depend on the other canonical coordinate $\Tilde{y}\equiv\frac{1}{\sqrt{2}}\left(\omega_{2}\phi_{1}+\omega_{1}\phi_{2}\right)$. One can hence integrated out the $\Tilde{y}$ degrees of freedom for $m\neq n$ in Eq.~\eqref{Nt3} as:

\begin{align}
&\sum_{m\neq n}\ \int\limits_{\text{BZ}}\bm{d^{2}\phi}\left|f\left(\bm{\phi}\right)\right|^{2}\left|u^{*}_{m}u_{n}\right|^{2}e^{-i\left(\epsilon_{n}-\epsilon_{m}\right)T}\\
&=\sum_{m\neq n}\int d\Tilde{x}\left(\int d\Tilde{y}\left|f\left(\Tilde{x},\Tilde{y}\right)\right|^{2}\left|u_{m}^{*}u_{n}\right|^{2}\right)e^{-i\left(\epsilon_{n}-\epsilon_{m}\right)T},
\end{align}

\noindent which reduces to a $1$-dimensional stationary phase problem. The non-degenerate condition $\text{Hess}\left[\epsilon_{n}-\epsilon_{m}\right]\neq0$ becomes $\partial_{\Tilde{x}}^{2}\left(\epsilon_{n}-\epsilon_{m}\right)\neq0$. From Eq.~\eqref{C5}, in the limit $T\rightarrow\infty$, the normalization factor converges to:

\begin{equation}
    \mathcal{N}^{2}\equiv\lim\limits_{T\rightarrow\infty}\mathcal{N}_{T}^{2}=\sum\limits_{n=1}^{d}\ \int\limits_{\text{BZ}}\bm{d^{2}\phi}\left|f\left(\bm{\phi}\right)\right|^{2}\left|u_{n}\right|^{4}.
\end{equation}

 To see a $T^2$ scaling of the QFI, we now compute $\frac{1}{T^{2}}F_{q}\left[ \ket{\text{PES}}_{T}, \hat{J}_{z}\right]$ and show it asymptotically takes a $T$-independent, positive value at $T\rightarrow\infty$. For a pure state, the QFI is simply $4$ times the variance of the generator:

\begin{equation}
 \frac{1}{T^{2}}F_{q}\left[ \ket{\text{PES}}_{T}, \hat{J}_{z}\right]=\frac{4}{T^{2}}\left(\braket{\hat{J}_{z}^{2}}_{\text{PES}}-\braket{\hat{J}_{z}}_{\text{PES}}^{2}\right), 
 \label{QFIappendix1}
\end{equation}

\noindent with $\braket{\square}_{\text{PES}}$ the expectation value of $\square$ with respect to the normalized path-entangled state $\ket{\text{PES}}_{T}$. Taking $\hat{J}_{z}=\frac{1}{2}\left(\hat{a}_{1}^{\dagger}\hat{a}_{1}-\hat{a}_{2}^{\dagger}\hat{a}_{2}\right)=-\frac{i}{2}\left(\partial_{\phi_{1}}-\partial_{\phi_{2}}\right)$, we evaluate the Eq.~\eqref{QFIappendix1} term by term:

\begin{align}
\frac{4}{T^{2}}\braket{\hat{J}_{z}^{2}}_{\text{PES}}=&\frac{4}{T^{2}}\bra{\text{PES}}_{T}\hat{J}_{z}^{2}\ket{\text{PES}}_{T}\label{C12}\\
=&-\frac{1}{T^{2}\mathcal{N}_{T}^{2}}\sum_{m,n=1}^{d}\ \int\limits_{\text{BZ}}\bm{d^{2}\Bar{\phi}d^{2}\phi}f^{*}\left(\bm{\Bar{\phi}}\right)f\left(\bm{\phi}\right)\left|u_{m}^{*}u_{n}\right|^{2}e^{-i\left(\epsilon_{n}-\epsilon_{m}\right)T}\braket{\bm{\Bar{\phi}}|\partial_{\Delta\phi}^{2}|\bm{\phi}}\\
=&\frac{1}{\mathcal{N}_{T}^{2}}\sum_{m,n=1}^{d}\ \int\limits_{\text{BZ}}\bm{d^{2}\phi}\left|f\left(\bm{\phi}\right)\right|^{2}\left|u_{m}^{*}u_{n}\right|^{2}\left[\partial_{\Delta\phi}\epsilon_{n}\right]^{2}e^{-i\left(\epsilon_{n}-\epsilon_{m}\right)T}+O\left(\frac{1}{T}\right)\\
=&\frac{1}{\mathcal{N}_{T}^{2}}\sum\limits_{n=1}^{d}\ \int\limits_{\text{BZ}}\bm{d^{2}\phi}\left|f\left(\bm{\phi}\right)\right|^{2}\left|u_{n}\right|^{4}\left[\partial_{\Delta\phi}\epsilon_{n}\right]^{2}+O\left(\frac{1}{\sqrt{T}}\right)\\
=&\frac{1}{\mathcal{N}^{2}}\sum\limits_{n=1}^{d}\ \int\limits_{\text{BZ}}\bm{d^{2}\phi}\left|f\left(\bm{\phi}\right)\right|^{2}\left|u_{n}\right|^{4}\left[\partial_{\Delta\phi}\epsilon_{n}\right]^{2}+O\left(\frac{1}{\sqrt{T}}\right),
\end{align}

\noindent with $\partial_{\Delta\phi}\equiv\partial_{\phi_{1}}-\partial_{\phi_{2}}$. From the $2$-nd to $3$-rd line we have integrated by parts, noting the fact that a factor of $T$ comes out only if the partial derivatives are applied to the exponential $e^{-i\left(\epsilon_{n}-\epsilon_{m}\right)T}$. We also assume that we are observing the system for a sufficiently long time such that $T \gg \left|N_{1} - N_{2}\right|$, where $N_{1}$ and $N_{2}$ are the initial numbers of bosons in two drives. This allows terms involving $\partial_{\Delta\phi} f(\bm{\phi})$ to be absorbed into $O\left(\frac{1}{T}\right)$. From the $3$-rd to $4$-th line we again use the stationary phase approximation and drop the fast-oscillating exponentials when $m\neq n$. Finally from the $4$-th to the last line we have merged the $T$-dependent terms of $\mathcal{N}_{T}$ into $O\left(\frac{1}{\sqrt{T}}\right)$. A similar calculation can be conducted for the second term:

    \begin{align}
        \frac{2}{T}\braket{\hat{J}_{z}}_{\text{PES}}=&\frac{2}{T}\bra{\text{PES}}_{T}\hat{J}_{z}\ket{\text{PES}}_{T}\\
        =&-\frac{i}{T\mathcal{N}_{T}^{2}}\sum_{m,n=1}^{d}\ \int\limits_{\text{BZ}}\bm{d^{2}\Bar{\phi}d^{2}\phi}f^{*}\left(\bm{\Bar{\phi}}\right)f\left(\bm{\phi}\right)\left|u_{m}^{*}u_{n}\right|^{2}e^{-i\left(\epsilon_{n}-\epsilon_{m}\right)T}\braket{\bm{\Bar{\phi}}|\partial_{\Delta\phi}|\bm{\phi}}\\
        =&\frac{1}{\mathcal{N}_{T}^{2}}\sum_{m,n=1}^{d}\ \int\limits_{\text{BZ}}\bm{d^{2}\phi}\left|f\left(\bm{\phi}\right)\right|^{2}\left|u_{m}^{*}u_{n}\right|^{2}\left[\partial_{\Delta\phi}\epsilon_{n}\right]e^{-i\left(\epsilon_{n}-\epsilon_{m}\right)T}+O\left(\frac{1}{T}\right)\\
        =&\frac{1}{\mathcal{N}_{T}^{2}}\sum\limits_{n=1}^{d}\ \int\limits_{\text{BZ}}\bm{d^{2}\phi}\left|f\left(\bm{\phi}\right)\right|^{2}\left|u_{n}\right|^{4}\left[\partial_{\Delta\phi}\epsilon_{n}\right]+O\left(\frac{1}{\sqrt{T}}\right)\\
        =&\frac{1}{\mathcal{N}^{2}}\sum\limits_{n=1}^{d}\ \int\limits_{\text{BZ}}\bm{d^{2}\phi}\left|f\left(\bm{\phi}\right)\right|^{2}\left|u_{n}\right|^{4}\left[\partial_{\Delta\phi}\epsilon_{n}\right]+O\left(\frac{1}{\sqrt{T}}\right)\label{C21},
    \end{align}

\noindent where again we have used the stationary phase approximation to replace the double summation to a single one. Note that the expectation value of $\hat{J}_{z}$ for the probe state $\ket{\text{PES}}_{T}$ after the projection $\hat{\Pi}_{\beta}$ does not vanish asymptotically in general. To prove an asymptotically positive $\frac{1}{T^{2}}F_{q}\left[ \ket{\text{PES}}_{T}, \hat{J}_{z}\right]$ at $T\rightarrow\infty$, one hence needs to prove the following proposition:

\begin{proposition}[Asymptotic coefficient of QFI]
For a path-entangled probe state defined in Eq.~\eqref{path-entangled state qudit}, the following inequality holds:
\begin{align}
&\frac{1}{\mathcal{N}^{2}}\sum\limits_{n=1}^{d}\ \int\limits_{\text{BZ}}\bm{d^{2}\phi}\left|f\left(\bm{\phi}\right)\right|^{2}\left|u_{n}\right|^{4}\left[\partial_{\Delta\phi}\epsilon_{n}\right]^{2}\\
&\geq\left[\frac{1}{\mathcal{N}^{2}}\sum\limits_{n=1}^{d}\ \int\limits_{\text{BZ}}\bm{d^{2}\phi}\left|f\left(\bm{\phi}\right)\right|^{2}\left|u_{n}\right|^{4}\left[\partial_{\Delta\phi}\epsilon_{n}\right]\right]^{2},
\label{main inequality}
\end{align}
\noindent with $\sum\limits_{n=1}^{d}\left|u_{n}\right|^{2}=1$, $\int\limits_{\text{BZ}} \bm{d^{2}\phi}\ \left|f(\bm{\phi})\right|^{2}=1$, $\sum\limits_{n=1}^{d}\ \int\limits_{\text{BZ}}\bm{d^{2}\phi}\left|f\left(\bm{\phi}\right)\right|^{2}\left|u_{n}\right|^{2}\partial_{\Delta\phi}\epsilon_{n}=0$, and $\mathcal{N}^{2}=\sum\limits_{n=1}^{d}\ \int\limits_{\text{BZ}}\bm{d^{2}\phi}\left|f\left(\bm{\phi}\right)\right|^{2}\left|u_{n}\right|^{4}$. The equality holds only if $\hat{P}_{j}\left[f\right]=0$ for $j=1,2$, given the initial field distribution $f\left(\bm{\phi}\right)$.
\label{theorem: Asymptotic coefficient of QFI}
\end{proposition}

\noindent The constraint $\sum\limits_{n=1}^{d}\ \int\limits_{\text{BZ}}\bm{d^{2}\phi}\left|f\left(\bm{\phi}\right)\right|^{2}\left|u_{n}\right|^{2}\partial_{\Delta\phi}\epsilon_{n}=0$ comes from the definition of $\ket{\beta,+}_{f}$ such that $\bra{\beta,+}_{f}\hat{P}_{j}[f]\ket{\beta,+}_{f}=0$. We start with some manipulation of the above inequality. Using the positiveness of $\mathcal{N}^{2}$, it is equivalent to prove the following inequality:

\begin{align}
&\sum\limits_{n=1}^{d}\ \int\limits_{\text{BZ}}\bm{d^{2}\phi}\left|f\left(\bm{\phi}\right)\right|^{2}\left|u_{n}\right|^{4}\ \sum\limits_{n=1}^{d}\ \int\limits_{\text{BZ}}\bm{d^{2}\phi}\left|f\left(\bm{\phi}\right)\right|^{2}\left|u_{n}\right|^{4}\left(\partial_{\Delta\phi}\epsilon_{n}\right)^{2}\\
&\geq\left[\sum\limits_{n=1}^{d}\ \int\limits_{\text{BZ}}\bm{d^{2}\phi}\left|f\left(\bm{\phi}\right)\right|^{2}\left|u_{n}\right|^{4}\partial_{\Delta\phi}\epsilon_{n}\right]^{2}.
\end{align}

\noindent For two integrable functions $x\left(\bm{\phi}\right),y\left(\bm{\phi}\right)$, the integral form of Cauchy-Schwarz inequality states $\left[\int_{\text{BZ}}\bm{d^{2}\phi}\ x\left(\bm{\phi}\right)y\left(\bm{\phi}\right)\right]^{2}\leq\int_{\text{BZ}}\bm{d^{2}\phi}\ x^{2}\left(\bm{\phi}\right)\int_{\text{BZ}}\bm{d^{2}\phi}\ y^{2}\left(\bm{\phi}\right)$. By choosing $x\left(\bm{\phi}\right)=\left|f\left(\bm{\phi}\right)\right|\sqrt{\sum\limits_{n=1}^{d}\left|u_{n}\right|^{4}}$ and $y\left(\bm{\phi}\right)=\left|f\left(\bm{\phi}\right)\right|\sqrt{\sum\limits_{n=1}^{d} \left|u_{n}\right|^{4}\left[\partial_{\Delta\phi}\epsilon_{n}\right]^{2}}$, we obtain:

\begin{align}
&\int\limits_{\text{BZ}}\bm{d^{2}\phi}\left|f\left(\bm{\phi}\right)\right|^{2}\sum\limits_{n=1}^{d}\left|u_{n}\right|^{4}\ \ \int\limits_{\text{BZ}}\bm{d^{2}\phi}\left|f\left(\bm{\phi}\right)\right|^{2}\sum\limits_{n=1}^{d}\left|u_{n}\right|^{4}\left(\partial_{\Delta\phi}\epsilon_{n}\right)^{2}\geq \left\{\int\limits_{\text{BZ}}\bm{d^{2}\phi}\left|f\left(\bm{\phi}\right)\right|^{2}\sqrt{\left[\sum\limits_{n=1}^{d}\left|u_{n}\right|^{4}\right]\left[\sum\limits_{n=1}^{d}\left|u_{n}\right|^{4}\left(\partial_{\Delta\phi}\epsilon_{n}\right)^{2}\right]}\right\}^{2}.
\label{first CS inequality}
\end{align}

\noindent The inequality in Eq.~\eqref{first CS inequality} is saturated if and only if $\sqrt{\sum\limits_{n=1}^{d}\left|u_{n}\right|^{4}}$ and $\sqrt{\sum\limits_{n=1}^{d} \left|u_{n}\right|^{4}\left[\partial_{\Delta\phi}\epsilon_{n}\right]^{2}}$ are linearly dependent, which dictates $\sqrt{\sum\limits_{n=1}^{d} \left|u_{n}\right|^{4}\left[\partial_{\Delta\phi}\epsilon_{n}\right]^{2}}=\lambda_{1}\sqrt{\sum\limits_{n=1}^{d}\left|u_{n}\right|^{4}}$, with $\lambda_{1}$ a $\bm{\phi}$-independent constant. We now use the summation form of Cauchy-Schwarz inequality for two indexed sets $\left\{p_{n}\right\}$ and $\left\{q_{n}\right\}$, which states $\left(\sum\limits_{n=1}^{d}p_{n}q_{n}\right)^{2}\leq\sum\limits_{n=1}^{d}p_{n}^{2}\ \sum\limits_{n=1}^{d}q_{n}^{2}$. By identifying $p_{n}=\left|u_{n}\right|^{2}$ and $q_{n}=\left|u_{n}\right|^{2}\partial_{\Delta\phi}\epsilon_{n}$, we obtain:

\begin{equation}
\left[\sum\limits_{n=1}^{d}\left|u_{n}\right|^{4}\right]\left[\sum\limits_{n=1}^{d}\left|u_{n}\right|^{4}\left(\partial_{\Delta\phi}\epsilon_{n}\right)^{2}\right]\geq\left[\sum\limits_{n=1}^{d}\left|u_{n}\right|^{4}\partial_{\Delta\phi}\epsilon_{n}\right]^{2}.
\label{local inequality}
\end{equation}

\noindent The inequality in Eq.~\eqref{local inequality} is saturated if and only if vectors $\left|u_{n}\right|^{2}$ and $\left|u_{n}\right|^{2}\partial_{\Delta\phi}\epsilon_{n}$ are linearly dependent, dictating $\left|u_{n}\right|^{2}\partial_{\Delta\phi}\epsilon_{n}=\lambda_{2}(\bm{\phi})\left|u_{n}\right|^{2}$ for $\forall n=1,...,d$, with $\lambda_{2}(\bm{\phi})$ only a (smooth) function of $\bm{\phi}$. Consequently, to saturate the inequality in Eq.~\eqref{main inequality}, we must satisfy the following two equations:

\begin{align}
    \sqrt{\sum\limits_{n=1}^{d} \left|u_{n}\right|^{4}\left[\partial_{\Delta\phi}\epsilon_{n}\right]^{2}}&=\lambda_{1}\sqrt{\sum\limits_{n=1}^{d}\left|u_{n}\right|^{4}}, \quad\left|u_{n}\right|^{2}\partial_{\Delta\phi}\epsilon_{n}=\lambda_{2}(\bm{\phi})\left|u_{n}\right|^{2}\label{equalitycondition2}.
\end{align}

\noindent The solution follows as:

\begin{align}
\lambda_{1}=\left|\lambda\right|,\quad\lambda_{2}(\bm{\phi})=\lambda,
\end{align}

\noindent with $\lambda$ a $\bm{\phi}$-independent constant. We next show that $\lambda\neq0$ if $\hat{P}_{j}[f]\neq0$, which ultimately contradicts the definition of $\ket{\beta,+}_{f}$, dictating that $\bra{\beta,+}_{f}\hat{P}_{j}[f]\ket{\beta,+}_{f}=0$. We first note that the trace norm of $\|\hat{P}_{j}[f]\|_{F}\equiv\sqrt{\text{Tr}\left(\hat{P}_{j}^{\dagger}[f]\hat{P}_{j}[f]\right)}>0$ if $\hat{P}_{j}[f]\neq0$. From energy conservation $\omega_{1}\hat{P}_{1}\left[f\right]+\omega_{2}\hat{P}_{2}\left[f\right]=0$, we also know $\|\hat{P}_{1}[f]-\hat{P}_{2}[f]\|_{F}>0$. From the definition of $\ket{\beta,+}_{f}$ [Eq.~\eqref{ancilla states}], the trace norm can be compactly written as:

\begin{align}
    \|\hat{P}_{1}[f]-\hat{P}_{2}[f]\|_{F}^{2}&=\left(W_{+}[f]+W_{-}[f]\right)\bra{\beta,+}_{f}\left(\hat{P}_{1}\left[f\right]-\hat{P}_{2}\left[f\right]\right)^{2}\ket{\beta,+}_{f}\\
    &=\left(W_{+}[f]+W_{-}[f]\right)\bra{\beta,+}_{f}\left[\sum\limits_{n=1}^{d}\ \int\limits_{BZ}\bm{d^{2}\phi}\left|f\left(\bm{\phi}\right)\right|^{2}\left(\partial_{\Delta\phi}\epsilon_{n}\right)\ket{n}\bra{n}\right]^{2}\ket{\beta,+}_{f}\\
    &=\left(W_{+}[f]+W_{-}[f]\right)\bra{\beta,+}_{f}\left[\int\limits_{BZ}\bm{d^{2}\phi}\left|f\left(\bm{\phi}\right)\right|^{2}\delta\hat{P}(\bm{\phi})\right]^{2}\ket{\beta,+}_{f}\\
    &=\left(W_{+}[f]+W_{-}[f]\right)\int\limits_{BZ}\bm{d^{2}\phi_{1}}\bm{d^{2}\phi_{2}}\left|f\left(\bm{\phi_{1}}\right)\right|^{2}\left|f\left(\bm{\phi_{2}}\right)\right|^{2}\bra{\beta,+}_{f}\delta\hat{P}(\bm{\phi_{1}})\delta\hat{P}(\bm{\phi_{2}})\ket{\beta,+}_{f},\label{norm last}
\end{align}

\noindent where in the third line we have introduced $\delta\hat{P}(\bm{\phi})\equiv\hat{P}_{1}(\bm{\phi})-\hat{P}_{2}(\bm{\phi})=\sum\limits_{n=1}^{d}\ \left[\partial_{\Delta\phi}\epsilon_{n}\left(\bm{\phi}\right)\right]\ket{n\left(\bm{\phi}\right)}\bra{n\left(\bm{\phi}\right)}$ [Eq.~\eqref{qudit power operator}] to simplify the notation. Using the covariance form of the Cauchy-Schwarz inequality $\left|\braket{\hat{X}\hat{Y}}_{\beta}\right|^{2}\leq\braket{\hat{X}^{2}}_{\beta}\braket{\hat{Y}^{2}}_{\beta}$, with $\braket{\square}_{\beta}\equiv{\bra{\beta,+}_{f}\square\ket{\beta,+}_{f}}$, the expectation in Eq.~\eqref{norm last} satisfies the following inequality:

\begin{align}
\left|\braket{\delta\hat{P}(\bm{\phi_{1}})\delta\hat{P}(\bm{\phi_{2}})}_{\beta}\right|&\leq\sqrt{\braket{\left[\delta\hat{P}(\bm{\phi_{1}})\right]^{2}}_{\beta}\braket{\left[\delta\hat{P}(\bm{\phi_{2}})\right]^{2}}_{\beta}}\\
&=\sqrt{\left(\sum\limits_{n=1}^{d}\ \left[\partial_{\Delta\phi}\epsilon_{n}\left(\bm{\phi_{1}}\right)\right]^{2}\left|u_{n}(\bm{\phi_{1}})\right|^2\right)\left(\sum\limits_{n=1}^{d}\ \left[\partial_{\Delta\phi}\epsilon_{n}\left(\bm{\phi_{2}}\right)\right]^{2}\left|u_{n}(\bm{\phi_{2}})\right|^2\right)}.\\
\end{align}

\noindent The trace norm is hence bounded by:

\begin{align}
    \|\hat{P}_{1}[f]-\hat{P}_{2}[f]\|_{F}&\leq\sqrt{W_{+}[f]+W_{-}[f]}\int\limits_{BZ}\bm{d^{2}\phi}\left|f\left(\bm{\phi}\right)\right|^{2}\sqrt{\sum\limits_{n=1}^{d}\ \left(\partial_{\Delta\phi}\epsilon_{n}\right)^{2}\left|u_{n}\right|^2}\\
    &=\left|\lambda\right|\sqrt{W_{+}[f]+W_{-}[f]},
\end{align}

\noindent where in the last line we have used Eq.~\eqref{equalitycondition2} and two normalization conditions: $\sum\limits_{n=1}^{d}\left|u_{n}\right|^{2}=1$ and $\int\limits_{\text{BZ}} \bm{d^{2}\phi}\ \left|f(\bm{\phi})\right|^{2}=1$. It is then clear that $\left|\lambda\right|>0$.

This, however, contradicts the requirement that $\bra{\beta,+}_{f}\hat{P}_{j}[f]\ket{\beta,+}_{f}=0$:

\begin{align}
    \sum\limits_{n=1}^{d}\ \int\limits_{\text{BZ}}\bm{d^{2}\phi}\left|f\left(\bm{\phi}\right)\right|^{2}\left|u_{n}\right|^{2}\partial_{\Delta\phi}\epsilon_{n}=\lambda\sum\limits_{n=1}^{d}\ \int\limits_{\text{BZ}}\bm{d^{2}\phi}\left|f\left(\bm{\phi}\right)\right|^{2}\left|u_{n}\right|^{2}=\lambda,
\end{align}

\noindent where we have used $\sum\limits_{n=1}^{d}\left|u_{n}\right|^{2}=1$ and $\int\limits_{\text{BZ}} \bm{d^{2}\phi}\ \left|f(\bm{\phi})\right|^{2}=1$. We then prove that the equlity in Eq.~\eqref{main inequality} holds only if $\hat{P}_{j}[f]=0$.

Finally, note that all the above calculations are $T$-independent. In fact, we have taken the asymptotic limit $T\rightarrow\infty$ and dropped the $O\left(\frac{1}{T}\right)$ contributions. As a result, for non-vanishing functional power operators $\hat{P}_{j}\left[f\right]$, the asymptotic coefficient of the QFI $\lim\limits_{T\rightarrow\infty}\frac{1}{T^{2}}F_{q}\left[ \ket{\text{PES}}_{T}, \hat{J}_{z}\right]$ takes a finite, $T$-independent and positive value.\\

We now show that the finite, $T$-independent, and positive asymptotic coefficient of the QFI for a $2$-level ancilla can be further bounded within $\left[\frac{1}{2}P^{2}\left[f\right],2P^{2}\left[f\right]\right]$, as given in Eq.~\eqref{QFIbounds}. Note that for a $2$-level ancilla, the analysis above
applies by setting $d=2$. Specifically, we have the following two constraints:

\begin{align}
&\left|u_{1}\right|^{2}+\left|u_{2}\right|^{2}=1,\label{property1: 2 level ancilla}\\ &\int\limits_{\text{BZ}}\bm{d^{2}\phi}\left|f\left(\bm{\phi}\right)\right|^{2}\left[\left|u_{1}\right|^{2}\partial_{\Delta\phi}\epsilon_{1}+\left|u_{2}\right|^{2}\partial_{\Delta\phi}\epsilon_{2}\right]=0.\label{property2: 2 level ancilla}
\end{align}
\noindent The special property of a $2$-level ancilla is that the expectation value of $\frac{2}{T}\braket{\hat{J}_{z}}_{\text{PES}}$ asymptotically vanishes:

\begin{align}
\lim\limits_{T\rightarrow\infty}\frac{2}{T}\braket{\hat{J}_{z}}_{\text{PES}}&=\frac{1}{\mathcal{N}^{2}}\int\limits_{\text{BZ}}\bm{d^{2}\phi}\left|f\left(\bm{\phi}\right)\right|^{2}\left[\left|u_{1}\right|^{4}\partial_{\Delta\phi}\epsilon_{1}+\left|u_{2}\right|^{4}\partial_{\Delta\phi}\epsilon_{2}\right]\label{Jz1: 2 level ancilla}\\
&=\frac{1}{\mathcal{N}^{2}}\int\limits_{\text{BZ}}\bm{d^{2}\phi}\left|f\left(\bm{\phi}\right)\right|^{2}\left[\left|u_{1}\right|^{4}\partial_{\Delta\phi}\epsilon_{1}+\left(1-\left|u_{1}\right|^{2}\right)^{2}\partial_{\Delta\phi}\epsilon_{2}\right]\label{Jz2: 2 level ancilla}\\
&=\frac{1}{\mathcal{N}^{2}}\int\limits_{\text{BZ}}\bm{d^{2}\phi}\left|f\left(\bm{\phi}\right)\right|^{2}\left[\left|u_{1}\right|^{4}\left(\partial_{\Delta\phi}\epsilon_{1}+\partial_{\Delta\phi}\epsilon_{2}\right)+\left(1-2\left|u_{1}\right|^{2}\right)\partial_{\Delta\phi}\epsilon_{2}\right]\label{Jz3: 2 level ancilla}\\
&=\frac{1}{\mathcal{N}^{2}}\int\limits_{\text{BZ}}\bm{d^{2}\phi}\left|f\left(\bm{\phi}\right)\right|^{2}\left[\left(1-2\left|u_{1}\right|^{2}\right)\partial_{\Delta\phi}\epsilon_{2}\right]\label{Jz4: 2 level ancilla}\\
&=\frac{1}{\mathcal{N}^{2}}\int\limits_{\text{BZ}}\bm{d^{2}\phi}\left|f\left(\bm{\phi}\right)\right|^{2}\left[\left(\left|u_{2}\right|^{2}-\left|u_{1}\right|^{2}\right)\partial_{\Delta\phi}\epsilon_{2}\right]\label{Jz5: 2 level ancilla}\\
&=\frac{1}{\mathcal{N}^{2}}\int\limits_{\text{BZ}}\bm{d^{2}\phi}\left|f\left(\bm{\phi}\right)\right|^{2}\left[\left|u_{1}\right|^{2}\partial_{\Delta\phi}\epsilon_{1}+\left|u_{2}\right|^{2}\partial_{\Delta\phi}\epsilon_{2}\right]=0\label{Jz6: 2 level ancilla},
\end{align}

\noindent where from Eq.~\eqref{Jz1: 2 level ancilla} to Eq.~\eqref{Jz2: 2 level ancilla} and Eq.~\eqref{Jz4: 2 level ancilla} to Eq.~\eqref{Jz5: 2 level ancilla} we have used the property Eq.~\eqref{property1: 2 level ancilla}; from Eq.~\eqref{Jz3: 2 level ancilla} to Eq.~\eqref{Jz4: 2 level ancilla} and Eq.~\eqref{Jz5: 2 level ancilla} to Eq.~\eqref{Jz6: 2 level ancilla} we have used the tracelessness of $\hat{P}_{1,2}$, which gives $\partial_{\Delta\phi}\epsilon_{1}+\partial_{\Delta\phi}\epsilon_{2}=0$; finally, in Eq.~\eqref{Jz6: 2 level ancilla}, we have used the property Eq.~\eqref{property2: 2 level ancilla}. As a result, to find an expression for $\lim\limits_{T\rightarrow\infty}\frac{1}{T^{2}}F_{q}\left[ \ket{\text{PES}}_{T}, \hat{J}_{z}\right]$, we only need to evaluate the following term:

\begin{align}
\lim\limits_{T\rightarrow\infty}\frac{1}{T^{2}}F_{q}\left[ \ket{\text{PES}}_{T}, \hat{J}_{z}\right]=&\frac{1}{\mathcal{N}^{2}}\int\limits_{\text{BZ}}\bm{d^{2}\phi}\left|f\left(\bm{\phi}\right)\right|^{2}\left[\left|u_{1}\right|^{4}\left(\partial_{\Delta\phi}\epsilon_{1}\right)^{2}+\left|u_{2}\right|^{4}\left(\partial_{\Delta\phi}\epsilon_{2}\right)^{2}\right]\label{QFI1: 2 level ancilla}\\
=&\frac{1}{\mathcal{N}^{2}}\int\limits_{\text{BZ}}\bm{d^{2}\phi}\left|f\left(\bm{\phi}\right)\right|^{2}\left(\left|u_{1}\right|^{4}+\left|u_{2}\right|^{4}\right)\left(\partial_{\Delta\phi}\epsilon_{1}\right)^{2},\label{QFI2: 2 level ancilla}\\
\mathcal{N}^{2}=&\int\limits_{\text{BZ}}\bm{d^{2}\phi}\left|f\left(\bm{\phi}\right)\right|^{2}\left(\left|u_{1}\right|^{4}+\left|u_{2}\right|^{4}\right),
\end{align}

\noindent where from Eq.~\eqref{QFI1: 2 level ancilla} to Eq.~\eqref{QFI2: 2 level ancilla}
we have used the tracelessness of the power operators that gives $\partial_{\Delta\phi}\epsilon_{1}+\partial_{\Delta\phi}\epsilon_{2}=0$. Noting that $\left|u_{1}\right|^{4}+\left|u_{2}\right|^{4}$ is bounded between $\left[\frac{1}{2},1\right]$ under the constraint $\left|u_{1}\right|^{2}+\left|u_{2}\right|^{2}=1$, we find the following bounds for the numerator and denominator of Eq.~\eqref{QFI2: 2 level ancilla}:

\begin{gather}
\frac{1}{2}P^{2}\left[f\right]\leq\int\limits_{\text{BZ}}\bm{d^{2}\phi}\left|f\left(\bm{\phi}\right)\right|^{2}\left(\left|u_{1}\right|^{4}+\left|u_{2}\right|^{4}\right)\left(\partial_{\Delta\phi}\epsilon_{1}\right)^{2}\leq P^{2}\left[f\right],\\
\frac{1}{2}\leq\mathcal{N}^{2}\leq1,
\end{gather}

\noindent with $P^{2}\left[f\right]\equiv\frac{1}{2}\int\limits_{\text{BZ}}\bm{d^{2}\phi}\left|f^{2}\left(\bm{\phi}\right)\right|\left[\left(\partial_{\Delta\phi}\epsilon_{1}\right)^{2}+\left(\partial_{\Delta\phi}\epsilon_{2}\right)^{2}\right]$. The asymptotic coefficient of the QFI is hence bounded by:

\begin{equation}
\frac{1}{2}P^{2}\left[f\right]\leq\lim\limits_{T\rightarrow\infty}\frac{1}{T^{2}}F_{q}\left[ \ket{\text{PES}}_{T}, \hat{J}_{z}\right]\leq2P^{2}\left[f\right].
\label{QFI bound appendix}
\end{equation}

\section{Path entanglement}
\label{mode entanglement appendix}

\begin{figure*}
    \includegraphics[width=\linewidth]{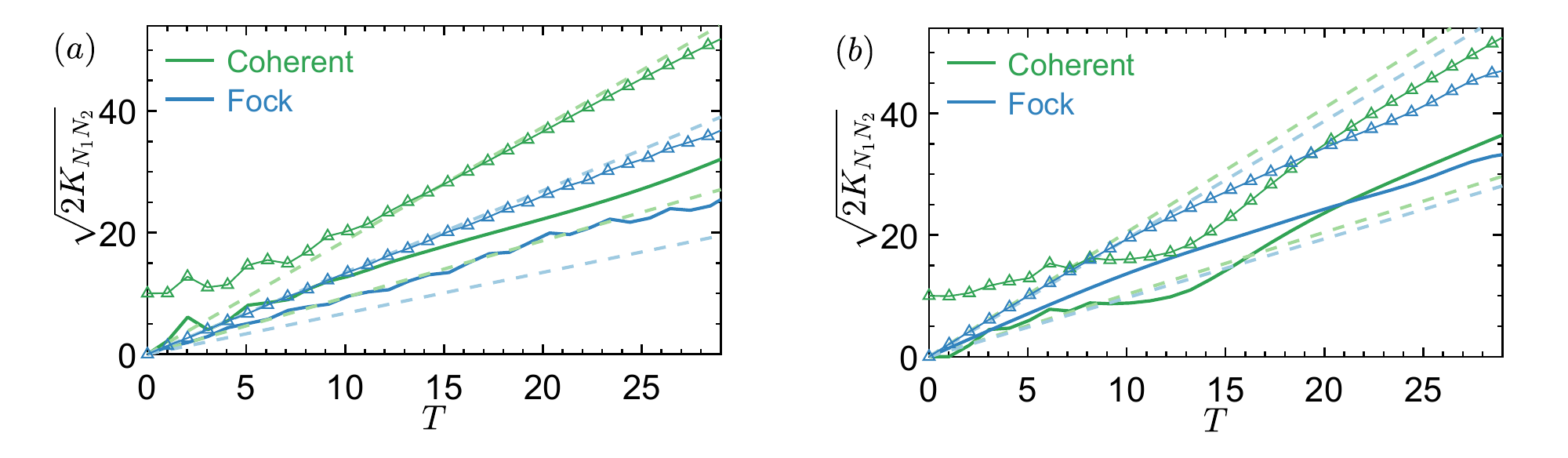}
    \caption{Time evolution of the covariance $K_{N_{1}N_{2}}\equiv-2\left[\braket{\hat{a}_{1}^{\dagger
    }\hat{a}_{1}\hat{a}_{2}^{\dagger
    }\hat{a}_{2}}_{\text{PES}}-\braket{\hat{a}_{1}^{\dagger
    }\hat{a}_{1}}_{\text{PES}}    \braket{\hat{a}_{2}^{\dagger
    }\hat{a}_{2}}_{\text{PES}}\right]$ for different models and inputs. Specifically, we select two models discussed in the main text: (a) qubit under circularly-polarized drives with same chiralities [Eq.~\eqref{2-level with circularly-polarized drives}] and (b) qubit under circularly-polarized drives with opposite chiralities [Eq.~\eqref{polarization conversion}]. Green and blue curves represent coherent-state and Fock-state inputs initially. Bounds in Eq.~\eqref{entanglement bound} are represented as colored dashed lines for both inputs. For comparison, we add the quantum Fisher information $\sqrt{F_{Q}}$ for both inputs ($-\Delta-$ curves). We start with an average of $50$ bosons in each drive. For coherent-state inputs, we introduce an initial phase difference $\phi_{20}-\phi_{10}=0.6\pi$ between two drives. Other parameters in different models are given as: (a) $\omega=\frac{\omega_{0}}{4}$, $\mathcal{A}=\frac{\omega_{0}}{8}$; and (b) $\omega=\omega_{0}$, $\mathcal{A}=\frac{\omega_{0}}{2}$.}
    \label{covariance plot}
\end{figure*}

We now show that $\ket{\text{PES}}_{T}$ with $F_{q}\propto T^{2}$, given by Eq.~\eqref{QFI bound appendix}, exhibits path entanglement for large conversion time $T$. Specifically, such entanglement is quantified by a similar bound $Q[f]$ in the long-time limit. For clarity, we explain the terminology 'path entanglement' (Sec.~\ref{entanglement generation}). It is borrowed from optical interferometry \cite{Jin2013}, where two spatially distinguishable modes of photons enter different interferometric arms or paths. Path entanglement simply refers to the mode entanglement between arms or paths. In order to show that the state $\ket{\text{PES}}_{T}$ [Eq.~\eqref{path-entangled state qudit}] is path-entangled,  we need to show the following inequality holds \cite{Gessner2017entanglement}:

\begin{equation}
    \text{Var}\left(\hat{J}_{z}\right)>\text{Var}\left(\frac{1}{2}\hat{a}_{1}^{\dagger}\hat{a}_{1}\right)+\text{Var}\left(\frac{1}{2}\hat{a}_{2}^{\dagger}\hat{a}_{2}\right),
\end{equation}

\noindent with $\text{Var}\left(\square\right)$ the variance of $\square$ with respect to $\ket{\text{PES}}_{T}$. We can equivalently show:

\begin{equation}
    \braket{\hat{a}_{1}^{\dagger
    }\hat{a}_{1}\hat{a}_{2}^{\dagger
    }\hat{a}_{2}}_{\text{PES}}<    \braket{\hat{a}_{1}^{\dagger
    }\hat{a}_{1}}_{\text{PES}}    \braket{\hat{a}_{2}^{\dagger
    }\hat{a}_{2}}_{\text{PES}}.
\label{entanglement condition}
\end{equation}

\noindent We stick to an ancilla qubit ($d=2$) and calculate the left-hand side of Eq.~\eqref{entanglement condition} in the large $T$ limit. Using the phase representation of the number operator $\hat{a}_{j}^{\dagger}\hat{a}_{j}=-i\partial_{\phi_{j}}$, we have

\begin{align}
\braket{\hat{a}_{1}^{\dagger
    }\hat{a}_{1}\hat{a}_{2}^{\dagger
    }\hat{a}_{2}}_{\text{PES}}=&-\frac{1}{\mathcal{N}_{T}^{2}}\sum_{m,n=1,2}\ \int\limits_{\text{BZ}}\bm{d^{2}\Bar{\phi}d^{2}\phi}f^{*}\left(\bm{\Bar{\phi}}\right)f\left(\bm{\phi}\right)\left|u_{m}^{*}u_{n}\right|^{2}e^{-i\left(\epsilon_{n}-\epsilon_{m}\right)T}\braket{\bm{\Bar{\phi}}|\partial_{\phi_{1}}\partial_{\phi_{2}}|\bm{\phi}}\\
    =&T^{2}\frac{1}{\mathcal{N}^{2}}\sum_{n=1,2}\ \int\limits_{\text{BZ}}\bm{d^{2}\phi}\left|f(\bm{\phi})\right|^{2}\left|u_{n}\right|^{4}(\partial_{\phi_{1}}\epsilon_{n})(\partial_{\phi_{2}}\epsilon_{n})+O(T^{3/2}),
\end{align}

\noindent where we use the stationary phase approximation [see discussion around Eq.~\eqref{C5}] to replace the double summation to a single one. From the energy conservation: $\left(\omega_{1}\partial_{\phi_{1}}+\omega_{2}\partial_{\phi_{2}}\right)\epsilon_{n}=0$, we know $(\partial_{\phi_{1}}\epsilon_{n})(\partial_{\phi_{2}}\epsilon_{n})\leq0$. Specifically, for a dispersive band structure, we have $(\partial_{\phi_{1}}\epsilon_{n})(\partial_{\phi_{2}}\epsilon_{n})<0$ and hence

\begin{equation}
    \braket{\hat{a}_{1}^{\dagger
    }\hat{a}_{1}\hat{a}_{2}^{\dagger
    }\hat{a}_{2}}_{\text{PES}}=-\mathcal{B}T^{2}+O(T^{3/2}),
\end{equation}

\noindent with $\mathcal{B}\equiv\frac{1}{\mathcal{N}^{2}}\sum\limits_{n=1,2}\ \int\limits_{\text{BZ}}\bm{d^{2}\phi}\left|f(\bm{\phi})\right|^{2}\left|u_{n}\right|^{4}\left|(\partial_{\phi_{1}}\epsilon_{n})(\partial_{\phi_{2}}\epsilon_{n})\right|$. 

We next evaluate the right-hand side of Eq.~\eqref{entanglement condition}:

\begin{align}
    \braket{\hat{a}_{j}^{\dagger}\hat{a}_{j}}_{\text{PES}}=&-\frac{i}{\mathcal{N}_{T}^{2}}\sum_{m,n=1,2}\ \int\limits_{\text{BZ}}\bm{d^{2}\Bar{\phi}d^{2}\phi}f^{*}\left(\bm{\Bar{\phi}}\right)f\left(\bm{\phi}\right)\left|u_{m}^{*}u_{n}\right|^{2}e^{-i\left(\epsilon_{n}-\epsilon_{m}\right)T}\braket{\bm{\Bar{\phi}}|\partial_{\phi_{j}}|\bm{\phi}}\\
    =&T\frac{1}{\mathcal{N}^{2}}\sum_{n=1,2}\ \int\limits_{\text{BZ}}\bm{d^{2}\phi} \left|f(\bm{\phi})\right|^{2}\left|u_{n}\right|^{4}\partial_{\phi_{j}}\epsilon_{n}+O(T^{1/2})\\
    =&O(T^{1/2}).
\end{align}

\noindent From the $1$-st line to the $2$-nd line, we drop the fast oscillating terms using the stationary phase approximation. From the $2$-nd to the last line, we use the fact that

\begin{equation}
    \sum_{n=1,2}\ \int\limits_{\text{BZ}}\bm{d^{2}\phi}\left|f(\bm{\phi})\right|^{2}\left|u_{n}\right|^{4}\partial_{\phi_{j}}\epsilon_{n}=0,
\label{vanishing expectation}
\end{equation}

\noindent following the same argument as in Eq.~\eqref{Jz1: 2 level ancilla}-\eqref{Jz6: 2 level ancilla}. It follows that

\begin{equation}
    \braket{\hat{a}_{1}^{\dagger
    }\hat{a}_{1}\hat{a}_{2}^{\dagger
    }\hat{a}_{2}}_{\text{PES}}-\braket{\hat{a}_{1}^{\dagger
    }\hat{a}_{1}}_{\text{PES}}    \braket{\hat{a}_{2}^{\dagger
    }\hat{a}_{2}}_{\text{PES}}= -\mathcal{B}T^{2}+O(T^{3/2}).
\end{equation}

\noindent This shows that at sufficiently late times $T$, Eq.~\eqref{entanglement condition} is fulfilled and $\ket{\text{PES}}_T$ is path-entangled. 

We now comment on the relation of the quadratic scaling of the QFI, derived in the previous appendix, and the path entanglement shown here.  We first bound $\mathcal{B}$. Using $\left|u_{1}\right|^{4}+\left|u_{2}\right|^{4}\in\left[\frac{1}{2},1\right]$, we have

\begin{equation}
    \frac{1}{2}Q[f]\leq\mathcal{B}\leq 2Q[f],
\label{entanglement bound}
\end{equation}

\noindent with $Q[f]\equiv\int\limits_{\text{BZ}}\bm{d^{2}\phi}\left|f^{2}\left(\bm{\phi}\right)\right|\left|\left(\partial_{\phi_{1}}\epsilon_{1}\right)\left(\partial_{\phi_{2}}\epsilon_{2}\right)\right|$. Note that we can express $P^2[f]$, defined in Eq.~\eqref{QFI bound appendix}, as
\begin{align}
P^{2}[f]=&\int_{\text{BZ}}\bm{d^{2}\phi}\left|f^{2}\left(\bm{\phi}\right)\right|\left[(\partial_{\phi_{1}}\epsilon_{1})^{2}+(\partial_{\phi_{2}}\epsilon_{2})^{2}+2(\partial_{\phi_{1}}\epsilon_{1})(\partial_{\phi_{2}}\epsilon_{2})\right]\\
=&\int_{\text{BZ}}\bm{d^{2}\phi}\left|f^{2}\left(\bm{\phi}\right)\right|\left[(\partial_{\phi_{1}}\epsilon_{1})^{2}+(\partial_{\phi_{2}}\epsilon_{2})^{2}\right]+2Q[f],
\end{align}
using $\partial_{\phi_{j}}\epsilon_{1}+\partial_{\phi_{j}}\epsilon_{2}=0$ from the tracelessness of $\hat{P}_{j}[f]$. It is then clear that a non-zero $\mathcal{Q}[f]$ from path entanglement can enhance the QFI bounded by $P^{2}[f]$. Thus, the $T^2$ scaling is not an exclusive property for path (mode) entanglement, such entanglement enters as an enhancement in  $F_{q}$. This phenomenon is also present for the N00N state: the sum of QFI $F_{q}\left[\text{N00N}, \frac{1}{2}\hat{a}_{1}^{\dagger}\hat{a}_{1}\right]+F_{q}\left[\text{N00N}, \frac{1}{2}\hat{a}_{2}^{\dagger}\hat{a}_{2}\right]=\frac{N^2}{2}$ for two number operators (divided by $\frac{1}{2}$) equals half of the QFI $F_{q}\left[\text{N00N},\hat{J}_{z}\right]$ for $\hat{J}_{z}=\frac{1}{2}\hat{a}_{1}^{\dagger}\hat{a}_{1}-\frac{1}{2}\hat{a}_{2}^{\dagger}\hat{a}_{2}$. The missing half comes exactly from the mode entanglement. Here, we see a more general version of this. Labeling $K_{N_{1}N_{2}}\equiv-2\left[\braket{\hat{a}_{1}^{\dagger
    }\hat{a}_{1}\hat{a}_{2}^{\dagger
    }\hat{a}_{2}}_{\text{PES}}-\braket{\hat{a}_{1}^{\dagger
    }\hat{a}_{1}}_{\text{PES}}    \braket{\hat{a}_{2}^{\dagger
    }\hat{a}_{2}}_{\text{PES}}\right]$, we note that the contribution in the QFI from mode entanglement is given by $2K_{N_{1}N_{2}}$. In Fig.~\ref{covariance plot}, we plot $\sqrt{2K_{N_{1}N_{2}}}$ as functions of $T$ for models given in Eq.~\eqref{2-level with circularly-polarized drives} and Eq.~\eqref{polarization conversion}.

\section{Maximizing the quantum Fisher information in qudit systems}
\label{optimization and qudits}

\begin{figure}
    \centering
    \includegraphics[width=0.5\linewidth]{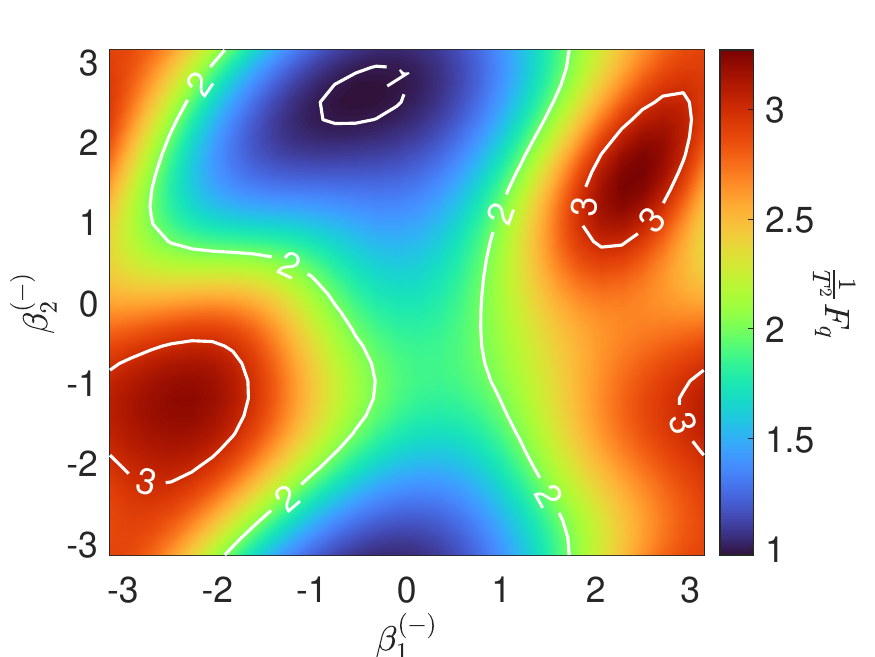}
    \caption{\textit{Optimization of the quantum Fisher information for the qutrit model.} For the qutrit model given in Eq.~\eqref{qutrit model}, there are $2$ degrees of freedom $\beta_{1}^{(-)},\beta_{2}^{(-)}$ in defining the initial qutrit state $\ket{\beta,+}_{f}$. The landscape of the quantum Fisher information $\frac{1}{T^2}F_{q}\left[ \ket{\text{PES}}_{T}, \hat{J}_{z}\right](\beta_{1}^{(-)},\beta_{2}^{(-)})$ shows strong dependence on these parameters. We initialize the drives in the Fock state with $30$ bosons in each mode choose the following parameters: $\omega_{12}=\omega$, $\omega_{23}=\frac{1}{2}\omega$, $\mathcal{A}=\frac{\omega}{2}$ and $T=10T_{\text{com}}$.}
    \label{QutritOptimization}
\end{figure}

 In this appendix, we generalize the above analysis to higher dimensional systems (qudits). Naturally, introducing a larger Hilbert space endows us with extra degrees of freedom in defining the initial state that optimizes the QFI. 
 
 We start with optimizaing the QFI for a qubit. According to our procedure of generating PES's, there is a phase degree of freedom $\beta$ in defining $\ket{\beta,+}_{f}$. While this initial phase does not affect the energy transfer since $\bra{\beta,+}_{f}\hat{P}_{j}\left[f\right]\ket{\beta,+}_{f}=0$ holds for $\forall \beta\in\left[0,2\pi\right)$, the scaling of QFI, $\lim\limits_{T\rightarrow\infty}\frac{1}{T^{2}}F_{q}\left[ \ket{\text{PES}}_{T}, \hat{J}_{z}\right]$, varies with $\beta$ in general. Hence, by optimizing the initial phase $\beta$, one maximizes the QFI of PES's. This, however, is only an improvement by a constant factor, compared with the improvement by a scaling factor $\frac{1}{T}$. Furthermore, the bound $P^{2}\left[f\right]$ on the QFI is independent of $\beta$, which means we can only maximize $\lim\limits_{T\rightarrow\infty}\frac{1}{T^{2}}F_{q}\left[ \ket{\text{PES}}_{T}, \hat{J}_{z}\right]$ within the interval $\left[\frac{1}{2}P^{2}\left[f\right],2P^{2}\left[f\right]\right]$. Following this procedure, we get $\sim10\%$ improvement of QFI for the polarization conversion model [Eq.~\eqref{polarization conversion}].

When promoting the qubit to a d-dimensional ancilla (qudit, $d>2$), such optimization becomes more significant. From the definition of $\ket{\beta,+}_{f}$ [Eq.~\eqref{ancilla states}], we have $D=W_{+}\left[f\right]+W_{-}\left[f\right]-1$ independent phases $\left\{\beta_{1}^{(+)},...,\beta_{W_{+}\left[f\right]-1}^{(+)},\beta_{1}^{(-)},...,\beta_{W_{-}\left[f\right]}^{(-)}\right\}$. Importantly, while the PES's in Eq.~\eqref{path-entangled state qudit} still exhibit the $F_{q}\propto T^2$ (App.~\ref{QFI appendix}),  they now generally have a non-vanishing expectation of $\hat{J}_{z}$ (where in the qubit case $\braket{\hat{J}_{z}}=0$ for all $\beta$).
Therefore, it is necessary to tune the $D$ phases such that the QFI is maximized.

As an example, we study the optimization of the QFI for the following driven-qutrit model:

\begin{equation}
\begin{split}
    \hat{H}_{\bm{\phi}}\left(t\right)=&\omega_{12}\ket{e1}\bra{e1}+\left(\omega_{12}+\omega_{23}\right)\ket{e2}\bra{e2}\\
    &+\frac{1}{2}\mathcal{A}\cos\left(\omega t+\phi_{1}\right)\left(\ket{e1}\bra{g}+\ket{g}\bra{e1}\right)\\
    &+\frac{1}{2}\mathcal{A}\cos\left(\omega t+\phi_{2}\right)\left(\ket{e2}\bra{e1}+\ket{e1}\bra{e2}\right)\\
    &+2\mathcal{A}\sin\left(\omega t+\phi_{1}\right)\left(\ket{e2}\bra{g}+\ket{g}\bra{e2}\right)\\
    &-2\mathcal{A}\sin\left(\omega t+\phi_{2}\right)\left(\ket{e2}\bra{g}+\ket{g}\bra{e2}\right),
\end{split}
\label{qutrit model}
\end{equation}

\noindent where $\ket{g},\ket{e1},\ket{e2}$ are the ground state, $1$-st excited state and $2$-nd excited state for the qutrit. For a symmetric Fock-state input $\left|f\left(\bm{\phi}\right)\right|^{2}=\frac{1}{4\pi^{2}}$, the functional power operator $P_{2}\left[f\right]$ with respect to the second drive has one positive eigenvalue $P_{+}$ and two negative eigenvalues $P_{-,1},P_{-,2}$ satisfying $P_{+}+P_{-,1}+P_{-,2}=0$. Denoting the correspoinding eigenvectors as $\ket{+}$, $\ket{-,1}$
, and $\ket{-,2}$, the initial qudit state $\ket{\beta,+}$ is hence defined as [Eq.~\eqref{ancilla states}]:

\begin{equation}
\begin{split}
\ket{\beta,+}=\frac{1}{\sqrt{3}}\left(\ket{+}+e^{i\beta_{1}^{(-)}}\ket{-,1}+e^{i\beta_{2}^{(-)}}\ket{-,2}\right).
\end{split}
\end{equation}

\noindent The optimization of the QFI goes over $\beta_{1}^{(-)},\beta_{2}^{(-)}\in\left[0,2\pi\right)$. In Fig.~\ref{QutritOptimization}, this optimization results in a significant increase in the QFI from $\frac{1}{T^2}F_{q}\left[ \ket{\text{PES}}_{T}, \hat{J}_{z}\right]<1$ to $\frac{1}{T^2}F_{q}\left[ \ket{\text{PES}}_{T}, \hat{J}_{z}\right]>3$.

\section{Derivation of the parity operator}
\label{parity operator appendix}

In this appendix, we derive the form of the parity operator following a Mach-Zehnder interferometer. We first recall how a $50-50$ beam splitter, represented as $\hat{\mathcal{U}}_{\text{de}}^{\text{MZI}}=e^{-i\hat{J}_{x}\frac{\pi}{2}}$, acts on the annihilation operators:

\begin{align}
\hat{\mathcal{U}}_{\text{de}}^{\text{MZI},\dagger}\hat{a}_{1}\hat{\mathcal{U}}_{\text{de}}^{\text{MZI}}=&e^{i\frac{\pi}{4}\left(\hat{a}_{1}^{\dagger}\hat{a}_{2}+\hat{a}_{2}^{\dagger}\hat{a}_{1}\right)}\hat{a}_{1} e^{-i\frac{\pi}{4}\left(\hat{a}_{1}^{\dagger}\hat{a}_{2}+\hat{a}_{2}^{\dagger}\hat{a}_{1}\right)}\\
=&\frac{1}{\sqrt{2}}\left(\hat{a}_{1}-i\hat{a}_{2}\right),\\
\hat{\mathcal{U}}_{\text{de}}^{\text{MZI},\dagger}\hat{a}_{2}\hat{\mathcal{U}}_{\text{de}}^{\text{MZI}}=&e^{i\frac{\pi}{4}\left(\hat{a}_{1}^{\dagger}\hat{a}_{2}+\hat{a}_{2}^{\dagger}\hat{a}_{1}\right)}\hat{a}_{2} e^{-i\frac{\pi}{4}\left(\hat{a}_{1}^{\dagger}\hat{a}_{2}+\hat{a}_{2}^{\dagger}\hat{a}_{1}\right)}\\
=&\frac{1}{\sqrt{2}}\left(\hat{a}_{2}-i\hat{a}_{1}\right),
\end{align}

\noindent where we have used the Baker-Campbell-Hausdorff (BCH) formula. Plugging these transformations into the parity operator, we obtain:

\begin{align}
\hat{\mathcal{U}}_{\text{de}}^{\text{MZI},\dagger}\hat{\Pi}\hat{\mathcal{U}}_{\text{de}}^{\text{MZI}}=&\hat{\mathcal{U}}_{\text{de}}^{\text{MZI},\dagger}\left(-1\right)^{\hat{a}_{2}\hat{a}_{2}}\hat{\mathcal{U}}_{\text{de}}^{\text{MZI}}\\
=&e^{i\frac{\pi}{4}\left(\hat{a}_{1}^{\dagger}\hat{a}_{2}+\hat{a}_{2}^{\dagger}\hat{a}_{1}\right)}e^{i\pi\hat{a}_{2}\hat{a}_{2}} e^{-i\frac{\pi}{4}\left(\hat{a}_{1}^{\dagger}\hat{a}_{2}+\hat{a}_{2}^{\dagger}\hat{a}_{1}\right)}\\
=&e^{i\frac{\pi}{2}\left(\hat{a}_{1}^{\dagger}\hat{a}_{1}+\hat{a}_{2}^{\dagger}\hat{a}_{2}\right)}e^{\frac{\pi}{2}\left(\hat{a}_{2}^{\dagger}\hat{a}_{1}-\hat{a}_{1}^{\dagger}\hat{a}_{2}\right)},\label{parityBS1}
\end{align}

\noindent where the first exponential in Eq.~\eqref{parityBS1} is a global phase factor in the SU($2$) interferometer given by the total boson number $\hat{N}=\hat{a}_{1}^{\dagger}\hat{a}_{1}+\hat{a}_{2}^{\dagger}\hat{a}_{2}$. The second term, which can be written as $e^{i\hat{J}_{y}\pi}$, has nontrivial action on the annihilation operators:

\begin{align}
e^{\frac{\pi}{2}\left(\hat{a}_{2}^{\dagger}\hat{a}_{1}-\hat{a}_{1}^{\dagger}\hat{a}_{2}\right)}\hat{a}_{1}e^{-\frac{\pi}{2}\left(\hat{a}_{2}^{\dagger}\hat{a}_{1}-\hat{a}_{1}^{\dagger}\hat{a}_{2}\right)}=&\hat{a}_{2},\\
e^{\frac{\pi}{2}\left(\hat{a}_{2}^{\dagger}\hat{a}_{1}-\hat{a}_{1}^{\dagger}\hat{a}_{2}\right)}\hat{a}_{2}e^{-\frac{\pi}{2}\left(\hat{a}_{2}^{\dagger}\hat{a}_{1}-\hat{a}_{2}^{\dagger}\hat{a}_{2}\right)}=&-\hat{a}_{1},
\end{align}

\noindent which hence behaves as a SWAP operator with an extra sign determined by the parity of bosons in the $2$nd mode:

\begin{equation}
 e^{\frac{\pi}{2}\left(\hat{a}_{2}^{\dagger}\hat{a}_{1}-\hat{a}_{1}^{\dagger}\hat{a}_{2}\right)}=\hat{S}\hat{\Pi},   
\end{equation}

\noindent with $\hat{S}=\sum\limits_{n,m=1}^{\infty}\ket{n}\bra{m}\otimes\ket{m}\bra{n}$ the SWAP operator over two bosonic modes. Finally we apply $\hat{\mathcal{U}}_{\theta}^{\text{MZI}}$ on the transformed parity operator and obtain:

\begin{align}
&\hat{\mathcal{U}}_{\theta}^{\text{MZI},\dagger}\ \hat{\mathcal{U}}_{\text{de}}^{\text{MZI},\dagger}\ \hat{\Pi}\ \hat{\mathcal{U}}_{\text{de}}^{\text{MZI}}\ \hat{\mathcal{U}}_{\theta}^{\text{MZI}}\\
&=\hat{\mathcal{U}}_{\theta}^{\text{MZI},\dagger}\ e^{i\frac{\pi}{2}\left(\hat{a}_{1}^{\dagger}\hat{a}_{1}+\hat{a}_{2}^{\dagger}\hat{a}_{2}\right)}\hat{S}\hat{\Pi}\ \hat{\mathcal{U}}_{\theta}^{\text{MZI}}\\
&=e^{i\frac{1}{2}\left(\hat{a}_{1}^{\dagger}\hat{a}_{1}-\hat{a}_{2}^{\dagger}\hat{a}_{2}\right)\theta}e^{i\frac{\pi}{2}\left(\hat{a}_{1}^{\dagger}\hat{a}_{1}+\hat{a}_{2}^{\dagger}\hat{a}_{2}\right)}\hat{S}\hat{\Pi}e^{-i\frac{1}{2}\left(\hat{a}_{1}^{\dagger}\hat{a}_{1}-\hat{a}_{2}^{\dagger}\hat{a}_{2}\right)\theta}\\
&=e^{i\frac{1}{2}\left(\hat{a}_{1}^{\dagger}\hat{a}_{1}-\hat{a}_{2}^{\dagger}\hat{a}_{2}\right)\theta}e^{i\frac{\pi}{2}\left(\hat{a}_{1}^{\dagger}\hat{a}_{1}+\hat{a}_{2}^{\dagger}\hat{a}_{2}\right)}\hat{S}e^{i\pi\hat{a}_{2}^{\dagger}\hat{a}_{2}}e^{-i\frac{1}{2}\left(\hat{a}_{1}^{\dagger}\hat{a}_{1}-\hat{a}_{2}^{\dagger}\hat{a}_{2}\right)\theta}\\
&=e^{i\left(\theta-\frac{\pi}{2}\right)\hat{a}_{1}^{\dagger}\hat{a}_{1}}\hat{S}e^{-i\left(\theta-\frac{\pi}{2}\right)\hat{a}_{1}^{\dagger}\hat{a}_{1}}\\
&=e^{i\left(\theta-\frac{\pi}{2}\right)\left(\hat{a}_{1}^{\dagger}\hat{a}_{1}-\hat{a}_{2}^{\dagger}\hat{a}_{2}\right)}\hat{S}\\
&=e^{2i\hat{J}_{z}\left(\theta-\frac{\pi}{2}\right)}\hat{S}\label{Parity_op},
\end{align}
\noindent where we have used $\hat{S}\ \hat{a}_{1}^{\dagger}\hat{a}_{1}\ \hat{S}=\hat{a}_{2}^{\dagger}\hat{a}_{2}$ and $\hat{S}\ \hat{a}_{2}^{\dagger}\hat{a}_{2}\ \hat{S}=\hat{a}_{1}^{\dagger}\hat{a}_{1}$.

\section{Sensitivity of parity measurement}
\label{characteristic functions appendix}

In this appendix, we prove two necessary conditions on driving frequencies and bandstructure for sub-SQL sensitivity from parity measurement on the path-entangled states $\ket{\text{PES}}_{T}$ defined in Eq.~\eqref{path-entangled state qudit}. We start with evaluating the expectation value of the parity operator $\hat{\Pi}_{\Tilde{\theta}}=e^{2i\hat{J}_{z}\Tilde{\theta}}\hat{S}$ [Eq.~\eqref{Parity_op}] for $\ket{\text{PES}}_{T}$. The action of $e^{2i\hat{J}_{z}\Tilde{\theta}}$ and $\hat{S}$ on a phase eigenstate $\ket{\bm{\phi}}=\ket{\phi_{1}}\otimes\ket{\phi_{2}}$ is given as:

\begin{align}
e^{2i\hat{J}_{z}\Tilde{\theta}}\ket{\bm{\phi}}&=\ket{\phi_{1}+\Tilde{\theta}}\otimes\ket{\phi_{2}-\Tilde{\theta}},\\
\hat{S}\ket{\bm{\phi}}&=\ket{\phi_{2}}\otimes\ket{\phi_{1}}.
\end{align}

\noindent Here, $e^{2i\hat{J}_{z}\Tilde{\theta}}$ acts as a translation in phase space, while $\hat{S}$ swaps two phases. We next evaluate the expectation explicitly:

\begin{align}
    \braket{\hat{\Pi}_{\Tilde{\theta}}}_{\text{PES}}=&\bra{\text{PES}}_{T}\hat{\Pi}_{\Tilde{\theta}}\ket{\text{PES}}_{T}\\
    =&\frac{1}{\mathcal{N}_{T}^{2}}\sum\limits_{m,n=1}^{d}\ \int\limits_{\text{BZ}}\bm{d^{2}\Bar{\phi}d^{2}\phi}f^{*}\left(\Bar{\bm{\phi}}\right)f\left(\bm{\phi}\right)u^{*}_{m}u_{n}e^{-i\left(\epsilon_{n}-\epsilon_{m}\right)T}\bra{m}\hat{\Pi}_{\beta}\ket{n}\bra{\Bar{\bm{\phi}}}\hat{\Pi}_{\Tilde{\theta}}\ket{\bm{\phi}}\\
    =&\frac{1}{\mathcal{N}_{T}^{2}}\sum\limits_{m,n=1}^{d}\ \int\limits_{\text{BZ}}\bm{d^{2}\Bar{\bm{\phi}}d^{2}\phi}f^{*}\left(\Bar{\bm{\phi}}\right)f\left(\bm{\phi}\right)\left|u^{*}_{m}u_{n}\right|^{2}e^{-i\left(\epsilon_{n}-\epsilon_{m}\right)T}\bra{\Bar{\bm{\phi}}}e^{2i\hat{J}_{z}\Tilde{\theta}}\hat{S}\ket{\bm{\phi}}\\
    =&\frac{1}{\mathcal{N}_{T}^{2}}\sum\limits_{m,n=1}^{d}\ \int\limits_{\text{BZ}}\bm{d^{2}\Bar{\bm{\phi}}d^{2}\phi}f^{*}\left(\Bar{\bm{\phi}}\right)f\left(\bm{\phi}\right)\left|u^{*}_{m}u_{n}\right|^{2}e^{-i\left(\epsilon_{n}-\epsilon_{m}\right)T}\braket{\Bar{\bm{\phi}}_{1}|\phi_{2}+\Tilde{\theta}}\braket{\Bar{\bm{\phi}}_{2}|\phi_{1}-\Tilde{\theta}}\\
    =&\frac{1}{\mathcal{N}_{T}^{2}}\sum\limits_{m,n=1}^{d}\ \int\limits_{\text{BZ}}\bm{d^{2}\Bar{\phi}d^{2}\phi}f^{*}\left(\Bar{\phi}\right)f\left(\bm{\phi}\right)\left|u^{*}_{m}u_{n}\right|^{2}e^{-i\left(\epsilon_{n}-\epsilon_{m}\right)T}\delta (\Bar{\phi}_{1}-\phi_{2}-\Tilde{\theta})\ \delta (\Bar{\phi}_{2}-\phi_{1}+\Tilde{\theta})\\
    =&\frac{1}{\mathcal{N}_{T}^{2}}\sum\limits_{m,n=1}^{d}\ \int\limits_{\text{BZ}}\bm{d^{2}\phi}f^{*}(\phi_{2}+\Tilde{\theta},\phi_{1}-\Tilde{\theta})f\left(\phi_{1},\phi_{2}\right)\left|u^{*}_{m}(\phi_{2}+\Tilde{\theta},\phi_{1}-\Tilde{\theta})u_{n}\left(\phi_{1},\phi_{2}\right)\right|^{2}e^{-i\left[\epsilon_{n}\left(\phi_{1},\phi_{2}\right)-\epsilon_{m}(\phi_{2}+\Tilde{\theta},\phi_{1}-\Tilde{\theta})\right]T},\label{Pi theta 1}
\end{align}

\noindent where in the last line we write the $\bm{\phi}$-dependence explicitly. To simplify the expression, we use the energy conservation $\epsilon_{1}\hat{P}_{1}+\epsilon_{2}\hat{P}_{2}=0$, which enables our treating of $\epsilon_{n}(\Tilde{x})=\epsilon_{n}\left(\phi_{1},\phi_{2}\right)$ as a single-parameter function of $\Tilde{x}=\frac{1}{\sqrt{2}}\left(\frac{\phi_{1}}{\omega_{1}}-\frac{\phi_{2}}{\omega_{2}}\right)$ that does not depend on $\Tilde{y}=\frac{1}{\sqrt{2}}\left(\omega_{2}\phi_{1}+\omega_{1}\phi_{2}\right)$ completely. In terms of $\Tilde{x},\Tilde{y}$, Eq.~\eqref{Pi theta 1} can be written as:

\begin{align}
  \braket{\hat{\Pi}_{\Tilde{\theta}}}_{\text{PES}}=&\frac{1}{\mathcal{N}_{T}^{2}}\sum\limits_{m,n=1}^{d}\ \int d\Tilde{x}d\Tilde{y}\mathcal{M}_{nm}(\Tilde{x},\Tilde{y},\Tilde{\theta})e^{-i\mathcal{\chi}_{nm}(\Tilde{x},\Tilde{y},\Tilde{\theta})T},\label{Pi theta 2}\\
  \mathcal{M}_{nm}(\Tilde{x},\Tilde{y},\Tilde{\theta})=&f^{*}(\phi_{2}+\Tilde{\theta},\phi_{1}-\Tilde{\theta})f\left(\phi_{1},\phi_{2}\right)\left|u^{*}_{m}(\phi_{2}+\Tilde{\theta},\phi_{1}-\Tilde{\theta})u_{n}\left(\phi_{1},\phi_{2}\right)\right|^{2},\\
   \mathcal{\chi}_{nm}(\Tilde{x},\Tilde{y},\Tilde{\theta})=&\epsilon_{n}(\Tilde{x})-\epsilon_{m}\left[-\frac{1}{2}\left(\frac{\omega_{1}}{\omega_{2}}+\frac{\omega_{2}}{\omega_{1}}\right)\Tilde{x}+\frac{1}{2}\left(\frac{1}{\omega_{1}^{2}}-\frac{1}{\omega_{2}^{2}}\right)\Tilde{y}+\frac{1}{\sqrt{2}}\left(\frac{1}{\omega_{1}}+\frac{1}{\omega_{2}}\right)\Tilde{\theta}\right],
\end{align}

\noindent where we have defined the characteristic function $\chi_{nm}$ as the difference between quasienergies from the $n$-th band at $(\phi_{1},\phi_{2})$ and the $m$-th band at $(\phi_{2}+\Tilde{\theta},\phi_{1}-\Tilde{\theta})$. A similar expression can be obtained for $\partial_{\Tilde{\theta}}\braket{\hat{\Pi}_{\Tilde{\theta}}}_{\text{PES}}$ by direct differentiation

\begin{align}
\partial_{\Tilde{\theta}}\braket{\hat{\Pi}_{\Tilde{\theta}}}_{\text{PES}}=&\frac{1}{\mathcal{N}_{T}^{2}}\sum\limits_{m,n=1}^{d}\ \int d\Tilde{x}d\Tilde{y}\left[\partial_{\Tilde{\theta}}\mathcal{M}_{nm}(\Tilde{x},\Tilde{y},\Tilde{\theta})\right]e^{-i\mathcal{\chi}_{nm}(\Tilde{x},\Tilde{y},\Tilde{\theta})T}\label{Pi theta 3}\\
&-\frac{iT}{\mathcal{N}_{T}^{2}}\sum\limits_{m,n=1}^{d}\ \int d\Tilde{x}d\Tilde{y}\mathcal{M}_{nm}(\Tilde{x},\Tilde{y},\Tilde{\theta})\left[\partial_{\Tilde{\theta}}\chi_{nm}(\Tilde{x},\Tilde{y},\Tilde{\theta})\right]e^{-i\mathcal{\chi}_{nm}(\Tilde{x},\Tilde{y},\Tilde{\theta})T}.\label{Pi theta 4}
\end{align}

\noindent Noting that at asymptotic time $T\rightarrow\infty$, the integrands in Eq.~\eqref{Pi theta 2}, \eqref{Pi theta 3}, \eqref{Pi theta 4} are fast oscillating and sharing same exponentials $e^{-i\chi_{nm}(\Tilde{x},\Tilde{y},\Tilde{\theta})T}$, these integrals are hence fully determined by their behaviors at critical points of the characteristic function $\chi_{nm}(\Tilde{x},\Tilde{y},\Tilde{\theta})$, known as the stationary phase approximation. To find all such critical points, we need to solve the following equation:

\begin{equation}
    \nabla\chi_{nm}(\Tilde{x},\Tilde{y},\Tilde{\theta})=0,
\end{equation}

\noindent where $\nabla=(\partial_{\Tilde{x}},\partial_{\Tilde{y}})$ is the gradient operator in the $\Tilde{x},\Tilde{y}$ basis. This gives:

\begin{align}
\partial_{\Tilde{x}}\chi_{nm}=0\quad&\rightarrow\quad\epsilon_{n}^{'}(\Tilde{x})+\frac{1}{2}\left(\frac{\omega_{1}}{\omega_{2}}+\frac{\omega_{2}}{\omega_{1}}\right)\epsilon_{m}^{'}\left[-\frac{1}{2}\left(\frac{\omega_{1}}{\omega_{2}}+\frac{\omega_{2}}{\omega_{1}}\right)\Tilde{x}+\frac{1}{2}\left(\frac{1}{\omega_{1}^{2}}-\frac{1}{\omega_{2}^{2}}\right)\Tilde{y}+\frac{1}{\sqrt{2}}\left(\frac{1}{\omega_{1}}+\frac{1}{\omega_{2}}\right)\Tilde{\theta}\right]=0,\\
\partial_{\Tilde{y}}\chi_{nm}=0\quad&\rightarrow\quad\frac{1}{2}\left(\frac{1}{\omega_{1}^{2}}-\frac{1}{\omega_{2}^{2}}\right)\epsilon_{m}^{'}\left[-\frac{1}{2}\left(\frac{\omega_{1}}{\omega_{2}}+\frac{\omega_{2}}{\omega_{1}}\right)\Tilde{x}+\frac{1}{2}\left(\frac{1}{\omega_{1}^{2}}-\frac{1}{\omega_{2}^{2}}\right)\Tilde{y}+\frac{1}{\sqrt{2}}\left(\frac{1}{\omega_{1}}+\frac{1}{\omega_{2}}\right)\Tilde{\theta}\right]=0,
\end{align}

\noindent where $\epsilon_{n,m}^{'}\left(t\right)\equiv\frac{d}{dt}\epsilon_{n,m}\left(t\right)$ is the differential notation. \\
\textbf{Case $\omega_{1}\neq\omega_{2}$:} For $\omega_{1}\neq\omega_{2}$, the above equations give:

\begin{align}
\epsilon_{n}^{'}(\Tilde{x})=\epsilon_{m}^{'}\left[-\frac{1}{2}\left(\frac{\omega_{1}}{\omega_{2}}+\frac{\omega_{2}}{\omega_{1}}\right)\Tilde{x}+\frac{1}{2}\left(\frac{1}{\omega_{1}^{2}}-\frac{1}{\omega_{2}^{2}}\right)\Tilde{y}+\frac{1}{\sqrt{2}}\left(\frac{1}{\omega_{1}}+\frac{1}{\omega_{2}}\right)\Tilde{\theta}\right]=0,
\label{critical points}
\end{align}

\noindent which gives rise to the following two cases: 1. no solution: there is no stationary phase and hence the expectation value of the parity operator $\braket{\hat{\Pi}_{\Tilde{\theta}}}_{\text{PES}}=o\left(\frac{1}{T}\right)$ decays faster than $T^{-1}$ as $T\rightarrow\infty$, which does not contain any information about the phase $\Tilde{\theta}$. A similar argument can be directly applied to $\partial_{\Tilde{\theta}}\braket{\hat{\Pi}_{\Tilde{\theta}}}_{\text{PES}}$ and gives $\partial_{\Tilde{\theta}}\braket{\hat{\Pi}_{\Tilde{\theta}}}_{\text{PES}}=o(1)$ as $T\rightarrow\infty$. From Eq.~\eqref{parity sensitivity}, we conclude that $\sqrt{F_{\Tilde{\theta}}}=o(1)$ as $T\rightarrow\infty$. 2. Solution exists: the stationary phases occur exactly at the non-pumping points $\epsilon_{n,m}^{'}=0$, which sets Eq.~\eqref{Pi theta 4} to $0$ at any moment as [Eq.~\eqref{critical points}]:

\begin{align}
 \partial_{\Tilde{\theta}}\chi_{nm}(\Tilde{x},\Tilde{y},\Tilde{\theta})=-\frac{1}{\sqrt{2}}\left(\frac{1}{\omega_{1}}+\frac{1}{\omega_{2}}\right)\epsilon_{m}^{'}\left[-\frac{1}{2}\left(\frac{\omega_{1}}{\omega_{2}}+\frac{\omega_{2}}{\omega_{1}}\right)\Tilde{x}+\frac{1}{2}\left(\frac{1}{\omega_{1}^{2}}-\frac{1}{\omega_{2}^{2}}\right)\Tilde{y}+\frac{1}{\sqrt{2}}\left(\frac{1}{\omega_{1}}+\frac{1}{\omega_{2}}\right)\Tilde{\theta}\right]=0.
\end{align}

\noindent We hence evaluate $\braket{\hat{\Pi}_{\Tilde{\theta}}}_{\text{PES}}$ according to Eq.~\eqref{Pi theta 2} and $\partial_{\Tilde{\theta}}\braket{\hat{\Pi}_{\Tilde{\theta}}}_{\text{PES}}$ according to Eq.~\eqref{Pi theta 3}. For an everywhere dispersive band structure ($\partial_{\Delta\phi}\epsilon_{n}$ and $\partial^{2}_{\Delta\phi}\epsilon_{n}$ not vanishing at the same time), the Hessian of $\chi_{nm}(\Tilde{x},\Tilde{y},\Tilde{\theta})$ is non-zero, dictating all critical points non-degenerate. In this case, we have $\braket{\hat{\Pi}_{\Tilde{\theta}}}_{\text{PES}}=O(\frac{1}{T})$ and $\partial_{\Tilde{\theta}}\braket{\hat{\Pi}_{\Tilde{\theta}}}_{\text{PES}}=O(\frac{1}{T})$, which necessarily gives rise to $\sqrt{F_{\Tilde{\theta}}}=O(\frac{1}{T})$. Together with the first case, we conclude that given an everywhere dispersive band structure, $\braket{\hat{\Pi}_{\Tilde{\theta}}}_{\text{PES}}=O(\frac{1}{T})$ and $\partial_{\Tilde{\theta}}\braket{\hat{\Pi}_{\Tilde{\theta}}}_{\text{PES}}=o(1)$, yielding $\sqrt{F_{\Tilde{\theta}}}=o(1)$ that fails to reach either the SQL or the HL. If the everywhere dispersive condition is lifted, the Hessian of $\chi_{nm}(\Tilde{x},\Tilde{y},\Tilde{\theta})$ is, if solution exists,
zero, which then gives rise to degenerate critical points. Here, we can bound both $\braket{\hat{\Pi}_{\Tilde{\theta}}}_{\text{PES}}=O(1)$ and $\partial_{\Tilde{\theta}}\braket{\hat{\Pi}_{\Tilde{\theta}}}_{\text{PES}}=O(1)$ by some $O(1)$ constant. In this case, $\braket{\hat{\Pi}_{\Tilde{\theta}}}_{\text{PES}}$ is still slowly oscillating, which in general gives $\sqrt{F_{\Tilde{\theta}}}=O(1)$. Potential exceptions arise when $\left|\braket{\hat{\Pi}_{\Tilde{\theta}}}_{\text{PES}}\right|\rightarrow1$ as $T\rightarrow\infty$, where the denominator of Eq.~\eqref{parity sensitivity} becomes singular, and no straightforward argument can be made for $F_{\Tilde{\theta}}$. Nevertheless, in the numerical examples studied in this work (see also the following discussions when $\omega_{1}=\omega_{2}$), we find no enhancement arising from these singularities. \\
\textbf{Case $\omega_{1}=\omega_{2}$:}
To enhance the sensitivity of the parity measurement from a non-vanishing Eq.~\eqref{Pi theta 4} as $T\rightarrow\infty$, we need to set the driving frequencies to be identical $\omega_{1}=\omega_{2}=\omega$. This constitutes our first necessary condition.

In this case, we relabel the characteristic functions $\chi_{nm}$ as follows:

\begin{align}
\mathcal{G}_{nm}(\Tilde{x},\Tilde{\theta})=&\epsilon_{n}(\Tilde{x})-\epsilon_{m}\left[-\Tilde{x}+\frac{\sqrt{2}}{\omega}\Tilde{\theta}\right],
\end{align}

\noindent which is the form given in Eq.~\eqref{characteristic g function}. The difference of $\mathcal{G}_{nm}$ from the general functions $\mathcal{\chi}_{nm}$ is that the former does not depend on the second variable $\Tilde{y}$. The critical points of $\mathcal{G}_{nm}$ are already degenerate as $\partial^{2}_{\Tilde{y}}\mathcal{G}_{nm}=\partial_{\Tilde{x}}\partial_{\Tilde{y}}\mathcal{G}_{nm}=0$. We need to integrate out this redundancy in evaluating $\braket{\hat{\Pi}_{\Tilde{\theta}}}_{\text{PES}}$ and $\partial_{\Tilde{\theta}}\braket{\hat{\Pi}_{\Tilde{\theta}}}_{\text{PES}}$

\begin{align}
\braket{\hat{\Pi}_{\Tilde{\theta}}}_{\text{PES}}=&\frac{1}{\mathcal{N}_{T}^{2}}\sum\limits_{m,n=1}^{d}\ \int d\Tilde{x}\Bar{\mathcal{M}}_{nm}(\Tilde{x},\Tilde{\theta})e^{-i\mathcal{G}_{nm}(\Tilde{x},\Tilde{\theta})T},\label{Pi theta id 1}\\
\partial_{\Tilde{\theta}}\braket{\hat{\Pi}_{\Tilde{\theta}}}_{\text{PES}}=&\frac{1}{\mathcal{N}_{T}^{2}}\sum\limits_{m,n=1}^{d}\ \int d\Tilde{x}\left[\partial_{\Tilde{\theta}}\Bar{\mathcal{M}}_{nm}(\Tilde{x},\Tilde{\theta})\right]e^{-i\mathcal{G}_{nm}(\Tilde{x},\Tilde{\theta})T}\label{Pi theta id3}\\
&-\frac{iT}{\mathcal{N}_{T}^{2}}\sum\limits_{m,n=1}^{d}\ \int d\Tilde{x}\Bar{\mathcal{M}}_{nm}(\Tilde{x},\Tilde{\theta})\left[\partial_{\Tilde{\theta}}\mathcal{G}_{nm}(\Tilde{x},\Tilde{\theta})\right]e^{-i\mathcal{G}_{nm}(\Tilde{x},\Tilde{\theta})T},\label{Pi theta id4}\\
  \Bar{\mathcal{M}}_{nm}(\Tilde{x},\Tilde{\theta})=&\int d\Tilde{y}f^{*}(\phi_{2}+\Tilde{\theta},\phi_{1}-\Tilde{\theta})f\left(\phi_{1},\phi_{2}\right)\left|u^{*}_{m}(\phi_{2}+\Tilde{\theta},\phi_{1}-\Tilde{\theta})u_{n}\left(\phi_{1},\phi_{2}\right)\right|^{2}.
\end{align}

\noindent As before, when $T\rightarrow\infty$, the integrands in Eq.~\eqref{Pi theta id 1}\eqref{Pi theta id3}\eqref{Pi theta id4} are fast-oscillating and the integrals are governed by their behaviors at critical points:

\begin{align}
\partial_{\Tilde{x}}\mathcal{G}_{nm}(\Tilde{x},\Tilde{\theta})=0.
\label{critical points id}
\end{align}

\noindent The existence of solutions to Eq.~\eqref{critical points id} is guaranteed, unlike in the case where $\omega_{1}\neq\omega_{2}$, as $\mathcal{G}_{nm}(\Tilde{x},\Tilde{\theta})$ is periodic in $\Tilde{x}$. By the mean value theorem, there must exist at least one $\Tilde{x}_{0}$ such that $\partial_{\Tilde{x}}\mathcal{G}_{nm}(\Tilde{x},\Tilde{\theta})|_{\Tilde{x}=\Tilde{x}_{0}}=0$. If these critical points are non-degenerate, the asymptotic behaviors of Eq.~\eqref{Pi theta id 1} and Eq.~\eqref{Pi theta id3} are both governed by $O(\frac{1}{\sqrt{T}})$, while that of Eq.~\eqref{Pi theta id4} by $O(\sqrt{T})$. The appearance of $\sqrt{T}$ instead of $T$ arises from the reduction of the integral's dimensionality from 2 to 1. In consequence, Eq.~\eqref{parity sensitivity} gives rise to $\sqrt{F_{\Tilde{\theta}}}=O(\sqrt{T})$, exhibiting a diffusive scaling as $T\rightarrow\infty$. We hence conclude that the sub-SQL sensitivity from parity measurement can be obtained only if the critical points of $\mathcal{G}_{n,m}$ are degenerate. This constitutes our second necessary condition.

As a final remark, the coefficient $\partial_{\Tilde{\theta}}\mathcal{G}_{nm}(\Tilde{x},\Tilde{\theta})$ of the integrand in Eq.~\eqref{Pi theta id4} need to be non-vanishing at stationary phases such the enhancement of $\partial_{\Tilde{\theta}}\braket{\hat{\Pi}}_{\text{PES}}$ from Eq.~\eqref{Pi theta id4} is non-zero. Otherwise, Eq.~\eqref{Pi theta id3} at most gives an $O(1)$ contribution in $\partial_{\Tilde{\theta}}\braket{\hat{\Pi}}_{\text{PES}}$ and $\braket{\hat{\Pi}}_{\text{PES}}$ again oscillates slowly. As before, we expect $\sqrt{F_{\Tilde{\theta}}}=O(1)$ in general. Strictly speaking, singularities may occur when $\left|\braket{\hat{\Pi}}_{\text{PES}}\right|\rightarrow1$, where it becomes difficult to make any straightforward statements about 
$\sqrt{F_{\Tilde{\theta}}}$. However, from extensive numerical studies, we see no enhancement arising from such singularities. An example can be seen from the driven model given in Eq.~\ref{2-level with circularly-polarized drives} with coherent-state inputs [Fig.~\ref{Paritysensitivity}.(b)].

\section{Global phase estimation}
\label{global phase estimation}

In this appendix, we review the standard Bayesian procedure \cite{Macieszczak_2014,Rubio_2019} for the global phase estimation, which defines the posterior variance $\Delta_{\text{p}}\theta^{2}$ in Sec.~\ref{Loss}. In a global quantum sensor, the unknown phase $\theta$ has a finite a priori uncertainty $\Delta\theta$. In this case, the a priori phase should be viewed as a random variable that takes values in $\left[0,2\pi\right)$ according to its distribution function $P\left(\theta\right)$. The task now becomes estimating a random outcome $\theta$ as precise as possible. The mathematical quantity to be minimized is the mean-square error (MSE) of the estimator \cite{PhysRevX.11.041045,Rubio_2019}:

\begin{equation}
    \mathcal{C}=\int_{0}^{2\pi}d\theta\ \mathcal{E}\left(\theta\right)P\left(\theta\right),
\end{equation}

\noindent where $\mathcal{E}\left(\theta\right)=\sum\limits_{\mu=1}^{M}\left(\theta_{\text{est}}\left(\pi_{\mu}\right)-\theta\right)^{2}p\left(\pi_{\mu}|\theta\right)$ is the squared error of the estimator for a given $\theta$. Similarly, the first minimization step is achieved by choosing a minimal MSE estimator whose estimated value coincides with the mean of $\theta$:

\begin{equation}
    \theta_{\text{est}}^{\text{MSE}}\left(\pi_{\mu}\right)=\int_{0}^{2\pi}d\theta\ \theta p\left(\theta|\pi_{\mu}\right),
\end{equation}

\noindent where the conditional probabilities $p\left(\theta|\pi_{\mu}\right)$ and $p\left(\pi_{\mu}|\theta\right)$ are related by the Bayes theorem $p\left(\theta|\pi_{\mu}\right)=p\left(\pi_{\mu}|\theta\right)\frac{P\left(\theta\right)}{p\left(\pi_{\mu}\right)}$ with the marginal probability distribution $p\left(\pi_{\mu}\right)$ of the outcome $\pi_{\mu}$ given by $p\left(\pi_{\mu}\right)=\int_{0}^{2\pi}d\theta\ p\left(\pi_{\mu}|\theta\right)P\left(\theta\right)$. As a result, the MSE to be minimized becomes the posterior variance of $\theta$:

\begin{equation}
    \Delta_{\text{p}}\theta^{2}=\sum\limits_{\mu=1}^{M}p\left(\pi_{\mu}\right)\int_{0}^{2\pi}d\theta \left(\theta-\theta_{\text{est}}^{MSE}\left(\pi_{\mu}\right)\right)^{2}p\left(\theta|\pi_{\mu}\right).
\label{MSE cost function}
\end{equation}

\noindent From classical information theory, for a single-shot measurement, $\Delta_{\text{p}}\theta^{2}$ is bounded by a generalized CRB \cite{PhysRevX.11.041045}:

\begin{equation}
    \Delta_{\text{p}}\theta^{2}\geq\frac{1}{\overline{F}_{\theta}+F_{0}},
\end{equation}

\noindent where $\overline{F}_{\theta}=\int_{0}^{2\pi}d\theta F_{\theta}P\left(\theta\right)$ is the averaged classical Fisher information [see Eq.~\eqref{classical Fisher information}] and $F_{0}=\int_{0}^{2\pi}d\theta \left[\frac{d}{d\theta}\log P\left(\theta\right)\right]^{2}P\left(\theta\right)$ is the classical Fisher information for the a priori distribution $P\left(\theta\right)$. As in the local phase estimation, the classical Fisher information is upper-bounded by the corresponding QFI $F_{q}$. The posterior variance $\Delta_{\text{p}}\theta^{2}$ is hence lower-bounded by:

\begin{equation}
    \Delta_{\text{p}}\theta^{2}\geq\frac{1}{F_{q}+F_{0}}.
\label{Bayes bound}
\end{equation}

\noindent Taking the limit $P\left(\theta\right)\rightarrow\delta\left(\theta-\theta_{0}\right)$, Eq.~\eqref{Bayes bound} converges to the qCRB for a local phase sensor [Eq.~\eqref{qCRB}].\\

 The equality in Eq.~\eqref{Bayes bound} holds if and only if $\overline{F}_{\theta}=F_{q}$ for any $\theta\in\left[0,2\pi\right)$, which is a stringent condition in experiments. For interferometer-like sensors, the sensitivity is, hence, much worse than the right-hand-side bound. It is also known that the optimal global quantum sensor deviates significantly from its local counterpart in both the probe states and sensing schemes \cite{Marciniak2022}. Recent efforts on optimizing protocols and minimizing the posterior variance $\Delta_{\text{p}}\theta^{2}$ can be found in \cite{Marciniak2022,PRXQuantum.4.020333,PhysRevX.11.041045}. In this work, we focus on the estimation of a local phase, treating the a priori uncertainty $\Delta\theta$ as a source of noise (Sec.~\ref{Effects of a priori uncertainty}).\\

\section{Other types of noise and imperfection}\label{Other noise}

\begin{figure*}[h]
\centering
     \includegraphics[width=0.91\textwidth]{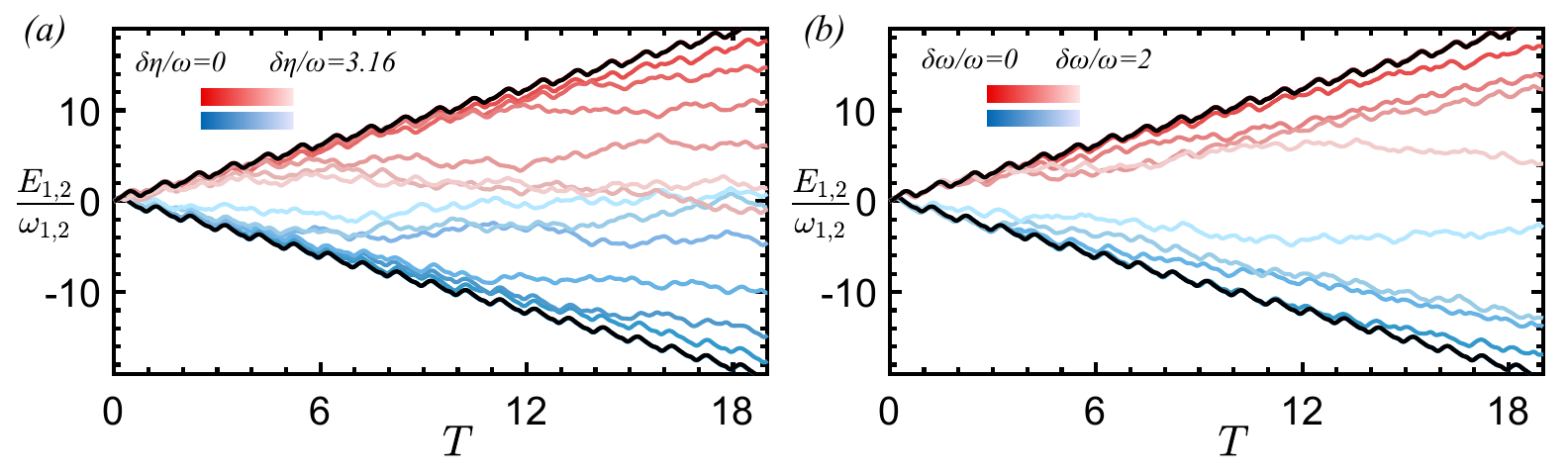}
        \caption{\textit{Energy transfer under dephasing and frequency noise.} In both plots, we show the work $E_{1,2}$ done by the 1st drive (red) and the 2nd drive (blue) under different strength of noise.}
        \label{Phase_Frequency_work}  
\end{figure*}

In the main text, we discuss how lossy interferometry and uncertain a priori phases compromise the performance of the Floquet sensor. In this appendix, we examine other common sources of noise relevant to any practical realization of the sensor. For qubit noise, we consider the detrimental effects of dephasing noise; for the drives, we focus instead on phase noise arising from randomness in the frequency.\\
\textbf{Qubit dephasing noise:} We consider the situation where the qubit is subject to a random dephasing noise $\delta\hat{H}(t)=\eta(t)\hat{\sigma}_{z}$, with the autocorrelation of $\eta(t)$ given in a white-Gaussian form
\begin{equation}
\braket{\eta(t)\eta(0)}=\delta\eta^{2}\tau\delta(t).
\end{equation}
Here, $\delta\eta$ is the standard deviation of the noise, $\tau=10^{-3} T_{com}$ is the time step for simulation and $\delta(t)$ is the Dirac delta function approximated by $\delta(0)=\frac{1}{\tau}$. The updated Hamiltonian therefore becomes $\hat{H}(t)\rightarrow\hat{H}(t)+\delta\hat{H}(t)$, where the bare part $\hat{H}(t)$ is given by Eq.~\eqref{two-drive model} in the main text. We expect such dephasing noise will eventually wash out any memory of the initial state of the qubit, which terminates the energy transfer in the long time limit. However, as such noise is much suppressed for qubits with long decoherence time $t_{2}$ (a good example is the fluxonium qubit that exhibits $t_{2}>100\mu s$ in recent designs). Therefore, we also expect the energy transfer is preserved at earlier times.

For concreteness, we numerically study the energy transfer for the polarization conversion model [Eq.~\eqref{polarization conversion}] under such dephasing noise. Since the expression for the qubit energy is given in the $\hat{\sigma}_{x}$ direction, such dephasing noise is, instead, given by $\delta\hat{H}(t)=\eta(t)\hat{\sigma}_{x}$. In Fig.\ref{Phase_Frequency_work}.(a), we illustrate the work done by two drives under varying noise strengths, $\delta\eta\in[0,3.16\omega]$. We observe that, as $\delta\eta$ increases, the energy transfer stops at earlier times. 

Let us estimate the $t_{2}$ for such dephasing noise. From fluctuation dissipation theory, it follows that  $t_{2}\sim\frac{1}{S(\omega)}$ (given $t_1\approx t_2$) with $S(\omega)=\int\limits_{-\infty}^{\infty}\braket{\eta(t)\eta(0)}e^{i\omega t}dt$. For $\delta\eta=3.16\omega$ and $\omega\sim 1GHz$, $t_{2}\sim100 ns$, which is much shorter than its realistic value $t_2>10\mu s$. For comparison, we added extra black curves corresponding to the realistic parameter $t_2=10\mu s$, which aligns well with noiseless ($\delta\eta=0$) case. As a result, dephasing noise is substantially suppressed under experimental conditions.

\textbf{Phase noise:} We next consider the phase noise in the drive, coming from a randomness in its frequency $\omega\rightarrow\omega+\Delta\omega$. We again assume a white-Gaussian noise in the frequency
\begin{equation}
\braket{\Delta\omega(t)\Delta\omega(0)}=\delta\omega^2\tau\delta(t).
\end{equation}
Here, $\delta\omega$ is the standard deviation of the noise. We note that such frequency noise induces a meander in the phase $\Delta\Phi(t)=\int\limits_{0}^{t}dt'\Delta\omega(t')$, satisfying $\braket{\Delta\Phi^{2}(t)}\propto t$. The Floquet Hamiltonian $\hat{H}(\omega t+\phi)$ is, therefore, replaced by $\hat{H}(\omega t+\phi+\Delta\Phi(t))$, in which this meandered phase $\Delta\Phi$ eventually washes out the memory of the initial phase and stops the energy conversion.

To see this, we assume that one drive in the polarization conversion model [Eq.~\eqref{polarization conversion}] is noisy and study energy transfer under different noise strengths $\delta\omega\in[0,2\omega]$. The result is given in Fig.\ref{Phase_Frequency_work}.(b). Similar to the dephasing noise, when increasing $\delta\omega$, the energy transfer is non-persistent due to the phase meander: At later times, this growing phase uncertainty overwhelms the initial phase $\phi$ completely. Nevertheless, in a practical setup, the frequency noise is significantly smaller than the bare frequency. For example, for a GHz microwave photon, the noise in frequency is on the order of KHz to MHz range. For comparison, we added extra black curves corresponding to the realistic parameter $\delta\omega/\omega=10^{-3}$, which aligns well with the noiseless ($\delta\omega=0$) case. Therefore, we conclude that frequency noise can be effectively suppressed as well.

\textbf{Detuning in frequency:} To form a Floquet system and to define two coin states $\ket{0}_{f},\ket{1}_{f}$ correctly, in the main text, we have assumed the frequency of two drives are exactly commensurate and fixed. However, in a real-time experiment, the actual frequency $\omega\rightarrow\omega+\delta\omega$ can be mistakenly detuned from the exact value we want. Here we use $\delta\omega$ to denote such detuning, rather than a random noise. Since we are initializing the qubit according to the exact value $\omega$, such detuning inevitably compromises the state preparation and the long time dynamics. Nevertheless, we expect the influence becomes significant only at $t\sim\frac{1}{\delta\omega}$, which is generally a much longer time scale than the period $T=\frac{2\pi}{\omega}$ upon assuming $\delta\omega\ll\omega$. 

\begin{figure*}[h]
\centering
     \includegraphics[width=0.94\textwidth]{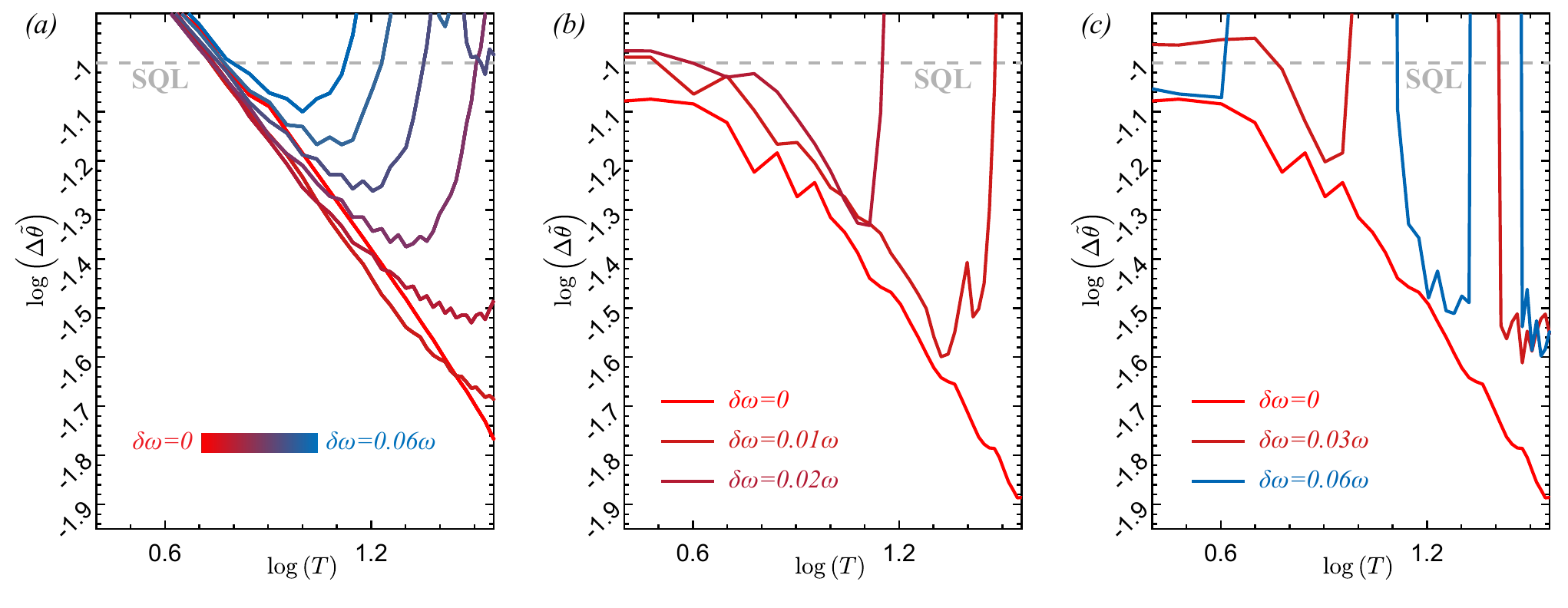}
        \caption{\textit{Sensitivity of the parity measurement with frequency detuning.} (a). Sensitivity of parity measurement for the Fock-state input with different detuned frequencies $\delta\omega$. (b),(c). Sensitivity of parity measurement for the coherent-state input with different detuned frequencies. It can be seen that as $\delta\omega$ increases, the deviation starts earlier, which is consistent with the rough estimation $t\sim\frac{1}{\delta\omega}$. For the coherent-state input, when such detuning is large, the parity sensitivity exhibits cyclic returns to the sub-SQL region after a sudden surge above the SQL. For all simulations, we choose the lattice version of Eq.~\eqref{polarization conversion} and start with 50 photons in each mode.}
        \label{Detuning_Parity}  
\end{figure*}

Meanwhile, we expect that the most significant influence is on the parity measurement, as it is extremely sensitive to the photon addition/reduction. From the energy transfer, an extra photon emerges at $t\sim\frac{1}{\delta\omega}$, changing its parity completely. To show this, in Fig. \ref{Detuning_Parity}, we show the sensitivity of the parity measurement under frequency detuning $\delta\omega$ for both Fock-state and coherent-state inputs. When increasing the detuning from $\delta\omega=0$ to $\delta\omega=0.06\omega$, we observe a decrease in $t$ when the sensitivity $\Delta\Tilde{\theta}$ deviates from the perfectly tuned case ($\delta\omega=0$), suggesting a relation $t\sim\frac{1}{\delta\omega}$. However, while the parity sensitivity leaves the sub-SQL region and never returns for the Fock-state input, we observe that the sensitivity for the coherent-state input cyclically returns to the sub-SQL region after experiencing a sudden surge above the SQL. Similarly, the surge arises when an extra photon is injected into the drive due to frequency detuning $\delta\omega$, which also follows the relation $t\sim\frac{1}{\delta\omega}$.

\section{Superconducting circuit realization}
\label{superconducting circuit appendix}
In this appendix, we give an example of establishing superconducting circuits for a specific two-tone Floquet system after quantizing the drives. We follow the standard procedure given in  \cite{PhysRevApplied.19.064043,PhysRevApplied.10.054062}.

The time-dependent Hamiltonian in Eq.~\eqref{SC realization} can be typically realized by the circuit showin in Fig.~\ref{superconducting circuit}, whose Lagrangian, $\mathcal{L}=T-U$, takes the following form

\begin{equation}
\begin{split}
    T=&\frac{1}{2}C_{q}\dot{\phi}_{q}^{2}+\frac{1}{2}C_{1}\dot{\phi}_{1}^{2}+\frac{1}{2}C_{2}\dot{\phi}_{2}^{2}+\frac{1}{2}C_{x_{1}}\dot{\phi}_{x{1}}^{2}+\frac{1}{2}C_{x_{2}}\dot{\phi}_{x{2}}^{2}+\frac{1}{2}C_{y_{1}}\dot{\phi}_{y{1}}^{2}+\frac{1}{2}C_{y_{2}}\dot{\phi}_{y{2}}^{2}
    +\frac{1}{2}C_{qc_{1}}\left(\dot{\phi}_{q}-\dot{\phi}_{y_{1}}\right)^{2}+\frac{1}{2}C_{qc_{2}}\left(\dot{\phi}_{q}-\dot{\phi}_{y_{2}}\right)^{2}\\
    &+\frac{1}{2}C_{c_{1}1}\left(\dot{\phi}_{1}-\dot{\phi}_{y_{1}}\right)^{2}+\frac{1}{2}C_{c_{2}2}\left(\dot{\phi}_{2}-\dot{\phi}_{y_{2}}\right)^{2}+\frac{1}{2}C_{q1}\left(\dot{\phi}_{q}-\dot{\phi}_{1}\right)^{2}+\frac{1}{2}C_{q2}\left(\dot{\phi}_{q}-\dot{\phi}_{2}\right)^{2},\\
    U=&U\left(\phi_{q}\right)+\frac{1}{2L_{1}}\phi_{1}^{2}+\frac{1}{2L_{2}}\phi_{2}^{2}+\frac{1}{2L_{q1}}\left(\phi_{q}-\phi_{1}\right)^{2}+\frac{1}{2L_{q2}}\left(\phi_{q}-\phi_{2}\right)^{2}+\frac{1}{2L_{qc_{1}}}\left(\phi_{q}-\phi_{c_{1}}\right)^{2}+\frac{1}{2L_{qc_{2}}}\left(\phi_{q}-\phi_{c_{2}}\right)^{2}\\
    &+\frac{1}{2L_{c_{1}1}}\left(\phi_{1}-\phi_{c_{1}}\right)^{2}+\frac{1}{2L_{2c_{2}}}\left(\phi_{2}-\phi_{c_{2}}\right)^{2}+E_{x_{1}}\left(1-\cos\phi_{x_{1}}\right)+E_{x_{2}}\left(1-\cos\phi_{x_{2}}\right)+E_{y_{1}}\left(1-\cos\phi_{y_{1}}\right)+E_{y_{2}}\left(1-\cos\phi_{y_{2}}\right)\\
    &+I\left(t\right)\phi_{q},
\end{split}
\end{equation}

\noindent where $U\left(\phi_{q}\right)$ is the potential for the qubit and $I\left(t\right)$ is the external current drive on the qubit. For notational simplicity, we implicitly incorporate the coupling strength into $I(t)$.
All other related quantities are marked in Fig.~\ref{superconducting circuit}. Introducing $\vec{\phi}=\left(\phi_{q},\phi_{1},\phi_{y_{1}},\phi_{x_{1}},\phi_{2},\phi_{y_{2}},\phi_{x_{2}}\right)^{T}$, the kinetic part $T$ has a compact form as $T=\frac{1}{2}\dot{\vec{\phi}}^{T}C\dot{\vec{\phi}}$, where the capacitance matrix is given by: 

\begin{center}
$C$$=$$\begin{pmatrix}
C_{q}+C_{qc_{1}}+C_{q1}+C_{qc_{2}}+C_{q2} & -C_{q1} & -C_{qc_{1}} & 0 & -C_{q2} & -C_{qc_{2}} & 0\\
-C_{q1} & C_{1}+C_{c_{1}1}+C_{q1} & -C_{c_{1}1} & 0 & 0 & 0 & 0\\
-C_{qc_{1}} & -C_{c_{1}1} & C_{y_{1}}+C_{qc_{1}}+C_{c_{1}1} & 0 & 0 & 0 & 0\\
0 & 0 & 0 & C_{x_{1}} & 0 & 0 & 0\\
-C_{q2} & 0 & 0 & 0 & C_{2}+C_{c_{2}2}+C_{q2} & -C_{c_{2}2} & 0\\
-C_{qc_{2}} & 0 & 0 & 0 & -C_{c_{2}2} & C_{y_{2}}+C_{qc_{2}}+C_{c_{2}2} & 0\\
0 & 0 & 0 & 0 & 0 & 0 & C_{x_{2}}\\

\end{pmatrix}$.	  
\end{center}

\noindent Assuming $C_{x_{1,2}},C_{y_{1,2}},C_{1,2},C_{q}\gg C_{qc_{1,2}}, C_{c_{1}1},C_{c_{2},2}\gg C_{q1},C_{q2}$, its inverse is given as

\begin{center}
$C^{-1}=\begin{pmatrix}
\frac{1}{C_{q}} & \frac{C_{q1}+\frac{C_{qc_{1}}C_{c_{1}1}}{C_{y{1}}}}{C_{q}C_{1}} & \frac{C_{qc_{1}}}{C_{q}C_{y_{1}}} & 0 & \frac{C_{q1}+\frac{C_{qc_{2}}C_{c_{2}2}}{C_{y{2}}}}{C_{q}C_{2}} & \frac{C_{qc_{2}}}{C_{q}C_{y_{2}}} & 0\\
\frac{C_{q1}+\frac{C_{qc_{1}}C_{c_{1}1}}{C_{y{1}}}}{C_{q}C_{1}} & \frac{1}{C_{1}} & \frac{C_{c_{1}1}}{C_{1}C_{y_{1}}} & 0 & 0 & 0 & 0\\
\frac{C_{qc_{1}}}{C_{q}C_{y_{1}}} & \frac{C_{c_{1}1}}{C_{1}C_{y_{1}}} & \frac{1}{C_{y_{1}}} & 0 & 0 & 0 & 0\\
0 & 0 & 0 & \frac{1}{C_{x_{1}}} & 0 & 0 & 0\\
\frac{C_{q1}+\frac{C_{qc_{2}}C_{c_{2}2}}{C_{y{2}}}}{C_{q}C_{2}} & 0 & 0 & 0 & \frac{1}{C_{2}} & \frac{C_{c_{2}2}}{C_{2}C_{y_{2}}} & 0\\
\frac{C_{qc_{2}}}{C_{q}C_{y_{2}}} & 0 & 0 & 0 & \frac{C_{c_{2}2}}{C_{2}C_{y_{2}}} & \frac{1}{C_{y_{2}}} & 0\\
0 & 0 & 0 & 0 & 0 & 0 & \frac{1}{C_{x_{2}}}\\

\end{pmatrix}$,	  
\end{center}

\noindent from which we can write down the Hamiltonian in terms of $\vec{\phi}$ and its canonical conjugate $\vec{n}=\frac{\partial\mathcal{L}}{\partial\dot{\vec{\phi}}}=C\dot{\vec{\phi}}$. After canonical quantization (i.e.\, promoting $\vec{\phi}$ and $\vec{n}$ to operators), we obtain:

\begin{equation}
\begin{split}
    \hat{H}=&4E_{Cq}\hat{n}_{q}^{2}+U\left(\phi_{q}\right)+\left(\frac{1}{2L_{qc_{1}}}+\frac{1}{2L_{q1}}+\frac{1}{2L_{qc_{2}}}+\frac{1}{2L_{q2}}\right)\phi_{q}^{2}\\
    &+4E_{C1}\hat{n}_{1}^{2}+\left(\frac{1}{2L_{1}}+\frac{1}{2L_{c_{1}1}}+\frac{1}{2L_{q1}}\right)\hat{\phi}_{1}^{2}\\
    &+4E_{C2}\hat{n}_{2}^{2}+\left(\frac{1}{2L_{1}}+\frac{1}{2L_{c_{2}2}}+\frac{1}{2L_{q2}}\right)\hat{\phi}_{2}^{2}\\
    &+4E_{Cy_{1}}\hat{n}_{y_{1}}^{2}-E_{y_{1}}\cos\hat{\phi}_{y_{1}}\\
    &+4E_{Cy_{2}}\hat{n}_{y_{2}}^{2}-E_{y_{2}}\cos\hat{\phi}_{y_{2}}\\
    &+4E_{Cx_{1}}\hat{n}_{x_{1}}^{2}-E_{x_{1}}\cos\hat{\phi}_{x_{1}}+\left(\frac{1}{2L_{qc_{1}}}+\frac{1}{2L_{c_{1}1}}\right)\hat{\phi}_{x_{1}}^{2}\\
    &+4E_{Cx_{2}}\hat{n}_{x_{2}}^{2}-E_{x_{2}}\cos\hat{\phi}_{x_{2}}+\left(\frac{1}{2L_{qc_{2}}}+\frac{1}{2L_{c_{2}2}}\right)\hat{\phi}_{x_{2}}^{2}\\
    &-\frac{1}{L_{qc_{1}}}\hat{\phi}_{q}\hat{\phi}_{x_{1}}-\frac{1}{L_{qc_{2}}}\hat{\phi}_{q}\hat{\phi}_{x_{2}}-\frac{1}{L_{c_{1}1}}\hat{\phi}_{1}\hat{\phi}_{x_{1}}-\frac{1}{L_{c_{2}2}}\hat{\phi}_{2}\hat{\phi}_{x_{2}}-\frac{1}{L_{q1}}\hat{\phi}_{q}\hat{\phi}_{1}-\frac{1}{L_{q2}}\hat{\phi}_{q}\hat{\phi}_{2}\\
    &+8\frac{C_{qc_{1}}}{\sqrt{C_{q}C_{y_{1}}}}\sqrt{E_{Cq}E_{Cy_{1}}}\hat{n}_{q}\hat{n}_{y_{1}}+8\frac{C_{qc_{2}}}{\sqrt{C_{q}C_{y_{2}}}}\sqrt{E_{Cq}E_{Cy_{2}}}\hat{n}_{q}\hat{n}_{y_{2}}\\
    &+8\frac{C_{c_{1}1}}{\sqrt{C_{1}C_{y_{1}}}}\sqrt{E_{C1}E_{Cy_{1}}}\hat{n}_{1}\hat{n}_{y_{1}}+8\frac{C_{c_{2}2}}{\sqrt{C_{1}C_{y_{2}}}}\sqrt{E_{C2}E_{Cy_{2}}}\hat{n}_{2}\hat{n}_{y_{2}}\\
    &+8\left(1+\frac{C_{qc_{1}}C_{c_{1}1}}{C_{q1}C_{y_{1}}}\right)\frac{C_{q1}}{\sqrt{C_{q}C_{1}}}\sqrt{E_{Cq}E_{C1}}\hat{n}_{q}\hat{n}_{1}+8\left(1+\frac{C_{qc_{2}}C_{c_{2}2}}{C_{q2}C_{y_{2}}}\right)\frac{C_{q2}}{\sqrt{C_{q}C_{2}}}\sqrt{E_{Cq}E_{C2}}\hat{n}_{q}\hat{n}_{2}\\
    &+I\left(t\right)\hat{\phi}_{q},
\end{split}
\end{equation}

\noindent with $E_{C\lambda}\equiv\frac{1}{8C_{\lambda}}$. Although the Hamiltonian looks formidable, each line has explicit physical meanings: the first line gives the renormalized qubit energy, the second and third lines give renormalized resonator energies, the fourth and fifth lines give the energy of two charge couplers, the sixth and seventh give the renormalized energies of two flux couplers, the eighth line gives all possible flux couplings, and the ninth to eleventh line give all possible charge couplings. Finally, the last line gives a drive on the qubit. Introducing creation and annihilation operators for each mode and approximating the qubit degrees of freedom by a 2-level system, the above Hamiltonian becomes:

\begin{figure*}
    \includegraphics[width=1\linewidth]{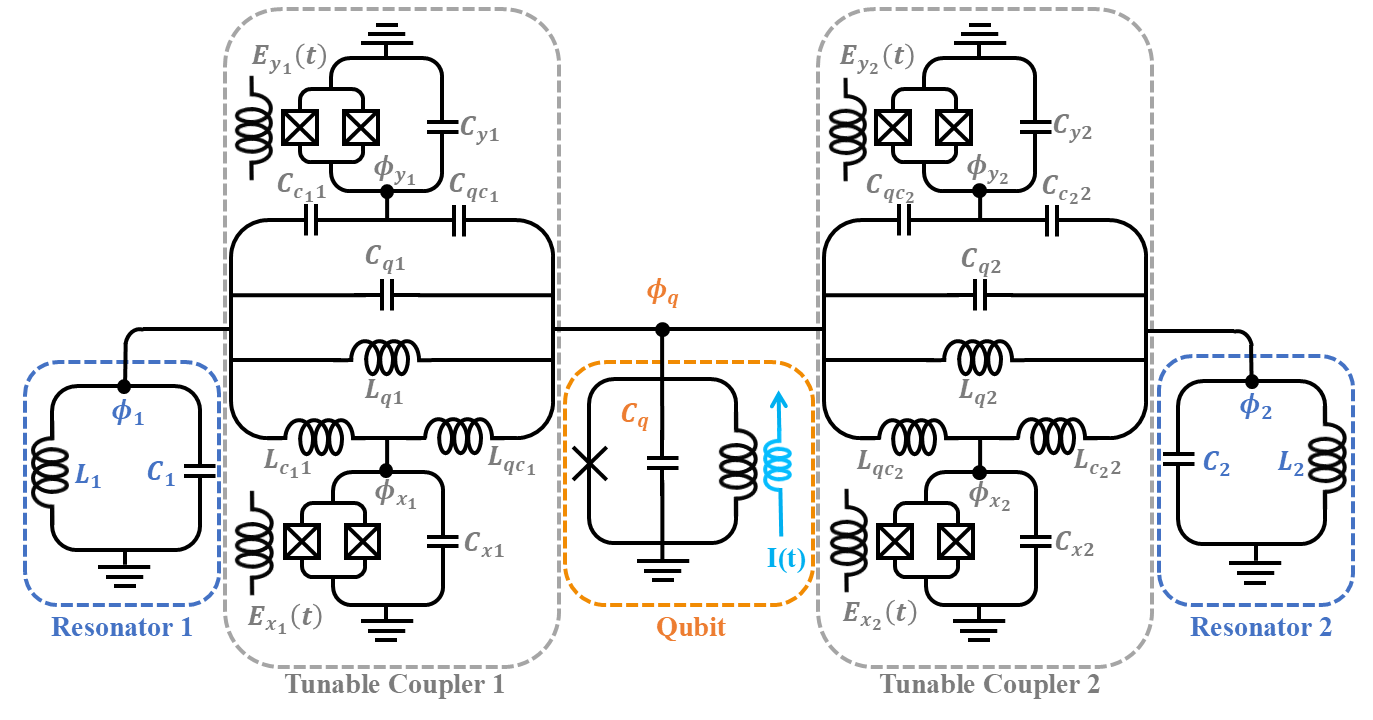}
    \caption{\textit{Superconducting circuit realization of the polarization conversion model.} We propose a superconducting circuit that implements the polarization conversion model [Eq.~\eqref{SC realization}] by tunably coupling two microwave resonators to a superconducting qubit. This tunability is achieved through passive couplings to idler modes under external flux drives. Additionally, we drive the qubit using an external current source $I(t)$, which acts as the qubit energy in the rotating frame.}
    \label{superconducting circuit}
\end{figure*}

\begin{equation}
\begin{split}
    \hat{H}=&\frac{\omega_{q}}{2}\hat{\sigma}_{z}+\omega_{1}\hat{a}_{1}^{\dagger}\hat{a}_{1}+\omega_{2}\hat{a}_{2}^{\dagger}\hat{a}_{2}+\omega_{y_{1}}\hat{b}_{y_{1}}^{\dagger}\hat{b}_{y_{1}}+\omega_{y_{2}}\hat{b}_{y_{2}}^{\dagger}\hat{b}_{y_{2}}+\omega_{x_{1}}\hat{b}_{x_{1}}^{\dagger}\hat{b}_{x_{1}}+\omega_{x_{2}}\hat{b}_{x_{2}}^{\dagger}\hat{b}_{x_{2}}\\
    &+g_{qx_{1}}\hat{\sigma}_{x}\left(\hat{b}_{x_{1}}+\hat{b}_{x_{1}}^{\dagger}\right)+g_{qx_{2}}\hat{\sigma}_{x}\left(\hat{b}_{x_{2}}+\hat{b}_{x_{2}}^{\dagger}\right)+g_{1x_{1}}\left(\hat{a}_{1}+\hat{a}_{1}^{\dagger}\right)\left(\hat{b}_{x_{1}}+\hat{b}_{x_{1}}^{\dagger}\right)\\
    &+g_{2x_{2}}\left(\hat{a}_{2}+\hat{a}_{2}^{\dagger}\right)\left(\hat{b}_{x_{2}}+\hat{b}_{x_{2}}^{\dagger}\right)+g_{q1}^{\left(x\right)}\hat{\sigma}_{x}\left(\hat{a}_{1}+\hat{a}_{1}^{\dagger}\right)+g_{q2}^{\left(x\right)}\hat{\sigma}_{x}\left(\hat{a}_{2}+\hat{a}_{2}^{\dagger}\right)\\
    &+ig_{qy_{1}}\hat{\sigma}_{y}\left(\hat{b}_{x_{1}}-\hat{b}_{x_{1}}^{\dagger}\right)+ig_{qy_{2}}\hat{\sigma}_{y}\left(\hat{b}_{y_{2}}-\hat{b}_{y_{2}}^{\dagger}\right)+g_{1y_{1}}\left(\hat{a}_{1}-\hat{a}_{1}^{\dagger}\right)\left(\hat{b}_{y_{1}}-\hat{b}_{y_{1}}^{\dagger}\right)\\
    &+g_{2y_{2}}\left(\hat{a}_{2}-\hat{a}_{2}^{\dagger}\right)\left(\hat{b}_{y_{2}}-\hat{b}_{y_{2}}^{\dagger}\right)+ig_{q1}^{\left(y\right)}\hat{\sigma}_{y}\left(\hat{a}_{1}-\hat{a}_{1}^{\dagger}\right)+ig_{q2}^{\left(y\right)}\hat{\sigma}_{y}\left(\hat{a}_{2}-\hat{a}_{2}^{\dagger}\right)\\
    &+\sqrt{\frac{\xi_{q}}{2}}I\left(t\right)\hat{\sigma}_{x}+\frac{\alpha_{x_{1}}}{12}\left(\hat{b}_{x_{1}}+\hat{b}_{x_{1}}^{\dagger}\right)^{4}+\frac{\alpha_{x_{2}}}{12}\left(\hat{b}_{x_{2}}+\hat{b}_{x_{2}}^{\dagger}\right)^{4}+\frac{\alpha_{y_{1}}}{12}\left(\hat{b}_{y_{1}}+\hat{b}_{y_{1}}^{\dagger}\right)^{4}+\frac{\alpha_{y_{2}}}{12}\left(\hat{b}_{y_{2}}+\hat{b}_{y_{2}}^{\dagger}\right)^{4},
\end{split}
\label{SC Hamiltonian}
\end{equation}

\noindent with $\omega_{q}$ and $\xi_{g}$ given by the potential $U\left(\phi_{q}\right)$ (i.e.\ the type) of the qubit (for example, this factor has similar form as that of $4$ couplers' if the qubit is transmon-like, or is given by the persistent current in a flux qubit), and are generally obtained by numerical study or direct measurement. In Fig.~\ref{superconducting circuit}, we diagrammatically draw a flux qubit to represent such degrees of freedom. Other parameters, on the other hand, can be written out explicitly:
   
\begin{equation}
\begin{split}
    \omega_{1,2}=&\sqrt{8E_{C1,C2}\left(\frac{1}{L_{1,2}}+\frac{1}{L_{c_{1}1,c_{2}2}}+\frac{1}{L_{q1,q2}}\right)},\\
    \omega_{y_{1,2}}=&\sqrt{8E_{Cy_{1},Cy_{2}}E_{y_{1},y_{2}}},\\
    \omega_{x_{1,2}}=&\sqrt{8E_{Cx_{1},Cx_{2}}\left(E_{x_{1},x_{2}}+\frac{1}{L_{qc_{1},qc_{2}}}+\frac{1}{L_{c_{1}1,c_{2}2}}\right)},\\
    g_{qx_{1},qx_{2}}=&-\frac{\sqrt{\xi_{q}\xi_{x_{1},x_{2}}}}{2L_{qc_{1},qc_{2}}},\\
    g_{1x_{1},2x_{2}}=&-\frac{\sqrt{\xi_{1,2}\xi_{x_{1},x_{2}}}}{2L_{c_{1}1,c_{2}2}},\\
    g_{q1,q2}^{\left(x\right)}=&-\frac{\sqrt{\xi_{q}\xi_{1,2}}}{2L_{q1,q2}},\\
    g_{qy_{1},q_{y2}}=&-4\frac{C_{qc_{1},qc_{2}}}{\sqrt{C_{q}C_{y_{1},y_{2}}}}\sqrt{\frac{E_{Cq}E_{Cy_{1},Cy_{2}}}{\xi_{q}\xi_{y_{1},y_{2}}}},\\
\end{split}
\quad\quad\quad
\begin{split}
    g_{1y_{1},2_{y2}}=&-4\frac{C_{c_{1}1,c_{2}2}}{\sqrt{C_{1,2}C_{y_{1},y_{2}}}}\sqrt{\frac{E_{C1,C2}E_{Cy_{1},Cy_{2}}}{\xi_{1,2}\xi_{y_{1},y_{2}}}},\\   
    g_{q1,q2}^{\left(y\right)}=&-4\left(1+\frac{C_{qc_{1},qc_{2}}}{C_{q1,q2}C_{y_{1}y_{2}}}\right)\frac{C_{q1,q2}}{\sqrt{C_{q}C_{1,2}}}\sqrt{\frac{E_{Cq}E_{C1,C2}}{\xi_{q}\xi_{1,2}}},\\
    \alpha_{y_{1},y_{2}}=&-\frac{\xi_{y_{1},y_{2}}^{2}}{8}E_{y_{1},y_{2}},\\
    \alpha_{x_{1},x_{2}}=&-\frac{\xi_{x_{1},x_{2}}^{2}}{8}E_{x_{1},x_{2}},\\
    \xi_{1,2}=&\sqrt{\frac{8E_{C1,C2}}{\frac{1}{L_{c_{1}1,c_{2}2}}+\frac{1}{L_{q1,q2}}}},\\
    \xi_{y_{1},y_{2}}=&\sqrt{\frac{8E_{Cy_{1},Cy_{2}}}{E_{y_{1},y_{2}}}},\\
    \xi_{x_{1},x_{2}}=&\sqrt{\frac{8E_{Cx_{1},Cx_{2}}}{E_{x_{1},x_{2}}+\frac{1}{L_{qc_{1},qc_{2}}}+\frac{1}{L_{c_{1}1,c_{2}2}}}}.
\end{split}
\end{equation}

\noindent We now apply a Schrieffer-Wolff transform $\hat{U}_{SW}=e^{\hat{S}_{x}+\hat{S}_{y}}$ on the Hamiltonian [Eq.~\eqref{SC Hamiltonian}], which is defined as:

\begin{equation}
\begin{split}
    \hat{S}_{x}=&\frac{1}{2}\frac{g_{qx_{1}}}{\omega_{q}-\omega_{x_{1}}}\left(\hat{\sigma}^{+}\hat{b}_{x_{1}}-\hat{\sigma}^{-}\hat{b}_{x_{1}}^{\dagger}\right)+\frac{1}{2}\frac{g_{qx_{2}}}{\omega_{q}-\omega_{x_{2}}}\left(\hat{\sigma}^{+}\hat{b}_{x_{2}}-\hat{\sigma}^{-}\hat{b}_{x_{2}}^{\dagger}\right)\\
    &+\frac{1}{2}\frac{g_{qx_{1}}}{\omega_{q}+\omega_{x_{1}}}\left(\hat{\sigma}^{+}\hat{b}_{x_{1}}^{\dagger}-\hat{\sigma}^{-}\hat{b}_{x_{1}}\right)+\frac{1}{2}\frac{g_{qx_{2}}}{\omega_{q}+\omega_{x_{2}}}\left(\hat{\sigma}^{+}\hat{b}_{x_{2}}^{\dagger}-\hat{\sigma}^{-}\hat{b}_{x_{2}}\right)\\
    &+\frac{g_{1x_{1}}}{\omega_{1}-\omega_{x_{1}}}\left(\hat{a}_{1}^{\dagger}\hat{b}_{x_{1}}-\hat{a}_{1}\hat{b}_{x_{1}}^{\dagger}\right)+\frac{g_{2x_{2}}}{\omega_{2}-\omega_{x_{2}}}\left(\hat{a}_{2}^{\dagger}\hat{b}_{x_{2}}-\hat{a}_{2}\hat{b}_{x_{2}}^{\dagger}\right)\\
    &+\frac{g_{1x_{1}}}{\omega_{1}+\omega_{x_{1}}}\left(\hat{a}_{1}^{\dagger}\hat{b}_{x_{1}}^{\dagger}-\hat{a}_{1}\hat{b}_{x_{1}}\right)+\frac{g_{2x_{2}}}{\omega_{2}+\omega_{x_{2}}}\left(\hat{a}_{2}^{\dagger}\hat{b}_{x_{2}}^{\dagger}-\hat{a}_{2}\hat{b}_{x_{2}}\right),\\
    \hat{S}_{y}=&\frac{1}{2}\frac{g_{qy_{1}}}{\omega_{q}-\omega_{y_{1}}}\left(\hat{\sigma}^{+}\hat{b}_{y_{1}}-\hat{\sigma}^{-}\hat{b}_{y_{1}}^{\dagger}\right)+\frac{1}{2}\frac{g_{qy_{2}}}{\omega_{q}-\omega_{y_{2}}}\left(\hat{\sigma}^{+}\hat{b}_{y_{2}}-\hat{\sigma}^{-}\hat{b}_{y_{2}}^{\dagger}\right)\\
    &-\frac{1}{2}\frac{g_{qy_{1}}}{\omega_{q}+\omega_{y_{1}}}\left(\hat{\sigma}^{+}\hat{b}_{y_{1}}^{\dagger}-\hat{\sigma}^{-}\hat{b}_{y_{1}}\right)-\frac{1}{2}\frac{g_{qy_{2}}}{\omega_{q}+\omega_{y_{2}}}\left(\hat{\sigma}^{+}\hat{b}_{y_{2}}^{\dagger}-\hat{\sigma}^{-}\hat{b}_{y_{2}}\right)\\
    &-\frac{g_{1y_{1}}}{\omega_{1}-\omega_{y_{1}}}\left(\hat{a}_{1}^{\dagger}\hat{b}_{y_{1}}-\hat{a}_{1}\hat{b}_{y_{1}}^{\dagger}\right)-\frac{g_{2y_{2}}}{\omega_{2}-\omega_{y_{2}}}\left(\hat{a}_{2}^{\dagger}\hat{b}_{y_{2}}-\hat{a}_{2}\hat{b}_{y_{2}}^{\dagger}\right)\\
    &+\frac{g_{1y_{1}}}{\omega_{1}+\omega_{y_{1}}}\left(\hat{a}_{1}^{\dagger}\hat{b}_{y_{1}}^{\dagger}-\hat{a}_{1}\hat{b}_{y_{1}}\right)+\frac{g_{2y_{2}}}{\omega_{2}+\omega_{y_{2}}}\left(\hat{a}_{2}^{\dagger}\hat{b}_{y_{2}}^{\dagger}-\hat{a}_{2}\hat{b}_{y_{2}}\right),
\end{split}
\end{equation}

\noindent with $\hat{\sigma}^{\pm}=\hat{\sigma}_{x}\pm i\hat{\sigma}_{y}$. By assuming that $4$ couplers remain in their ground states (i.e.\ the coupling is dispersive) and small nonlinearities $\left|\alpha_{x_{1},x_{2},y_{1},y_{2}}\right|\ll\left|\omega_{q,1,2}\pm\omega_{x_{1},x_{2},y_{1},y_{2}}\right|$, dropping all fast-oscillating terms, we finally obtain an effective Hamiltonian between the driven qubit and two resonators:

\begin{equation}
\begin{split}
   \hat{H}_{eff}=&\frac{\omega_{q}}{2}\hat{\sigma}_{z}+\sqrt{\frac{\xi_{q}}{2}}I\left(t\right)\hat{\sigma}_{x}+\omega_{1}\hat{a}_{1}^{\dagger}\hat{a}_{1}+\omega_{2}\hat{a}_{2}^{\dagger}\hat{a}_{2}\\
   &+\Tilde{g}_{q1}^{\left(x\right)}\hat{\sigma}_{x}\left(\hat{a}_{1}+\hat{a}_{1}^{\dagger}\right)+\Tilde{g}_{q2}^{\left(x\right)}\hat{\sigma}_{x}\left(\hat{a}_{2}+\hat{a}_{2}^{\dagger}\right)+i\Tilde{g}_{q1}^{\left(y\right)}\hat{\sigma}_{y}\left(\hat{a}_{1}-\hat{a}_{1}^{\dagger}\right)+i\Tilde{g}_{q2}^{\left(y\right)}\hat{\sigma}_{y}\left(\hat{a}_{2}-\hat{a}_{2}^{\dagger}\right),
\end{split}
\end{equation}

\noindent where 

\begin{equation}
\begin{split}
    \Tilde{g}_{q1}^{\left(x\right)}=&g_{q1}^{\left(x\right)}+\frac{1}{2}g_{qx_{1}}g_{1x_{1}}\left[\left(\frac{1}{\omega_{q}-\omega_{x_{1}}}-\frac{1}{\omega_{q}+\omega_{x_{1}}}\right)+\left(\frac{1}{\omega_{1}-\omega_{x_{1}}}-\frac{1}{\omega_{1}+\omega_{x_{1}}}\right)\right],\\
    \Tilde{g}_{q2}^{\left(x\right)}=&g_{q2}^{\left(x\right)}+\frac{1}{2}g_{qx_{2}}g_{1x_{2}}\left[\left(\frac{1}{\omega_{q}-\omega_{x_{2}}}-\frac{1}{\omega_{q}+\omega_{x_{2}}}\right)+\left(\frac{1}{\omega_{2}-\omega_{x_{2}}}-\frac{1}{\omega_{2}+\omega_{x_{2}}}\right)\right],\\
    \Tilde{g}_{q1}^{\left(y\right)}=&g_{q1}^{\left(y\right)}-\frac{1}{2}g_{qy_{1}}g_{1y_{1}}\left[\left(\frac{1}{\omega_{q}-\omega_{y_{1}}}-\frac{1}{\omega_{q}+\omega_{y_{1}}}\right)+\left(\frac{1}{\omega_{1}-\omega_{y_{1}}}-\frac{1}{\omega_{1}+\omega_{y_{1}}}\right)\right],\\
    \Tilde{g}_{q2}^{\left(y\right)}=&g_{q2}^{\left(y\right)}-\frac{1}{2}g_{qy_{2}}g_{1y_{2}}\left[\left(\frac{1}{\omega_{q}-\omega_{y_{2}}}-\frac{1}{\omega_{q}+\omega_{y_{2}}}\right)+\left(\frac{1}{\omega_{2}-\omega_{y_{2}}}-\frac{1}{\omega_{2}+\omega_{y_{2}}}\right)\right].
\end{split}
\end{equation}

\noindent By threading time-dependent external flux in the dc-SQUIDs of $4$ couplers, we can make $4$ couplers' frequencies time-dependent $\omega_{x_{1},x_{2},y_{1},y_{2}}\left(t\right)$, and hence $4$ time-dependent couplings $\Tilde{g}_{q1,q2}^{\left(x,y\right)}\left(t\right)$. Note that we leave a time-independent coupling $g_{q1,q2}^{\left(x,y\right)}$ in each of the couplings, such that the offset can be tuned exactly to zero. Setting $\omega_{1}=\omega_{2}=\omega$ and 

\begin{equation}
    \begin{split}
        \Tilde{g}_{q1}^{\left(x\right)}\left(t\right)=&\Tilde{g}_{q2}^{\left(x\right)}\left(t\right)=\mathcal{A}\cos\left(\omega_{q}t\right),\\
        \Tilde{g}_{q1}^{\left(y\right)}\left(t\right)=&-\Tilde{g}_{q2}^{\left(y\right)}\left(t\right)=-\mathcal{A}\cos\left(\omega_{q}t\right),\\
        \sqrt{\frac{\xi_{q}}{2}}I\left(t\right)=&\omega_{0}\cos\left(\omega_{q}t\right),
    \end{split}
\end{equation}

\noindent we recover the Hamiltonian in Eq.~\eqref{SC realization}.

\twocolumngrid
%


\end{document}